\documentclass[a4paper,11pt]{article}
\pdfoutput=0 % if your are submitting a pdflatex (i.e. if you have
\usepackage{jheppub} % for details on the use of the package, please
\usepackage[T1]{fontenc} % if needed

\newcommand{\bfm}[1]{\mbox{\boldmath$#1$}}

\newcommand{\gsim}{\;\rlap{\lower 3.5 pt \hbox{$\mathchar \sim$}} \raise 1pt
\hbox {$>$}\;}
\newcommand{\lsim}{\;\rlap{\lower 3.5 pt \hbox{$\mathchar \sim$}} \raise 1pt
\hbox {$<$}\;}

\title{\boldmath High-Energy  Limit of Mass-Suppressed Amplitudes in Gauge Theories}

\preprint{ALBERTA-THY-06-18}

\author[a]{Tao Liu,}
\author[a,b]{Alexander Penin}
\affiliation[a]{Department of Physics, University of Alberta,
Edmonton AB T6G 2J1, Canada
}
\affiliation[b]{Institute for Theoretical Physics, ETH Z\"urich, 8093 Z\"urich, Switzerland
}
\emailAdd{ltao@ualberta.ca}
\emailAdd{penin@itp.phys.ethz.ch}
\abstract{We present a detailed analysis of the factorization and  all-order
resummation of the double-logarithmic  radiative corrections which determine
the asymptotic behavior of the  gauge theory  amplitudes suppressed by the
leading power of the fermion mass in the limit of  high-energy fixed-angle
scattering. The result is applied to estimate the bottom quark mediated
contribution to the  Higgs boson  production in gluon fusion.}

\begin{document}
\maketitle
\flushbottom

\section{Introduction}
A distinct feature of the gauge theory scattering amplitudes in the
high-energy limit is the presence of the ``Sudakov'' radiative corrections
enhanced by the second power of the large logarithm of the energy ratio to
a characteristic infrared scale of the process per each power of the coupling
constant. These double-logarithmic corrections determine the leading
deviation in the asymptotic behavior of the quantum field theory  amplitudes
from the classical result. Since the original work~\cite{Sudakov:1954sw} on
the double-logarithmic approximation of the electron form factor  in QED the
analysis has been extended to  nonabelian gauge theories and to subleading
logarithms
\cite{Frenkel:1976bj,Smilga:1979uj,Mueller:1979ih,Collins:1980ih,Sen:1981sd,Sterman:1986aj}.
Sudakov logarithms exponentiate and result in a strong universal suppression
of the scattering amplitudes in the limit when all the kinematic invariants
of the process are large. This analysis however does not extend to the
amplitudes suppressed by a power of an infrared scale in the high energy
limit. The power-suppressed  contributions now attract  a lot of attention
in various contexts (see {\it e.g.}
\cite{Ferroglia:2009ep,Laenen:2010uz,Banfi:2013eda,Becher:2013iya,deFlorian:2014vta,
Anastasiou:2014lda,Penin:2014msa,Almasy:2015dyv,Melnikov:2016emg,Penin:2016wiw,
Bonocore:2016awd,Boughezal:2016zws,Moult:2017rpl,Liu:2017axv,Liu:2017vkm,Beneke:2017ztn,
Boughezal:2018mvf,Bruser:2018jnc,Moult:2018jjd,Alte:2018nbn,Ebert:2018lzn,Beneke:2018rbh}).
Incorporating the logarithmically enhanced power-suppressed terms can
significantly increase the accuracy and extend the region where  the
leading-power approximation is applicable. Besides their phenomenological
importance these contributions are very interesting from the general
effective field theory point of view since the structure of the
renormalization group evolution in this case becomes highly nontrivial
already in the leading logarithmic approximation.

We focus on the double-logarithmic corrections to the amplitudes suppressed
by the leading power of the fermion mass. In general very little is known  so
far about the  all-order structure of such corrections. In contrast to
Sudakov logarithms they  do not exponentiate and do not factorize into the
wave functions of scattering particle.   While the mass effects on  the
leading-power  contributions have been extensively studied in the context of
the high-order electroweak and QED radiative
corrections~\cite{Feucht:2004rp,Jantzen:2005az,Penin:2005eh,Penin:2005kf,
Bonciani:2007eh,Bonciani:2008ep,Kuhn:2007ca,Kuhn:2011mh,Penin:2011aa}, a few
known examples of  the all-order resummation for the power-suppressed terms
are restricted to  abelian gauge theory
\cite{Penin:2014msa,Melnikov:2016emg,Gorshkov:1966ht,Kotsky:1997rq}.
Extension of the analysis to QCD is not straightforward and requires a
systematic treatment of the factorization. Only recently the first QCD result
in the field has been reported  in a short  letter \cite{Liu:2017vkm}. Below
we present a detailed account of this analysis.

The paper is organized as follows. In the next section we use a simple
example of quark scattering by a scalar color-singlet gluon field operator to
outline the method and to derive the factorization formula for the
mass-suppressed double-logarithmic corrections. In Sect.~\ref{sec::3} we
apply the method to the analysis of the Higgs boson production in gluon
fusion mediated by a bottom quark loop. In Sect.~\ref{sec::4} we derive the
asymptotic behaviour of the leading power corrections to various massive
quark form factors. The universality of our solution for different amplitudes
and gauge models as well as the phenomenological applications are discussed
in Sect.~\ref{sec::5}.

\section{Massive quark scattering by a  gluon field operator}
\label{sec::2}
Throughout  this paper we deal with  a massive quark scattering  by various
external currents. To introduce the main idea  of our approach we start with
an amplitude ${\cal G}$ for the  scattering of a quark of mass $m_q$, initial
momentum $p_1$ and final momentum $p_2$,  by a local operator
$(G^a_{\mu\nu})^2$ of the gauge field strength tensor. The origin of such a
vertex is not relevant for our discussion and one may suggest  that it
describes the gluon field interaction to the Higgs boson mediated by an
infinitely heavy quark loop. This rather artificial amplitude  is a perfect
example to reveal the main features of the  general problem in the most
illustrative way and with minimal technical complications.

\subsection{The leading-order amplitude}
\label{sec::2.1}
\begin{figure}
\begin{center}
\begin{tabular}{cccc}
\includegraphics[width=1.5cm]{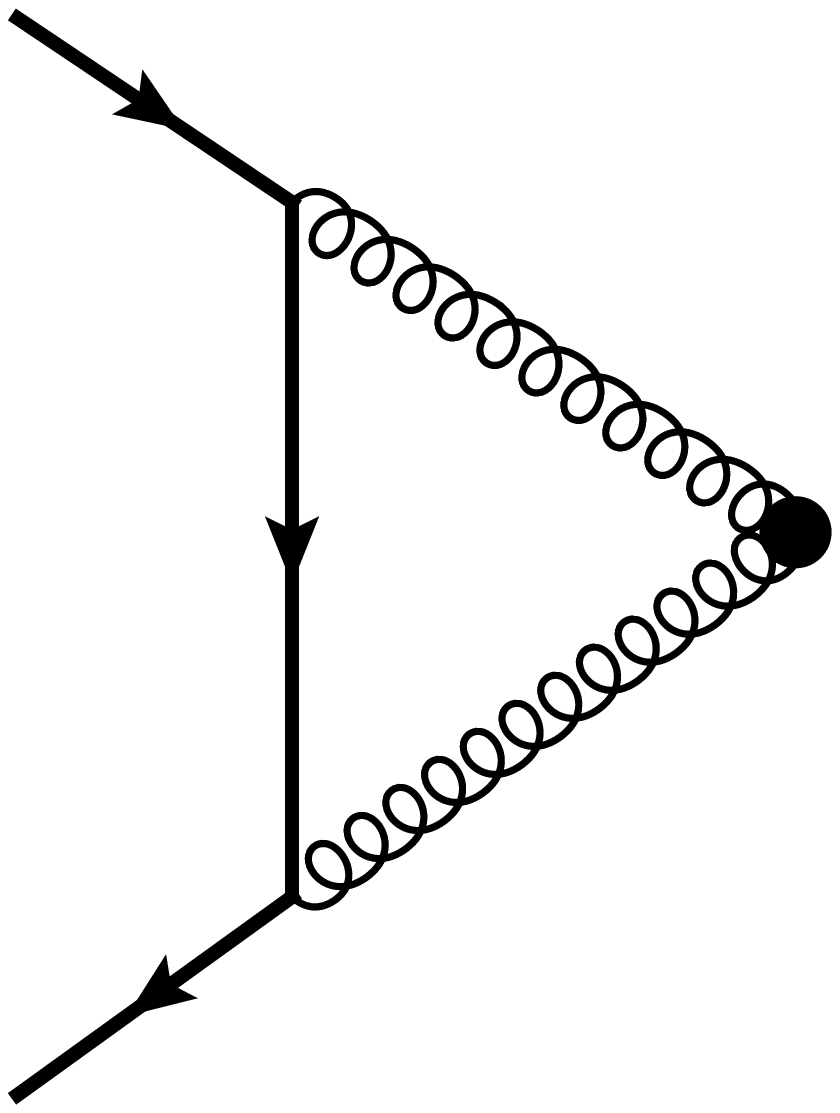}&
\hspace*{03mm}\includegraphics[width=1.5cm]{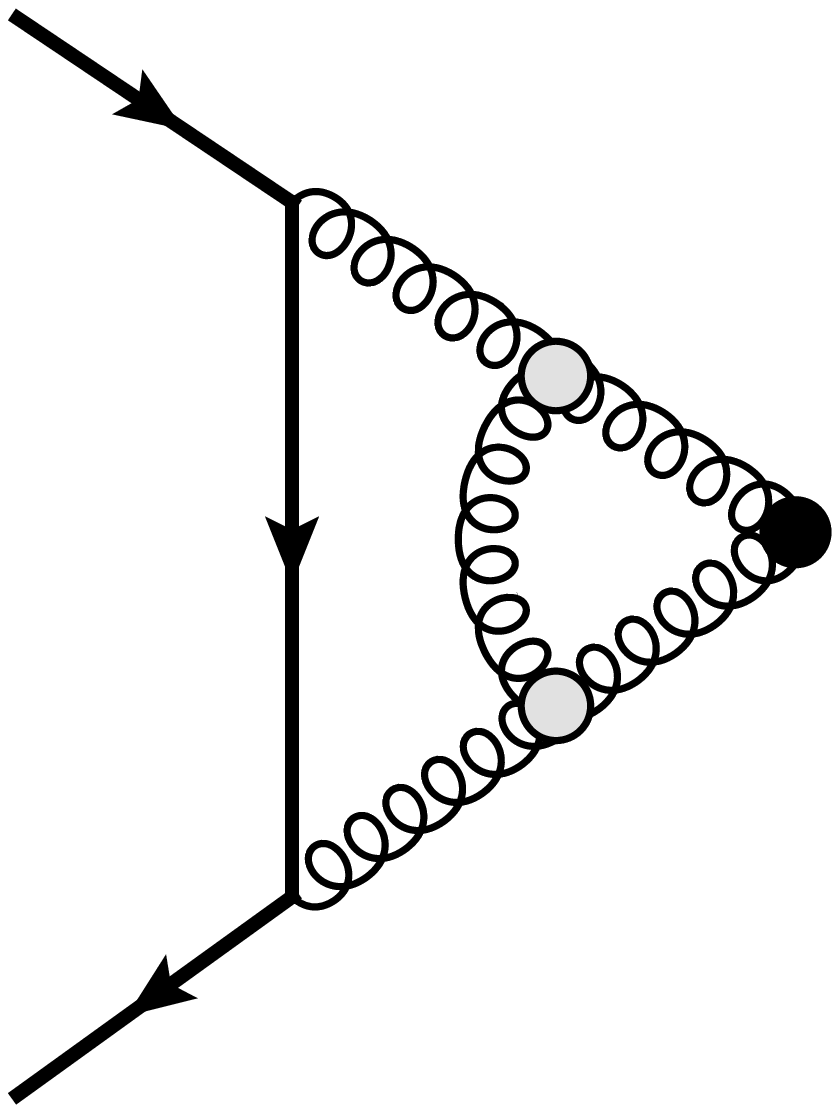}&
\hspace*{03mm}\includegraphics[width=1.8cm]{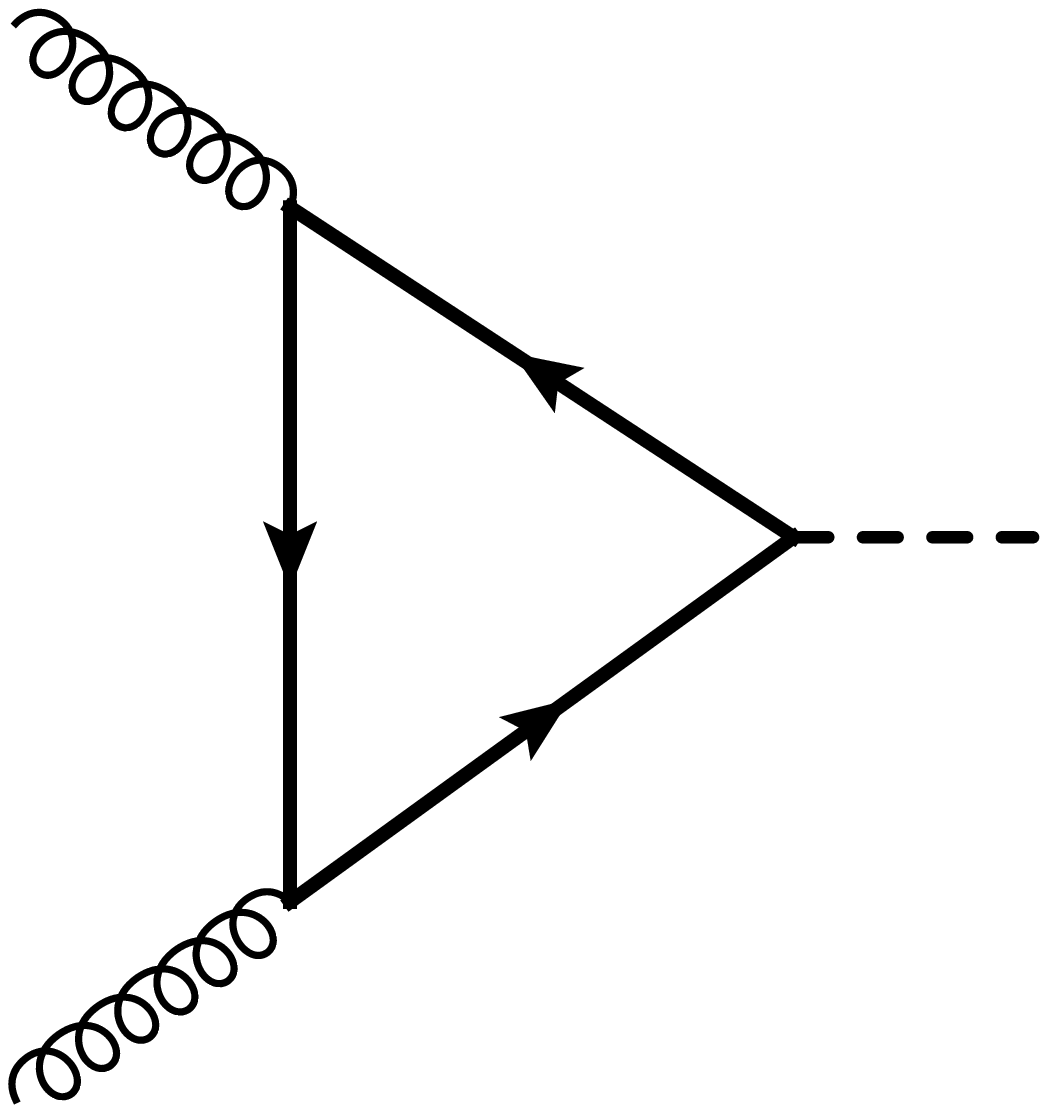}&
\hspace*{03mm}\includegraphics[width=1.8cm]{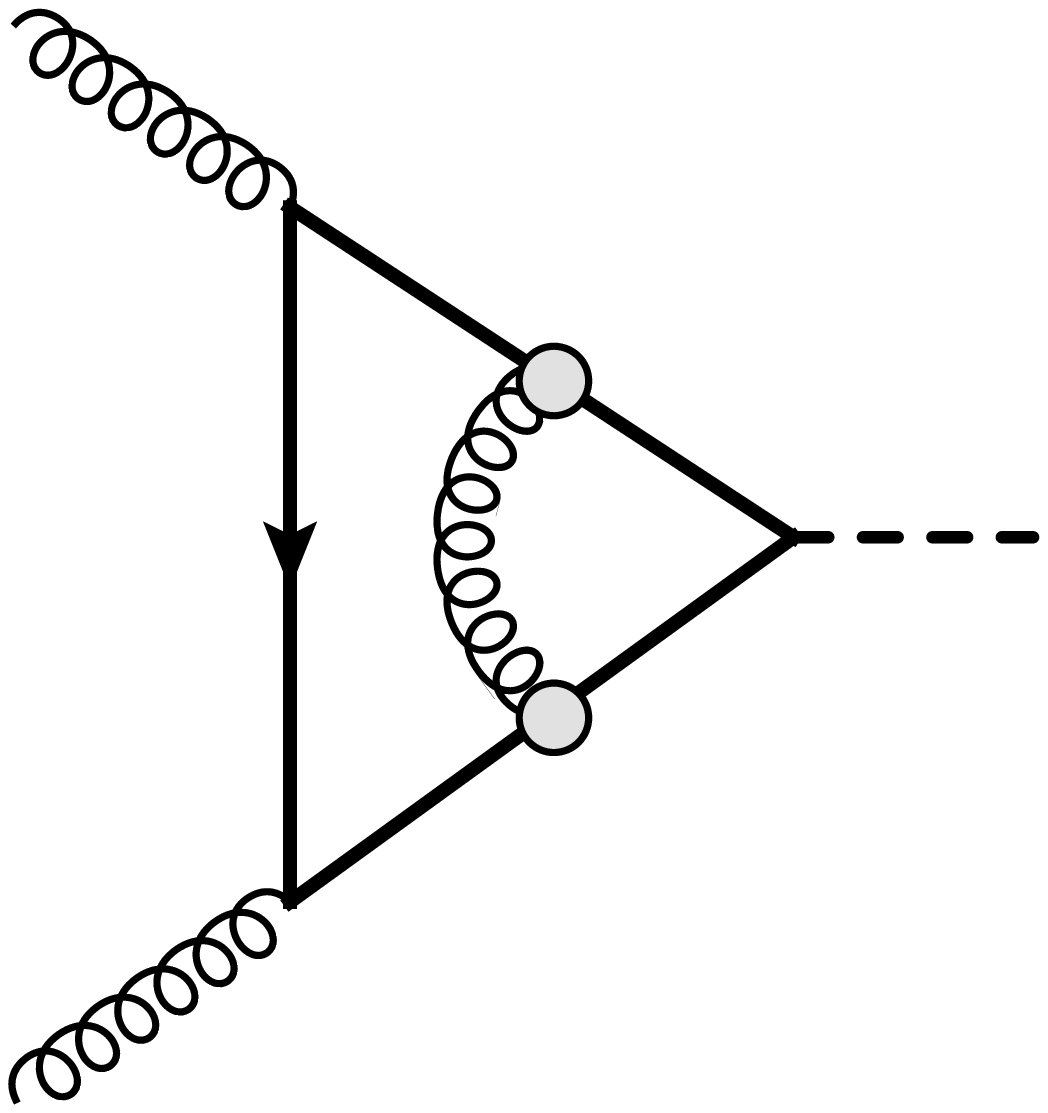}\\
(a)&\hspace*{03mm}(b)&\hspace*{03mm}(c)& \hspace*{03mm} (d)\\
\end{tabular}
\end{center}
\caption{\label{fig::1}  The leading-order one-loop Feynman diagrams for  (a)
quark scattering by the $(G_{\mu\nu}^a)^2$ vertex (black circle) and (c) the
Higgs boson production in gluon fusion. The diagrams (b) and (d) with the
effective vertices (gray circles) defined in the text represent the
non-Sudakov double-logarithmic corrections to the process  (a) and (c),
respectively.}
\end{figure}

We consider the limit of the on-shell quark $p_1^2=p_2^2=m_q^2$ and the large
Euclidean momentum transfer $Q^2=-(p_2-p_1)^2$ when the ratio $\rho\equiv
{m_q^2/Q^2}$ is positive and small.  In the light-cone coordinates
$p_1\approx p_1^-$ and $p_{2}\approx p_2^+$.  The leading-order scattering is
given by the one-loop diagram in Fig.~\ref{fig::1}(a). Conservation of
helicity  at high energy requires a helicity flip on the virtual quark line.
As a consequence at high energy the amplitude  is suppressed  by the first
power of $m_q$. The virtual quark propagator then can be approximated  as
follows $S(l)\approx{m_q\over l^2-m_q^2}$. Thus, the one-loop integral
reduces to
\begin{equation}
{2iQ^2\over \pi^2 }\int{{d^4l}\over (l^2-m_q^2) (p_1+l)^2
(p_2 +l)^2}\,,
\label{eq::intl}
\end{equation}
where the prefactor is introduce for convenience.  For the soft quark
momentum $m_q\ll l\ll Q$   the  gauge boson propagators are eikonal {\it
i.e.} proportional to ${1\over 2p_il}$, and the integral has  the
double-logarithmic scaling. To evaluate the double-logarithmic contribution
the propagators can be approximated as follows \cite{Sudakov:1954sw}
\begin{eqnarray}
&& {1\over l^2-m_q^2}  \approx
- i \pi \delta(Q^2uv + {l}_\perp^2-m_q^2)\,,
\nonumber\\
&&{1\over (p_1+l)^2}\approx \frac{1}{Q^2v}\,,
\nonumber\\
&&{1\over (p_2 +l)^2} \approx \frac{1}{Q^2u}\,,
\label{eq::prop}
\end{eqnarray}
where we introduce  the standard Sudakov parametrization  of the  soft quark
momentum $l=up_1+vp_2+l_\perp$. The validity of the eikonal approximation in
Eq.~(\ref{eq::prop}) requires $|u|,|v|<1$ and the additional kinematical
constraints $uv>\rho$  has to be imposed to ensure that the soft quark
propagator can go on the mass shell. After integrating Eq.~(\ref{eq::intl})
over ${l}_\perp$  with the double-logarithmic accuracy we get
\begin{equation}
2\int_{\rho}^{1}{{\rm d}v\over v}
\int_{\rho/v}^{1}{{\rm d}u\over u}=2\ln^2\!\rho\int_0^1
{\rm d}\xi \int_{0}^{1-\xi}{\rm d}\eta
=\ln^2\!\rho\,,
\label{eq::intluv}
\end{equation}
where the normalized logarithmic variables read $\eta=\ln v/\ln\rho $,
$\xi=\ln u/\ln\rho$.  This defines the leading order amplitude
\begin{equation}
{\cal G}^0= 2C_F x m_q\,\bar{q}q\,,
\label{eq::G0}
\end{equation}
where $x={\alpha_s\over 4\pi}\ln^2\!\rho$ is a double-logarithmic variable,
$C_F=(N_c^2-1)/(2N_c)$ is the quadratic Casimir operator of the fundamental
representation of the $SU(N_c)$ color group,  and $\alpha_s$ is the strong
coupling constant. Thus we have a typical situation when a soft quark
exchange generates the double-logarithmic contribution to the mass-suppressed
amplitude. As we see, the emission of the soft quark results in the change of
the  color group representation of a particle propagating along the eikonal
line, or  the {\it eikonal color charge nonconservation}. This is a crucial
feature of the process which plays an important role in further analysis.

\subsection{Factorization of  the double-logarithmic corrections}
\label{sec::2.2}
\begin{figure}
\begin{center}
\begin{tabular}{cccc}
\includegraphics[width=1.7cm]{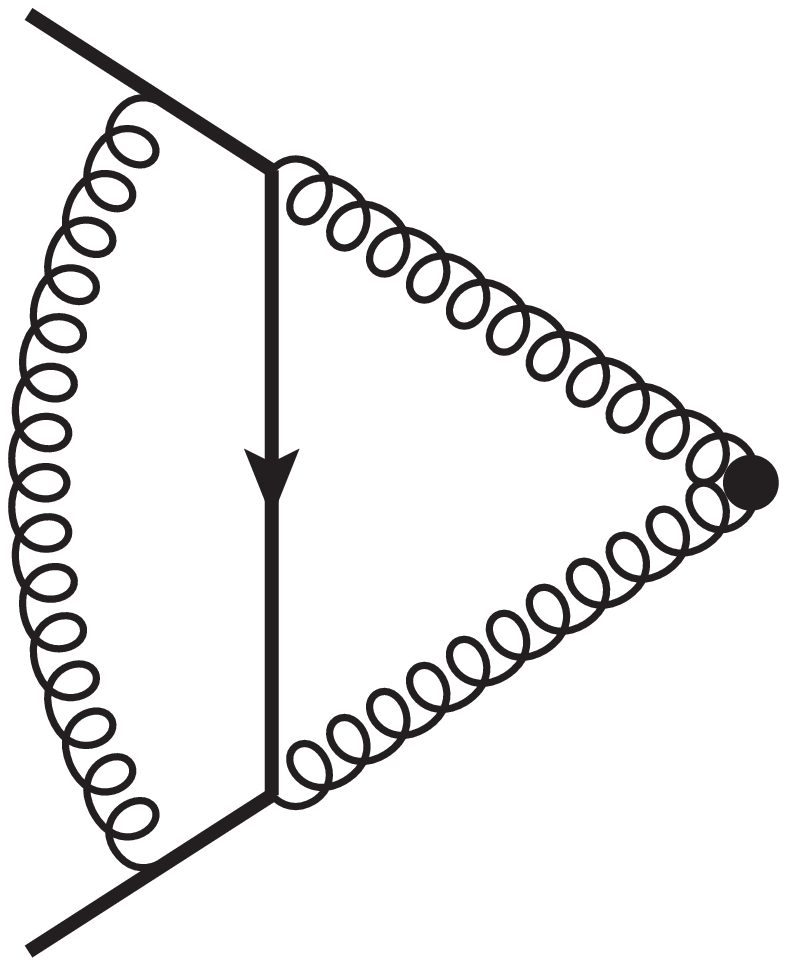}&
\hspace*{03mm}\includegraphics[width=1.7cm]{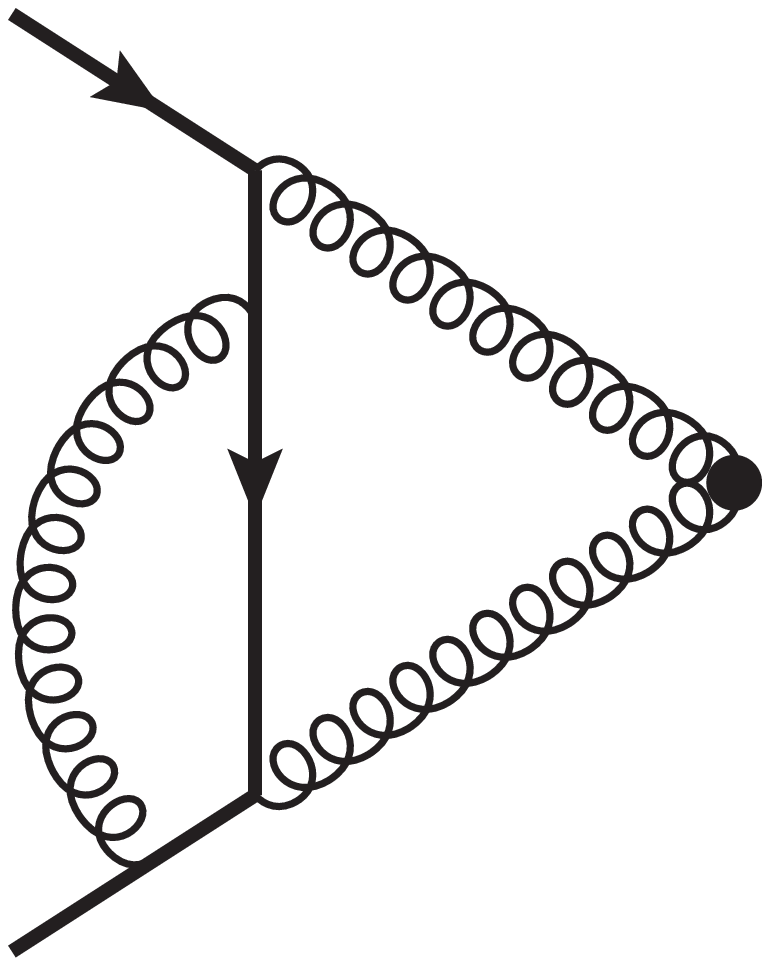}&
\hspace*{03mm}\includegraphics[width=1.7cm]{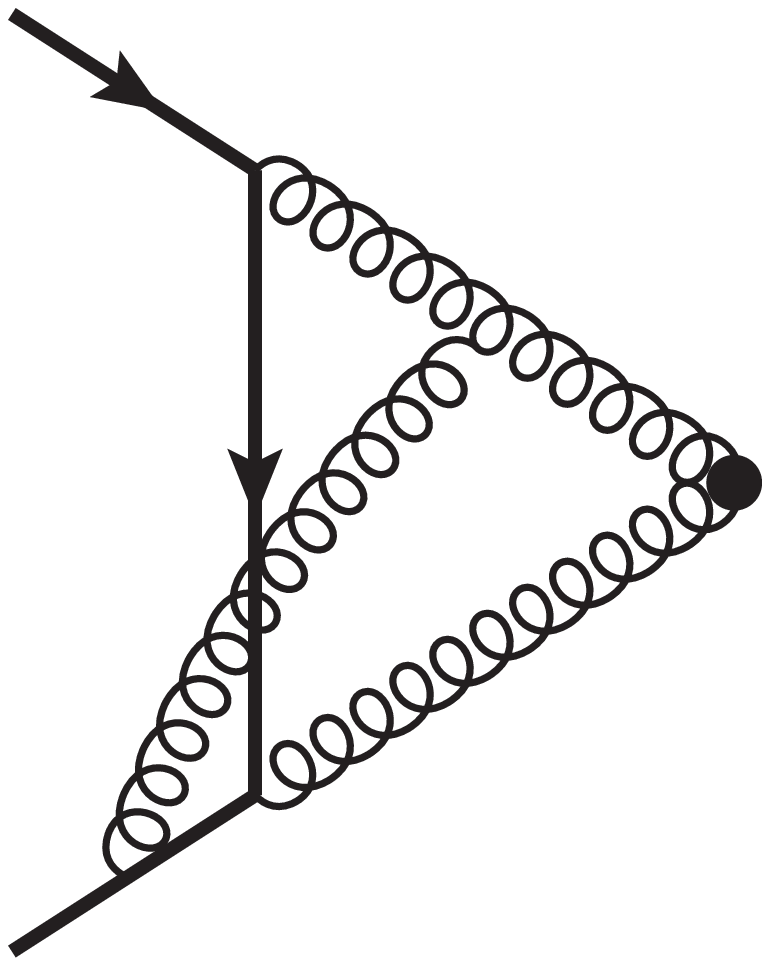}&
\hspace*{03mm}\includegraphics[width=1.7cm]{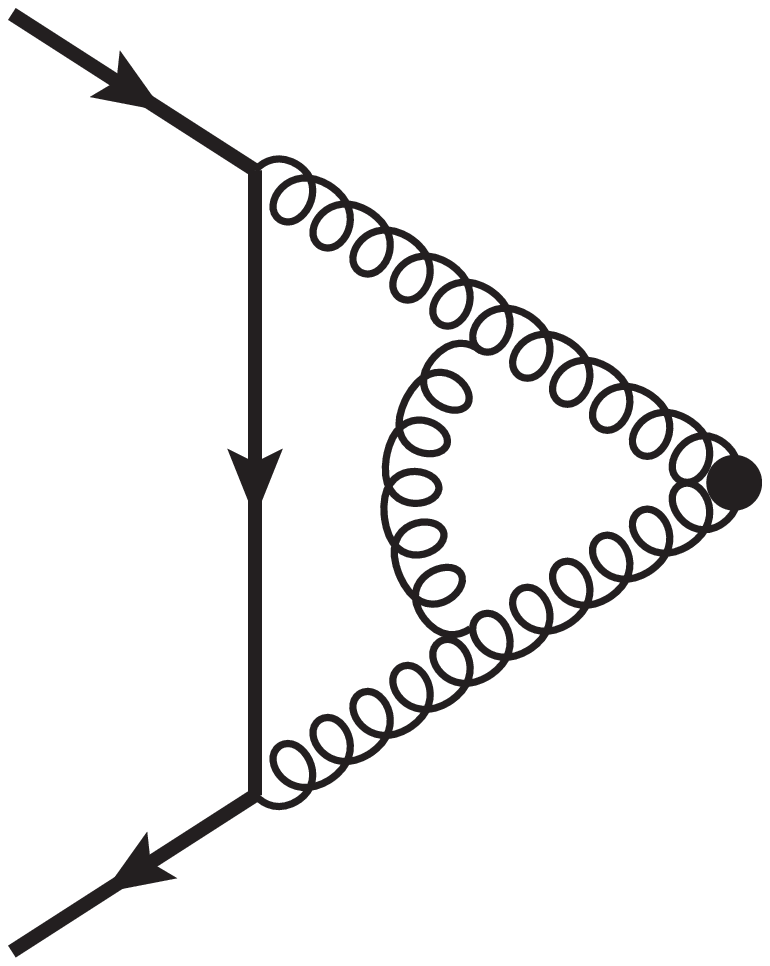}\\
(a)&\hspace*{03mm}(b)&\hspace*{03mm}(c)& \hspace*{03mm} (d)\\
\end{tabular}
\end{center}
\caption{\label{fig::2}  The two-loop  Feynman diagrams for the quark scattering
by the $(G_{\mu\nu}^a)^2$ vertex (black circle) which contribute in the
double-logarithmic approximation. Symmetric diagrams are not shown.}
\end{figure}

\begin{figure}
\begin{center}
\begin{tabular}{cccc}
\includegraphics[width=1.5cm]{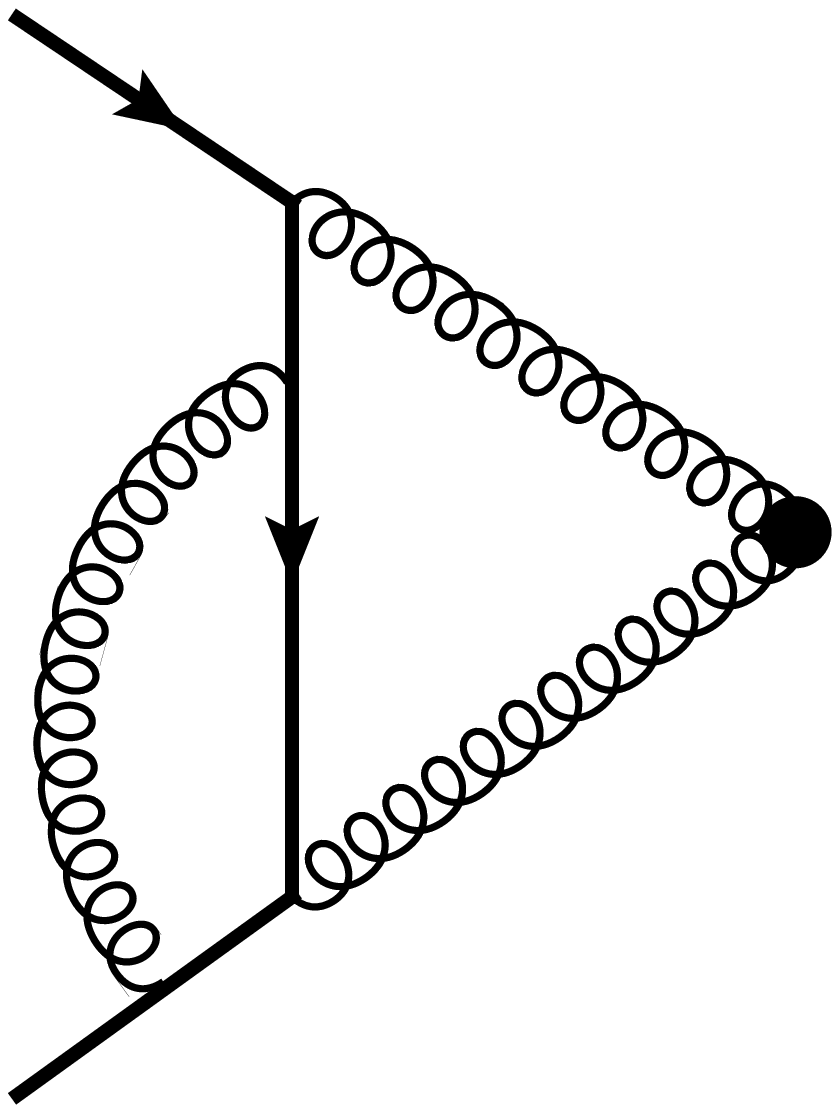}~\raisebox{9.5mm}{$\bfm{\to}$}&
\hspace*{00mm}\includegraphics[width=1.5cm]{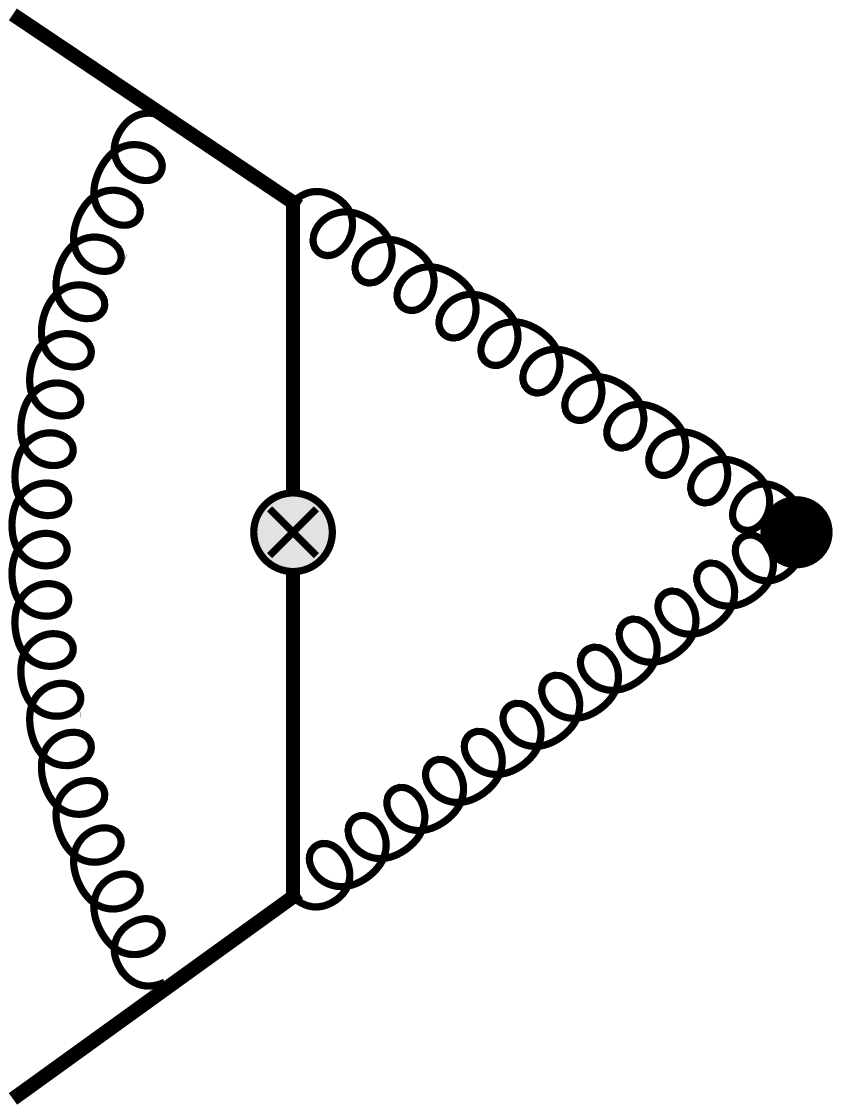}~\raisebox{9.5mm}{$\bfm{\to}$}&
\hspace*{00mm}\includegraphics[width=1.5cm]{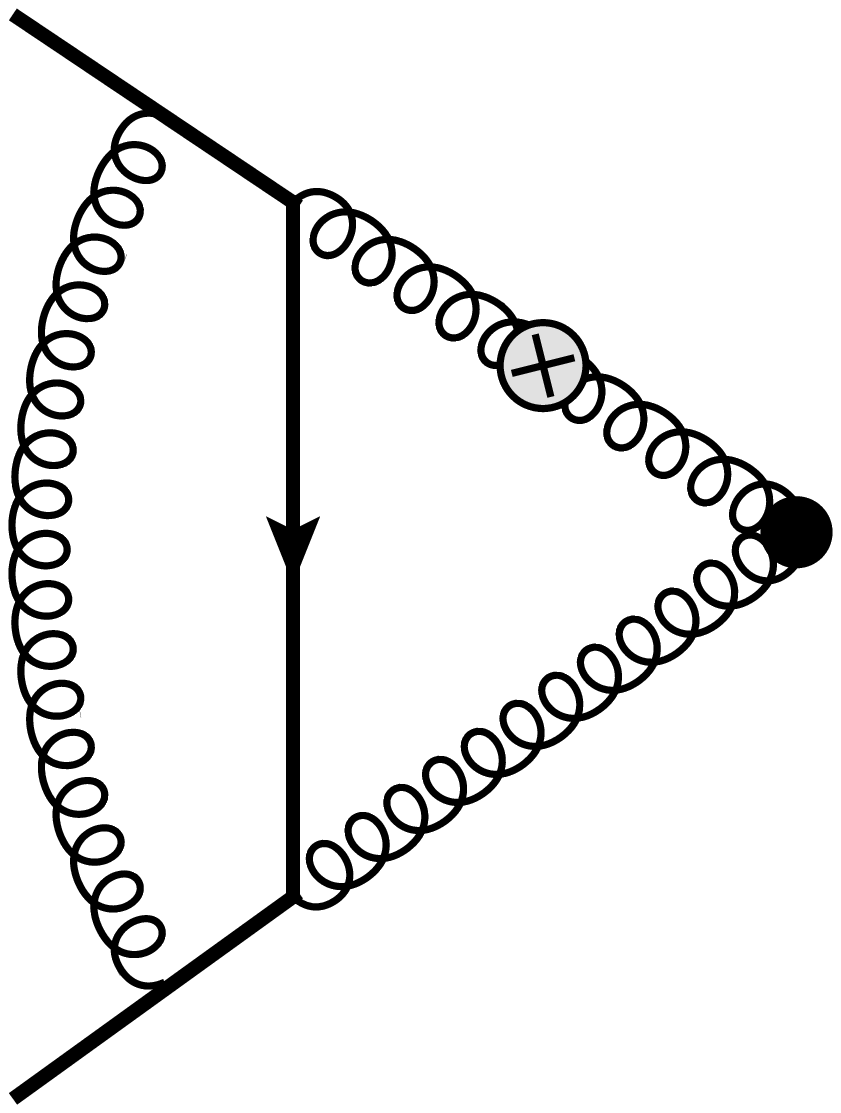}~\raisebox{9.5mm}{$\bfm{\to}$}&
\hspace*{00mm}\includegraphics[width=1.5cm]{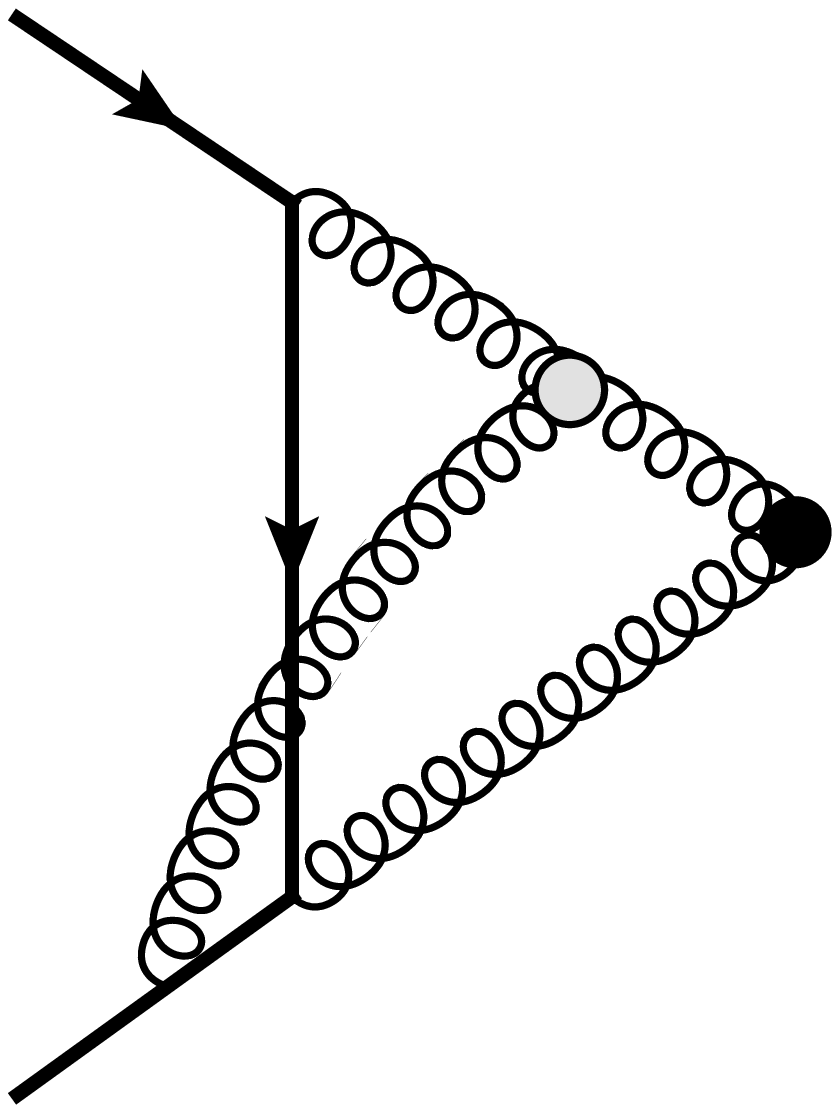}\\
(a)&\hspace*{00mm}(b)&\hspace*{00mm}(c)& \hspace*{00mm} (d)\\
\end{tabular}
\end{center}
\caption{\label{fig::3}  Diagramatic representation of the sequence of
identities which move the soft gauge boson vertex from the soft quark to the
eikonal gauge boson line, as explained in the text.}
\end{figure}

We start with the two-loop radiative corrections. In a covariant gauge the
two-loop double-logarithmic  contributions  are produced by the  Feynman diagrams
in Fig.~\ref{fig::2}. Let us consider first the abelian case of the
photon interaction corresponding to the
diagrams in  Figs.~\ref{fig::2}(a,b). The key  idea of our  approach is to
move the soft photon vertex from the virtual soft quark line in
Fig.~\ref{fig::2}(a) to an eikonal photon line through a sequence of
identities graphically represented in Fig.~\ref{fig::3}. Let us describe this
sequence in more detail. In a covariant gauge only $A^-$  light-cone
component of the photon field can be emitted by the eikonal quark line with
the momentum $p_{2}$, while the emission of the $A^+$ and transverse
components is suppressed. Since $A^-$ is not a physical polarization its
interaction to the quark line is completely determined by  gauge
invariance. For the soft quark line in Fig.~\ref{fig::3}(a) we therefore can
write the following relation
\begin{equation}
S(l)\gamma^{\mu}S(l+l_g)\approx S(l)\gamma^-S(l+l_g^+)={1\over l_g^+}
\left(S(l)-S(l+l_g^+)\right)\,,
\label{eq::wardident}
\end{equation}
with $l_g$ being  the soft photon momentum. Multiplying
Eq.~(\ref{eq::wardident})  with ${l_g}_\mu\approx {l_g}_-=l_g^+$  gives the
standard QED Ward identity. Note that we neglect $l_g^-={l_g}_+$ in
$S(l+l_g)$ to get the logarithmic scaling of the integral over  $l_g^-$ since
the lower eikonal quark propagator is proportional to  $1/l_g^-$. The  right
hand side of Eq.~(\ref{eq::wardident}) corresponds to the diagram
Fig.~\ref{fig::2}(b) where the crossed circle on the quark propagator
represents the replacement $S(l)\to S(l)-S(l+l_g^+)$ and the $1/l_g^+$ factor
is absorbed  into the upper eikonal quark propagator. By the momentum shift
$l\to l-l_g^+$ in the second term of the above expression the crossed circle
can be moved to the upper eikonal photon line  which becomes ${1\over
2p_1l}-{1\over 2p_1(l+l_g^+)}$, Fig.~\ref{fig::2}(c). The opposite eikonal
line is not sensitive to this shift since $p_2^-\approx 0$. On the final step
we use the  ``inverted Ward identity''  on the upper  eikonal photon line
\begin{equation}
{1\over l_g^+}\left({1\over 2p_1l}-{1\over 2p_1(l+l_g^+)}\right)
={1\over 2p_1l}2{p_1}^-{1\over 2p_1(l+l_g^+)}
\approx {1\over (p_1+l)^2}2p_1^\mu{1\over (p_1+l+l_g)^2}
\label{eq::invident}
\end{equation}
to  transform the diagram Fig.~\ref{fig::2}(c) into Fig.~\ref{fig::2}(d) with
an effective dipole coupling $2e_q p_1^\mu$ to the eikonal {\it photon},
where  $e_q$ is the {\it quark} charge.  Note that we can replace $2p_1(l+l_g^+)$
by $(p_1+l+l_g)^2$  in the gauge boson propagator as long as $l_g\ll Q$ since
$p_1^+\approx 0$.

By adding the symmetric diagram we  get a ``ladder'' structure
characteristic  to the standard eikonal factorization  for the Sudakov
form factor.  This factorization, however,  requires the summation over all
possible insertions of the soft photon vertex along each eikonal line while
in the case under consideration the diagram in Fig.~\ref{fig::1}(b) with the
soft exchange between the  photon  lines is missing. This diagram can be
added to complete the  factorization and then subtracted.
Note that the first Ward identity of the sequence  in Fig.~\ref{fig::1}  is
sufficient to prove the factorization of the soft photons with the momentum
$l_g\ll m_q$ as it has been done  in the original paper~\cite{Yennie:1961ad}.
This  algorithm however does not work for the momentum interval $m_q\ll
l_g\ll Q$ which does contribute to the double-logarithmic  corrections. Our
method extends the factorization to this region at the expense of introducing
the above subtraction term, which compensates the charge variation of the
eikonal line after the soft quark emission.

After adding the diagram Fig.~\ref{fig::1}(b) the integral over the soft
photon momentum in the double-logarithmic approximation factors out with
respect to the leading order amplitude and reads
\begin{equation}
-{e_q^2\over (4\pi)^2}{2iQ^2\over \pi^2 }
\int{{d^dl_g}\over l^2 ((p_1+l_g)^2-m_q^2)
((p_2 +l_g)^2-m_q^2)}= -{e_q^2\over (4\pi)^2}
\left(2{\ln\rho\over \varepsilon}+\ln^2\rho\right)\,,
\label{eq::1loopsudff}
\end{equation}
where the dimensional regularization with $d=4-2\varepsilon$ is used to deal
with the infrared divergence.  The  above equation  coincides with the
one-loop on-shell Sudakov form factor, which includes all the universal
Sudakov double logarithms for the amplitudes with the quark and antiquark
external lines. The remaining soft photon contribution is given by the
diagram Fig.~\ref{fig::1}(b)  with the coefficient  $-e_q^2$.  The
corresponding two-loop  integral reads
\begin{equation}
\left({2iQ^2\over \pi^2 }\right)^2
\int{{d^4l}\over (l^2-m_q^2) (p_1+l)^2 (p_2 +l)^2}
{{d^4l_g}\over l_g^2 (p_1+l_g+l)^2
(p_2 +l_g+l)^2}\,.
\label{eq::intg}
\end{equation}
The integration over the soft quark momentum $l$ is double-logarithmic if the
latter can be neglected in the eikonal propagators with the the soft  gluon
momentum $l_g$. This defines  the conditions $lp_1< l_gp_1$, $lp_2< l_gp_2$
corresponding to the ordering of the Sudakov parameters along the  eikonal
lines $v< v_g$, $u< u_g$. Then in the double-logarithmic approximation the
propagators take the following form
\begin{eqnarray}
&& {1\over l_g^2}  \approx
- i \pi \delta(Q^2u_gv_g + {l_g}_\perp^2)\,,
\quad
{1\over (p_1+l_g+l)^2}\approx \frac{1}{Q^2v_g}\,,
\quad
{1\over (p_2 +l_g+l)^2} \approx \frac{1}{Q^2u_g}\,,
\nonumber\\
&&
{1\over l^2-m_q^2} \approx
- i \pi \delta(Q^2uv + {l}_\perp^2-m_q^2)\,,
\quad
{1\over (p_1+l)^2}\approx \frac{1}{Q^2v}\,,
\quad {1\over (p_2 +l)^2} \approx \frac{1}{Q^2u}\,.
\label{eq::2loopprop}
\end{eqnarray}
After integrating    over the transverse momentum components
Eq.~(\ref{eq::intg}) reduces to
\begin{equation}
4\int_{\rho}^{1}{{\rm d}v\over v}
\int_{\rho/v}^{1}{{\rm d}u\over u}\
\int_{v}^{1}{{\rm d}v_g\over v_g}
\int_{u}^{1}{{\rm d}u_g\over u_g}\,.
\label{eq::intgsud}
\end{equation}
By subsequent  integrating over the parameters $u_g,~v_g$ and converting the
result  to the logarithmic variables we get the two-loop non-Sudakov
double-logarithmic correction  to the amplitude
\begin{equation}
{e_q^2\over (4\pi)^2}\ln^2\!\rho
\left(2\int_0^1 {\rm d}\xi \int_{0}^{1-\xi}{\rm d}\eta\,
\left(2\eta\xi\right)\right){\cal G}^0\,
={e_q^2\over (4\pi)^2}{\ln^2\!\rho \over 6}\,{\cal G}^0\,.
\label{eq::2loopg}
\end{equation}
Note that the result is  infrared finite since the quark mass regulates both
collinear and soft divergences.

The above result  can be generalized  to QCD in a straightforward way. The
difference with respect to the abelian case is that the factor $e_q^2/(4\pi)$
should be replaced by $C_F\alpha_s$. Moreover, the contribution similar to
Fig.~\ref{fig::1}(b) does exist in QCD  due to gluon  self-coupling and is
proportional to the quadratic Casimir operator of the adjoint  representation
$C_A=N_c$.  Thus the part of the soft gluon exchange which does not factorize
into external lines is given by the diagram Fig.~\ref{fig::1}(b) with the
color factor $C_A-C_F$, which directly links it to the variation of the color
charge along the eikonal lines.  Let us now demonstrate how the above
factorization emerges in the direct evaluation of the two-loop QCD diagrams
in Fig.~\ref{fig::2}. In general the calculation can be performed in the same
way as in Eq.~(\ref{eq::intg}) up to the treatment of the infrared
divergences not regulated by the quark mass.  Fig.~\ref{fig::2}(a) is the
only diagram with such a divergence in the final result.  The integration
over the soft gluon momentum $l_g$ in this diagram is double-logarithmic when
one can neglect it in the eikonal propagators with the soft  quark
momentum $l$. This defines  the conditions $l_gp_i\ll lp_i$ corresponding to
the ordering of the Sudakov parameters $v_g\ll v$, $u_g\ll u$.  Thus $l_g$
should be retained only in the propagators without the soft quark momenta and
the integral over the soft gluon momentum is reduced to
\begin{equation}
{2iQ^2\over \pi^2 }\int{{d^4l_g}\over l_g^2 ((p_1+l_g)^2-m_q^2)
((p_2 +l_g)^2-m_q^2)}\,,
\label{eq::intlg}
\end{equation}
with the above restriction on $l_g$. In the double-logarithmic approximation
the propagators in this expression take a form  slightly different from
Eq.~(\ref{eq::2loopprop})
\begin{eqnarray}
&& {1\over l_g^2}  \approx
- i \pi \delta(Q^2u_gv_g + {l_g}_\perp^2)\,,
\nonumber\\
&&{1\over (p_1+l_g)^2-m_q^2}\approx \frac{1}{Q^2(v_g+2\rho u_g)}\,,
\nonumber\\
&&{1\over (p_2 +l_g)^2-m_q^2} \approx \frac{1}{Q^2(u_g+2\rho v_g)}\,.
\label{eq::singprop}
\end{eqnarray}
After integrating  Eq.~(\ref{eq::intlg}) over ${l_g}_\perp$  with the
double-logarithmic accuracy we get
\begin{equation}
2\int_{\rho u_g}^{v}{{\rm d}v_g\over v_g}
\int_{\rho v_g}^{u}{{\rm d}u_g\over u_g}\,.
\label{eq::intluv2loop}
\end{equation}
Eq.~(\ref{eq::intluv2loop})  has soft  divergence  when $v_g$ and $u_g$
simultaneously become small.  This divergence can be removed by subtracting
the factorized expression
\begin{equation}
2\int_{\rho u_g}^{1}{{\rm d}v_g\over v_g}
\int_{\rho v_g}^{1}{{\rm d}u_g\over u_g}\,.
\label{eq::intlfac}
\end{equation}
The subtraction term does not depend on the soft quark momenta and is
equivalent to the factorized  one-loop Sudakov form factor in
Eq.~(\ref{eq::1loopsudff}), which does not have a nonabelian contribution.
The  subtracted expression reads
\begin{eqnarray}
&-&
2\left(\int_{v}^{1}{{\rm d}v_g\over v_g}\int_{\rho v_g}^{u}{{\rm d}u_g\over u_g}
+\int_{\rho u_g}^{v}{{\rm d}v_g\over v_g}\int_{u}^{1}{{\rm d}u_g\over u_g}
+\int_{v}^{1}{{\rm d}v_g\over v_g}\int_{u}^{1}{{\rm d}u_g\over u_g}\right)
\nonumber \\
&& \nonumber \\
&=& -\ln^2\!\rho \left((\eta-\xi)^2+2(\eta+\xi)\right)
\,.
\label{eq::intsub}
\end{eqnarray}
The  contributions of the infrared subtracted diagram Fig.~\ref{fig::2}(a)
along with the  remaining  infrared finite  diagrams can be written as
the integral over the soft quark momentum variables
\begin{equation}
x\left(2\sum_i c^{(1)}_\lambda \int_0^1 {\rm d}\xi \int_{0}^{1-\xi}
{\rm d}\eta\, w^{(1)}_\lambda(\eta,\xi)\right)\, {\cal G}^0\,,
\label{eq::G2loop}
 \end{equation}
where the  color factors $c^{(1)}_\lambda$ and the weight function
$w^{(1)}_\lambda$ resulting from the logarithmic integration over the soft
gluon momentum are collected in Table~\ref{tab::1}.\footnote{Further details
of the soft gluon momentum integration can be found in
Ref.~\cite{Liu:2017axv}.} Summing up the contributions we get
\begin{equation}
-z\left(2\int_0^1 {\rm d}\xi \int_{0}^{1-\xi}{\rm d}\eta\,
\left(2\eta\xi\right)\right){\cal G}^0
=-{z\over 6}\,{\cal G}^0\,,
\label{eq::2loopgna}
\end{equation}
where $z=(C_A-C_F)x$, which coincides with Eq.~(\ref{eq::2loopg}) up to the
modification of the  effective coupling  discussed above. Thus we observe the
relations between the diagrams imposed by the Ward identities at the
integrand level. The only new relation with respect to the abelian case
provides the  cancellation of the color space commutator of the soft gluon
vertex at the first step of the sequence  in Fig.~\ref{fig::3} by the diagram
Fig.~\ref{fig::2}(c) with the three-gluon coupling, as can be seen from the
second and the third lines of Table~\ref{tab::1}. It is equivalent to the
standard Ward identity for the factorization of the soft gluon emission which
provides the cancellation of the nonabelian contribution in the
double-logarithmic Sudakov form factor. We have verified the above result
diagram by diagram through the explicit evaluation of the two-loop  integrals
in the high-energy limit within the expansion by regions framework
\cite{Beneke:1997zp,Smirnov:1997gx,Smirnov:2002pj}.

\begin{table}
\begin{center}
 \begin{tabular}{|c|c|c|c|}
        \hline
        $\lambda$ & $w^{(1)}_\lambda$     &  $c^{(1)}_\lambda$ \\
        \hline
        a & $(\eta-\xi)^2+2(\eta+\xi)$    &  $C_F$\\
        b & $\eta^2 - 2\eta\xi + 2\eta $  &  $-C_F +\frac{1}{2}C_A$ \\
        c & $\eta^2 - 2\eta\xi + 2\eta $  &  $-\frac{1}{2}C_A$ \\
        d & $ 2\eta\xi $                  &  $-C_A$\\
        \hline
  \end{tabular}
\end{center}
\caption{\label{tab::1} The weights $w^{(1)}_\lambda$ and the color factors
$c^{(1)}_\lambda$ for the diagrams in Fig.~\ref{fig::2}. The weights for the
symmetric diagrams are obtained by  interchanging the $\eta$ and $\xi$
variables. The singular part of the infrared divergent diagram (a) is
subtracted as discussed in the text.}
\end{table}

\subsection{Resummation of the double-logarithmic corrections and the
asymptotic behavior of the amplitude}

With the established factorization structure  at hand it is straightforward
to perform the resummation of the double-logarithmic corrections to
all orders of perturbation theory. Indeed, the emission of the soft gluons
from an eikonal line of a given color charge factorizes and exponentiates
\cite{Frenkel:1984pz} so we can apply the procedure discussed in the previous
section  to an arbitrary number of gluons emitted from the soft quark line.
Then the factorized one-loop Sudakov logarithms exponentiate to the universal
factor for the quark-antiquark external on-shell lines
\begin{equation}
 Z_{q}^2=\exp\left[-C_F\left({\alpha_s\over 2\pi}
 {\ln\rho\over \varepsilon}+x\right)\right].
\label{eq::Zq}
\end{equation}
The same statement is true for the soft gluon exchange between the effective
vertices, Fig.~\ref{fig::1}(b). Thus the all-order non-Sudakov double
logarithms can be obtained by replacing the one-loop contribution
$-2z\eta\xi$ in Eq.~(\ref{eq::2loopg})  with its exponent inside the integral
over the logarithmic variables. Hence the all-order expression for the
double-logarithmic corrections  to the amplitude reads
\begin{equation}
{\cal G}= Z_{q}^2 g(-z)\,{\cal G}^{(0)}\,,
\label{eq::Gfac}
\end{equation}
where the function $g(-z)$ incorporates the non-Sudakov contribution of
Fig.~\ref{fig::1}(b) with an arbitrary number of the effective  soft gluon
exchanges.  The function $g(z)$ of the variable $z=(C_A-C_F)x$ is normalized
to $g(0)=1$ and is given by the  two-fold integral
\begin{equation}
g(z)=2\int_0^1 {\rm d}\xi \int_{0}^{1-\xi}{\rm d}\eta e^{2z\eta\xi}\,.
\label{eq::g}
\end{equation}
The integral Eq.~(\ref{eq::g}) can be solved in terms of  the  generalized
hypergeometric function
\begin{equation}
g(z)={}_2F_2\left(1,1;{3/2},2;{z/2}\right)=2\sum_0^\infty {n!\over (2n+2)!}(2z)^n
\label{eq::gseries}
\end{equation}
with the following asymptotic behavior at $z\to\infty$
\begin{equation}
g(-z)\sim {\ln(2z) +\gamma_E\over z}, \quad g(z)\sim
\left({2\pi e^{z}\over z^{3}}\right)^{1/2}\!\!,
\label{eq::gasymp}
\end{equation}
where $\gamma_E=0.577215\ldots$ is the Euler constant and both limits are
necessary since the variable $z$ is positive in QCD and negative in QED. The
above equations determine the amplitude  ${\cal G}$ in the  high-energy limit
in double-logarithmic approximation. In the next section we verify the
perturbative expansion of Eq.~(\ref{eq::Gfac}) to ${\cal O}(\alpha_s^3)$ by
explicit evaluation of the three-loop double-logarithmic term.

\newpage

\begin{figure}[h]
\begin{center}
\begin{tabular}{ccccc}
\includegraphics[width=1.8cm]{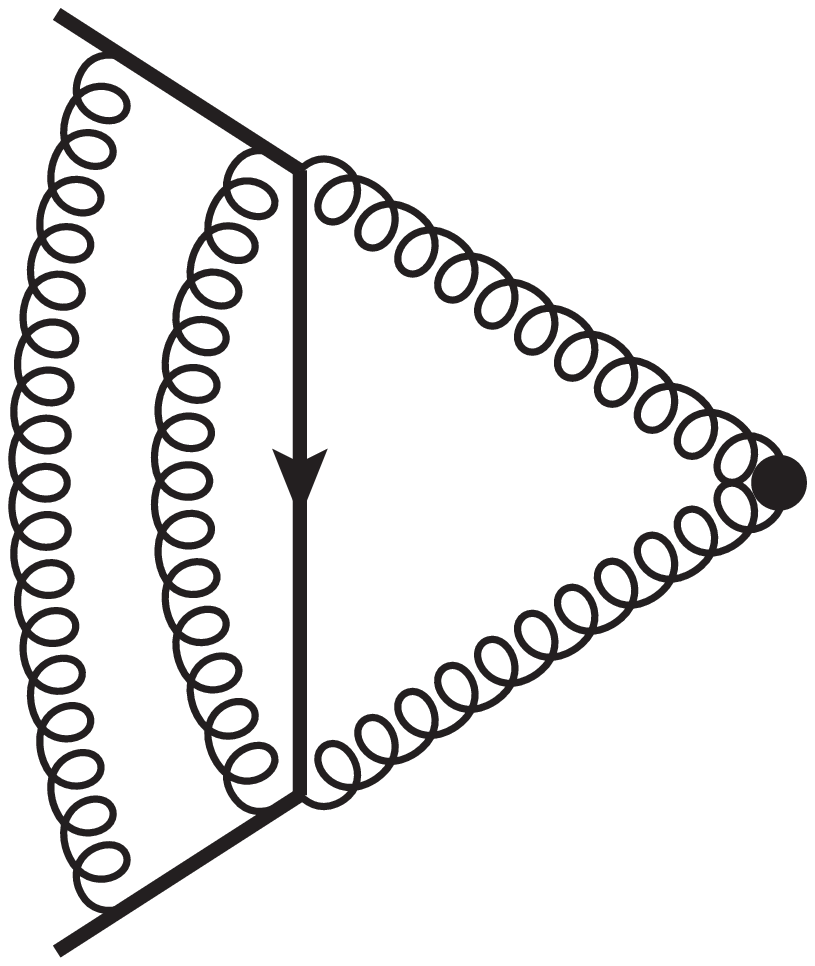} &
\hspace*{03mm}\includegraphics[width=1.8cm]{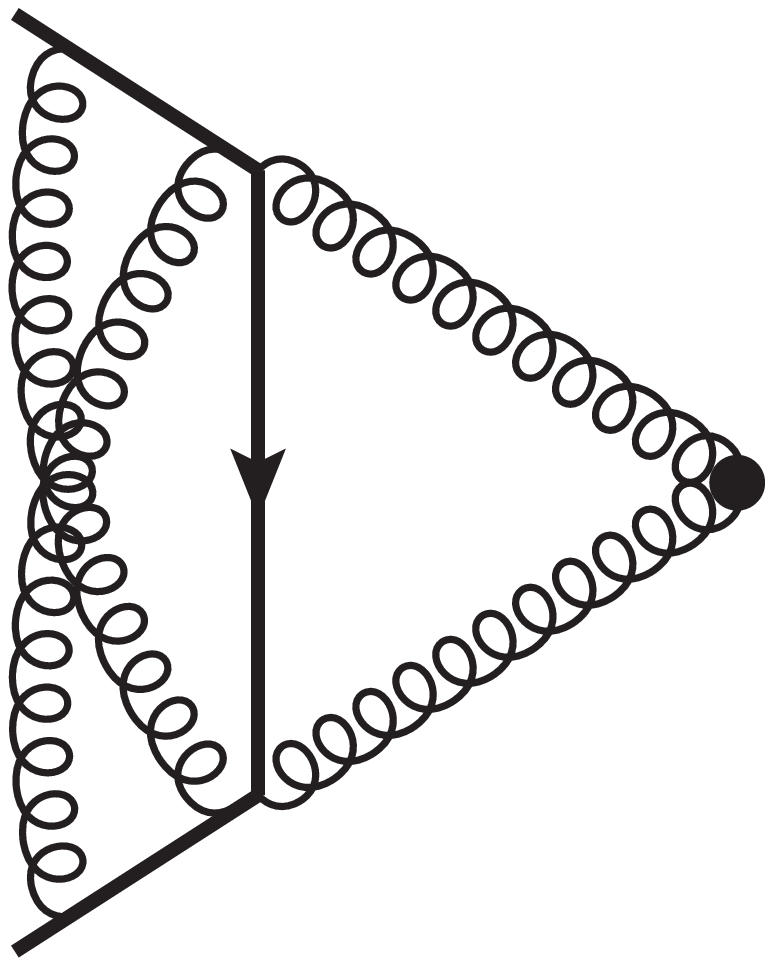} &
\hspace*{03mm}\includegraphics[width=1.8cm]{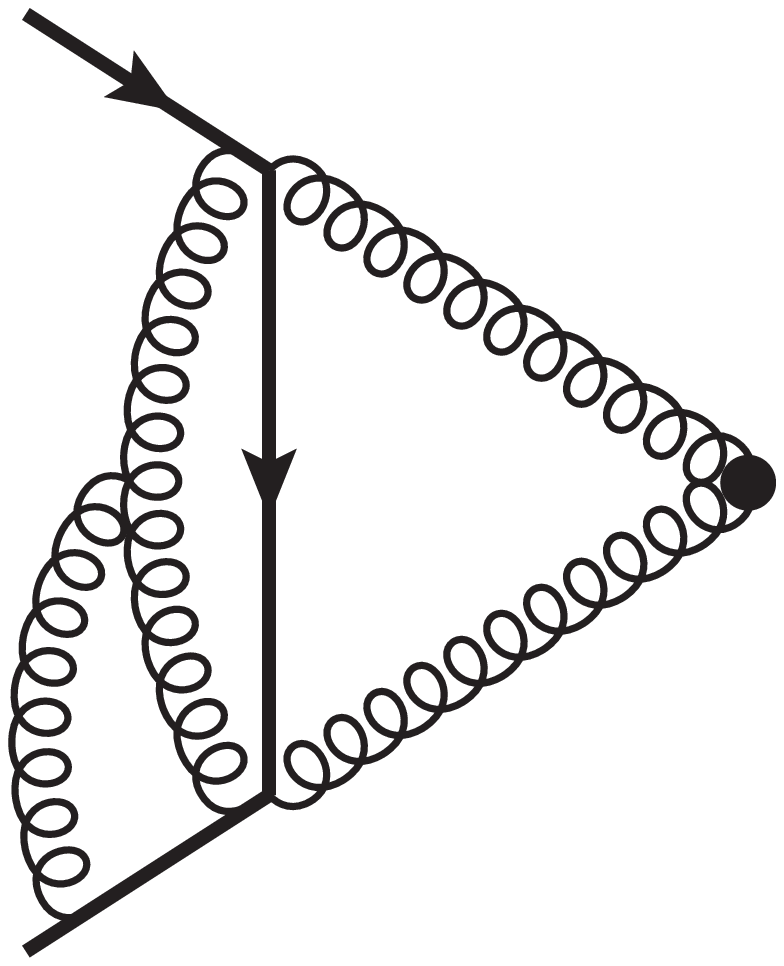} & \hspace*{03mm}\includegraphics[width=1.8cm]{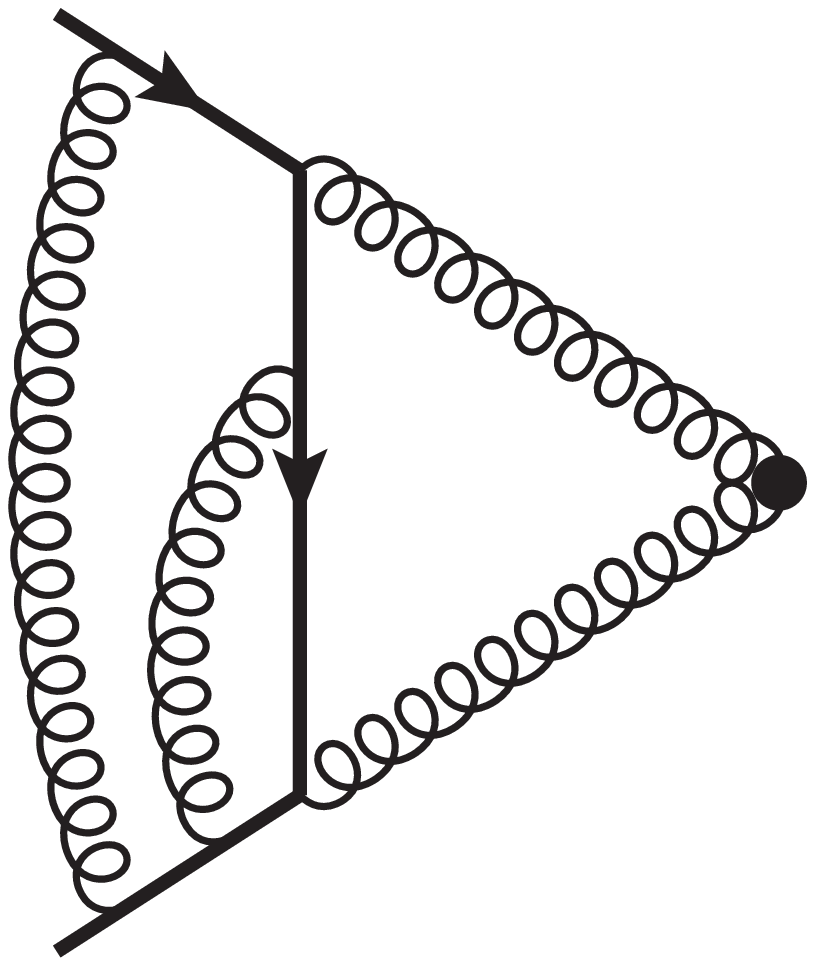} & \hspace*{03mm}\includegraphics[width=1.8cm]{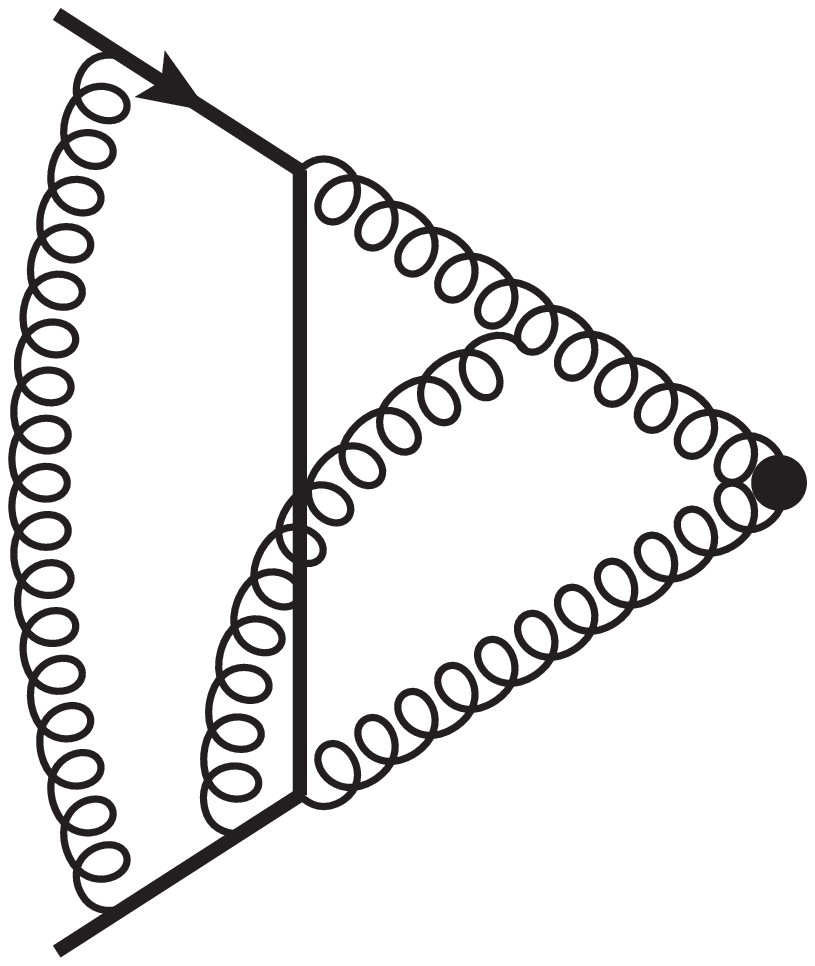}\\
~~~(a) & ~~~(b) & ~~~(c) & ~~~(d) & ~~~(e) \\
~~~ & ~~~ & ~~~ & ~~~ & ~~~ \\
\includegraphics[width=1.8cm]{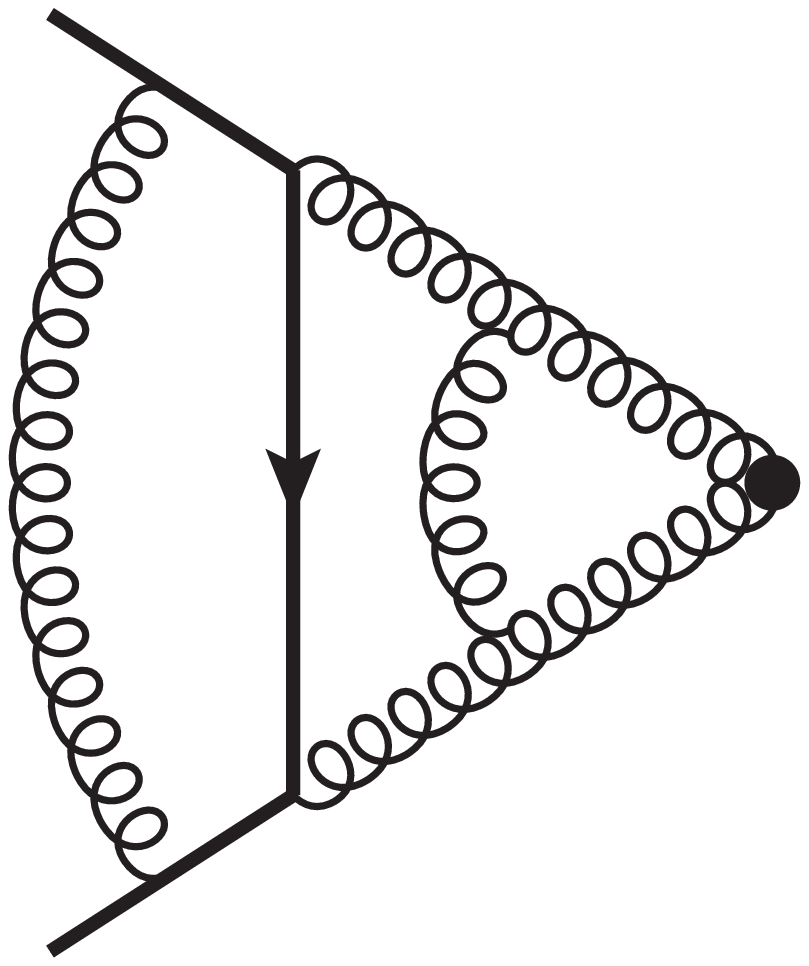} &
\hspace*{03mm}\includegraphics[width=1.8cm]{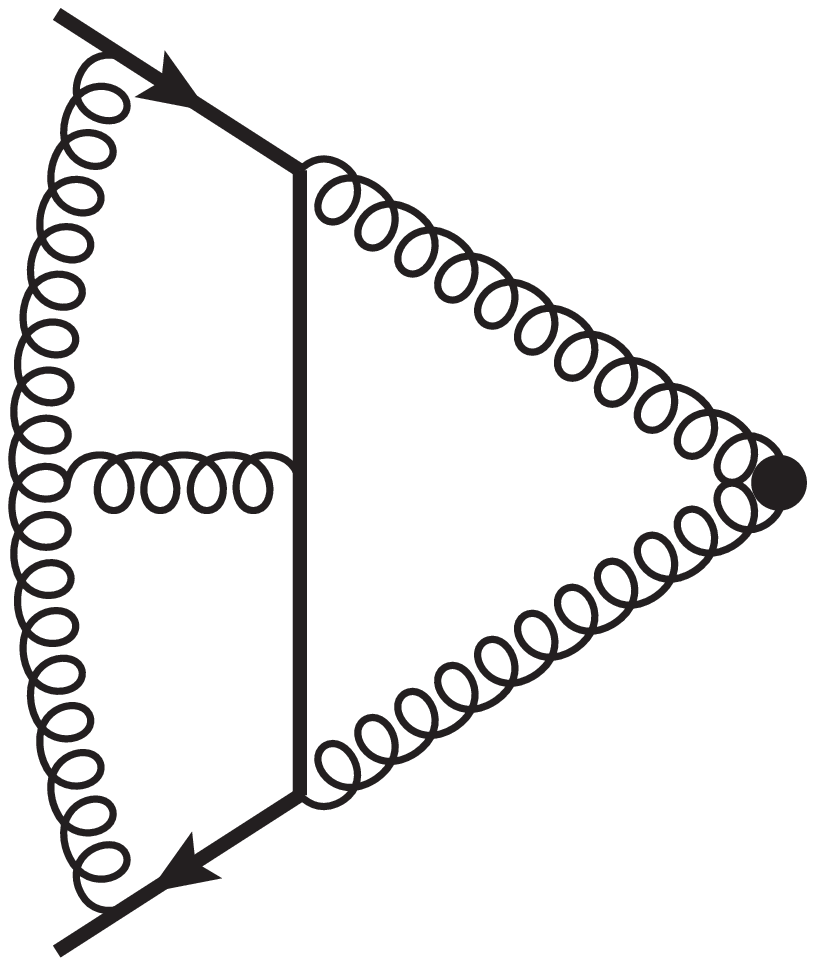} & \hspace*{03mm}\includegraphics[width=1.8cm]{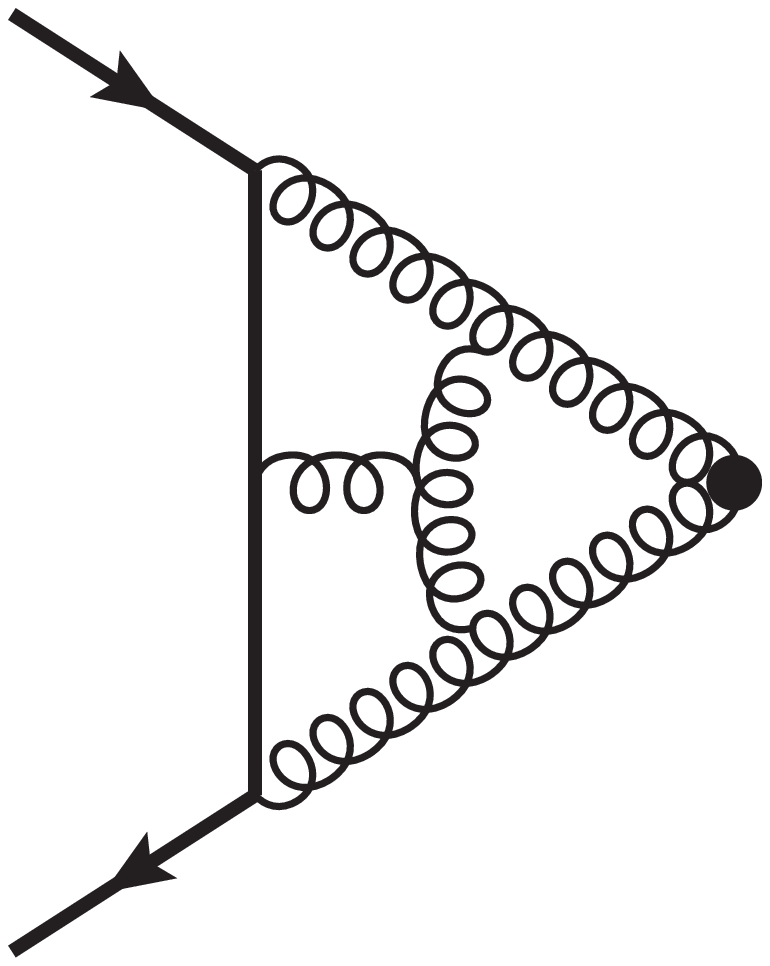} & \hspace*{03mm}\includegraphics[width=1.8cm]{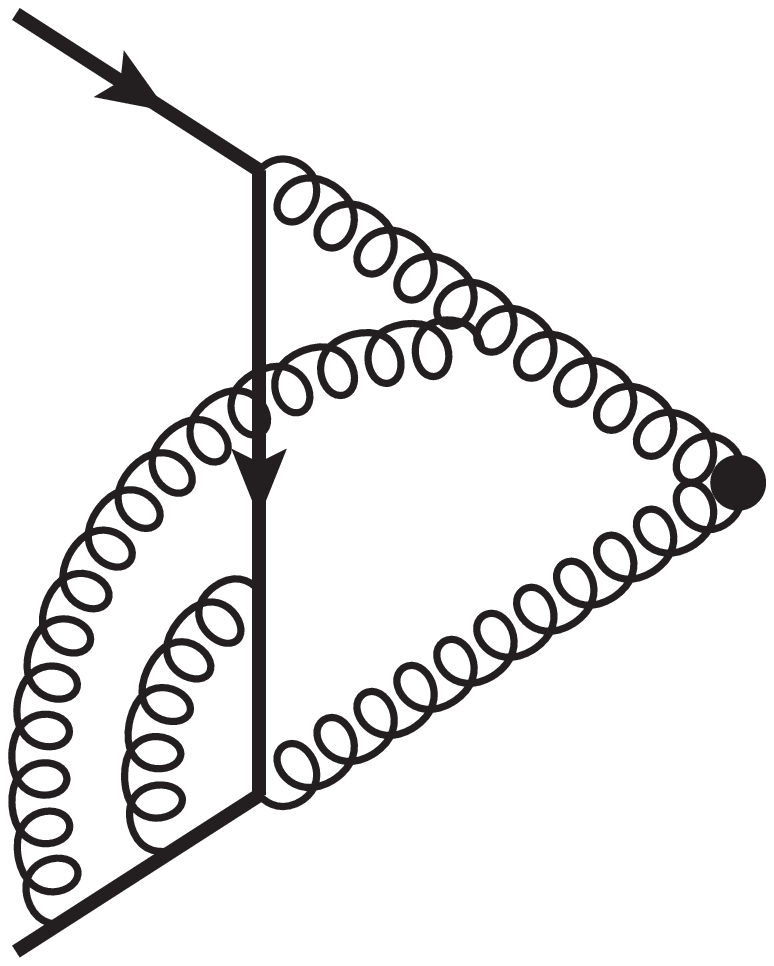} & \hspace*{03mm}\includegraphics[width=1.8cm]{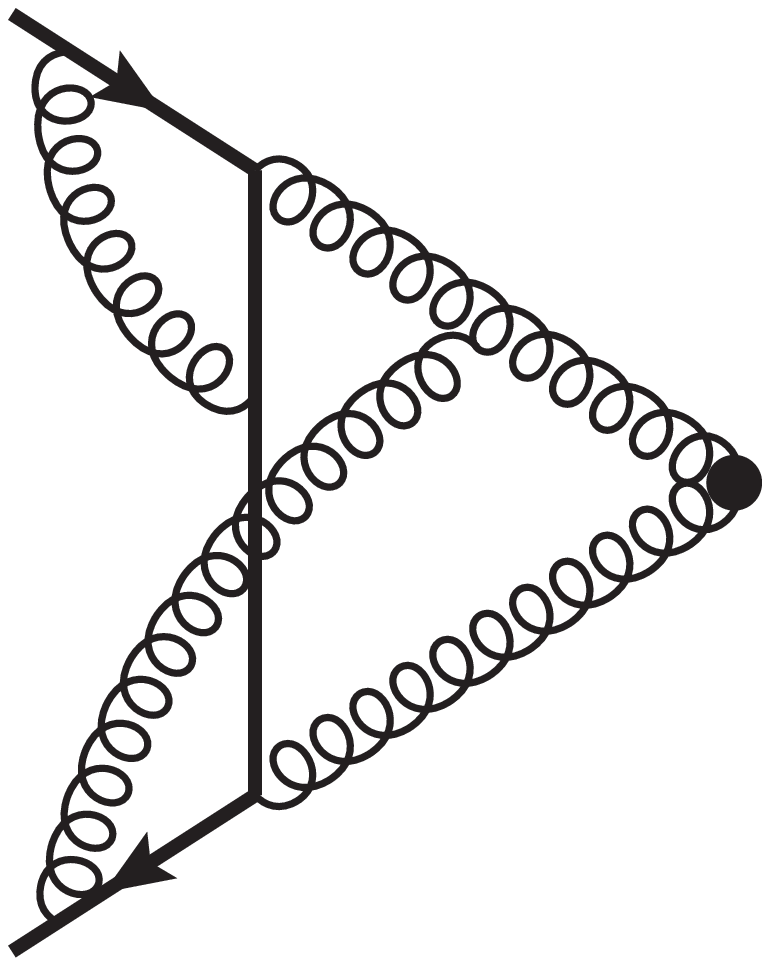}\\
~~~(f) & ~~~(g) & ~~~(h) & ~~~(i) & ~~~(j) \\
~~~ & ~~~ & ~~~ & ~~~ & ~~~ \\
\includegraphics[width=1.8cm]{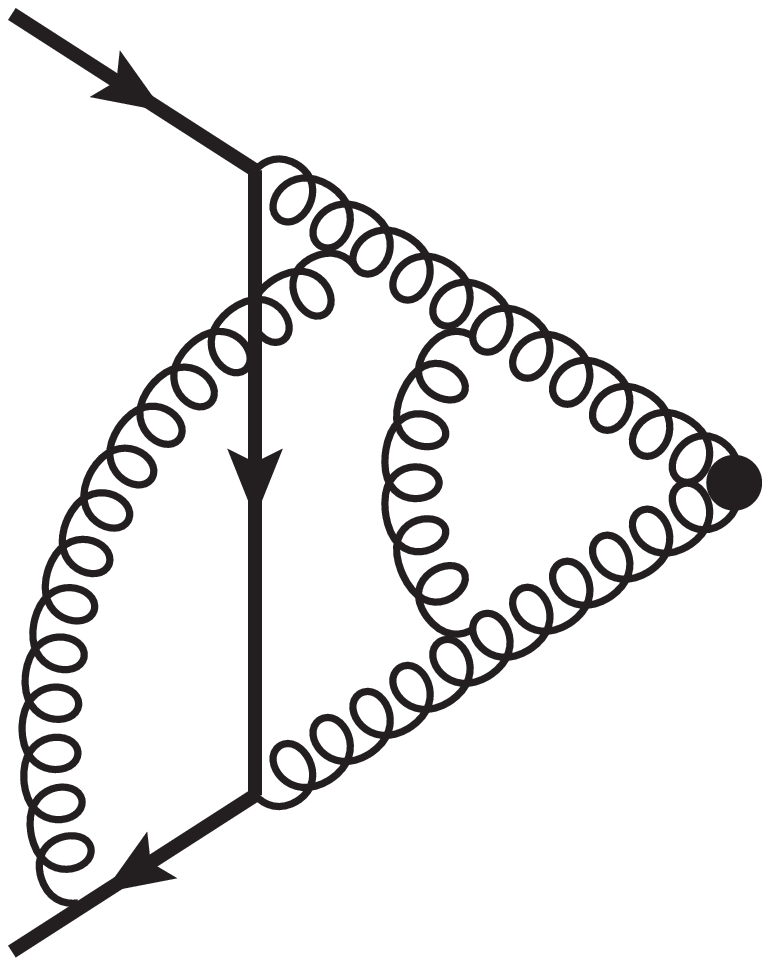} &
\hspace*{03mm}\includegraphics[width=1.8cm]{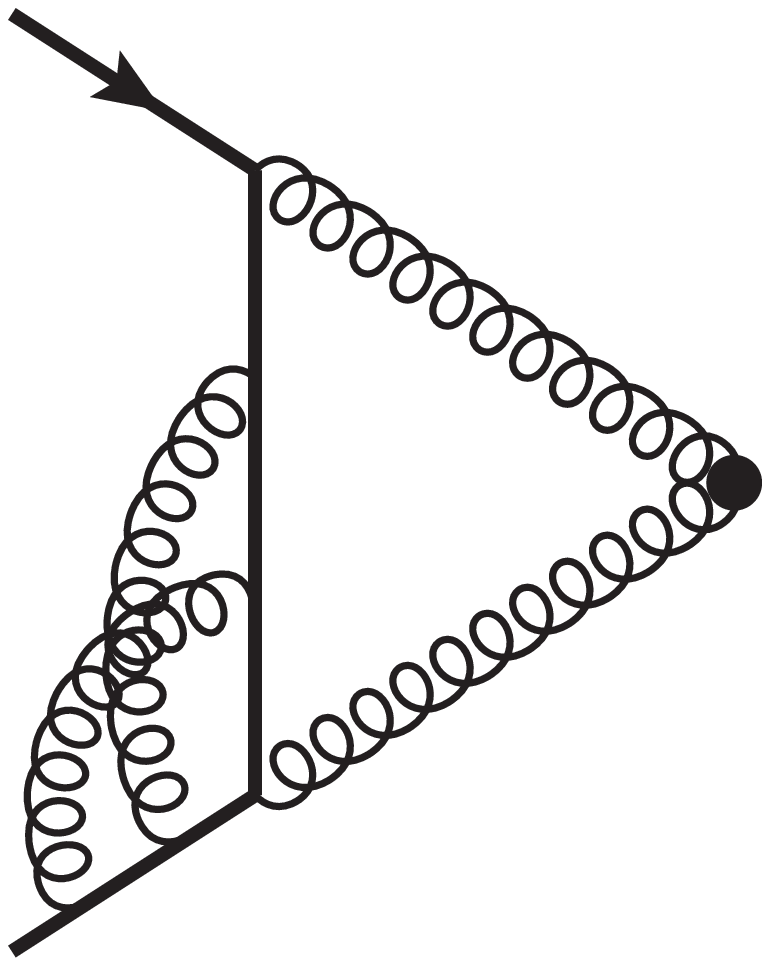} & \hspace*{03mm}\includegraphics[width=1.8cm]{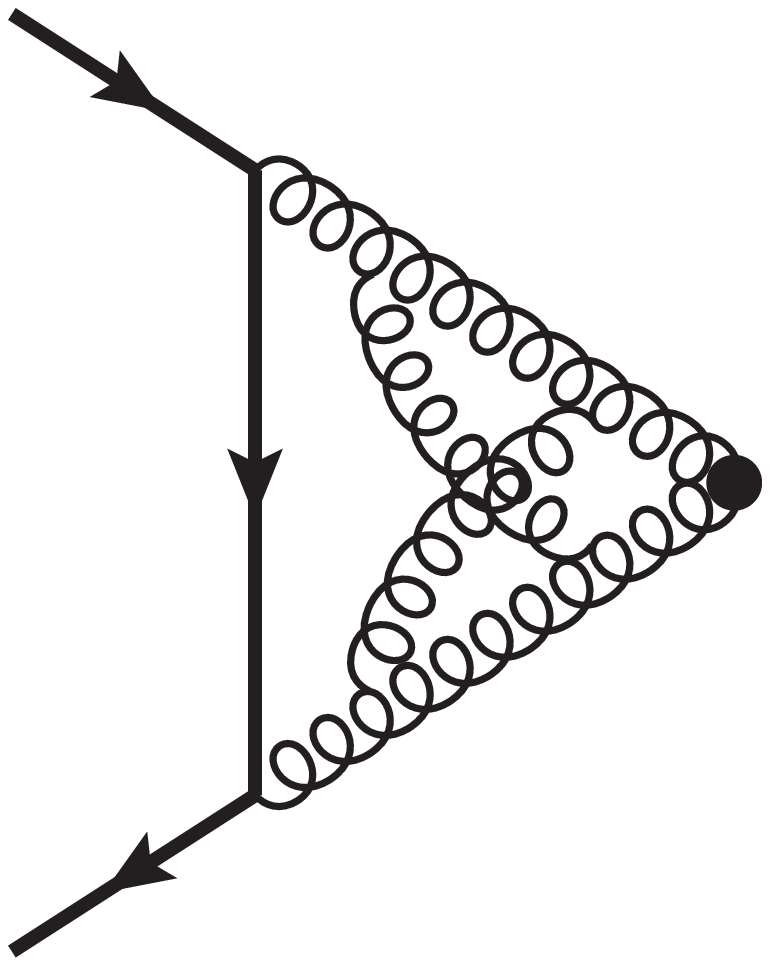} & \hspace*{03mm}\includegraphics[width=1.8cm]{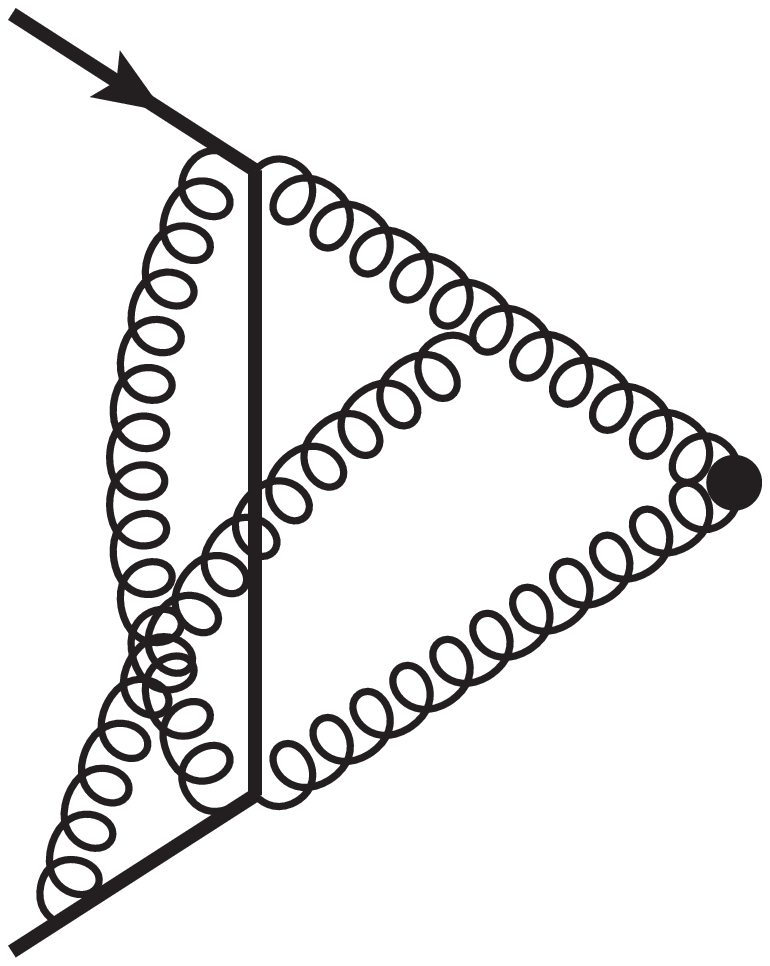} & \hspace*{03mm}\includegraphics[width=1.8cm]{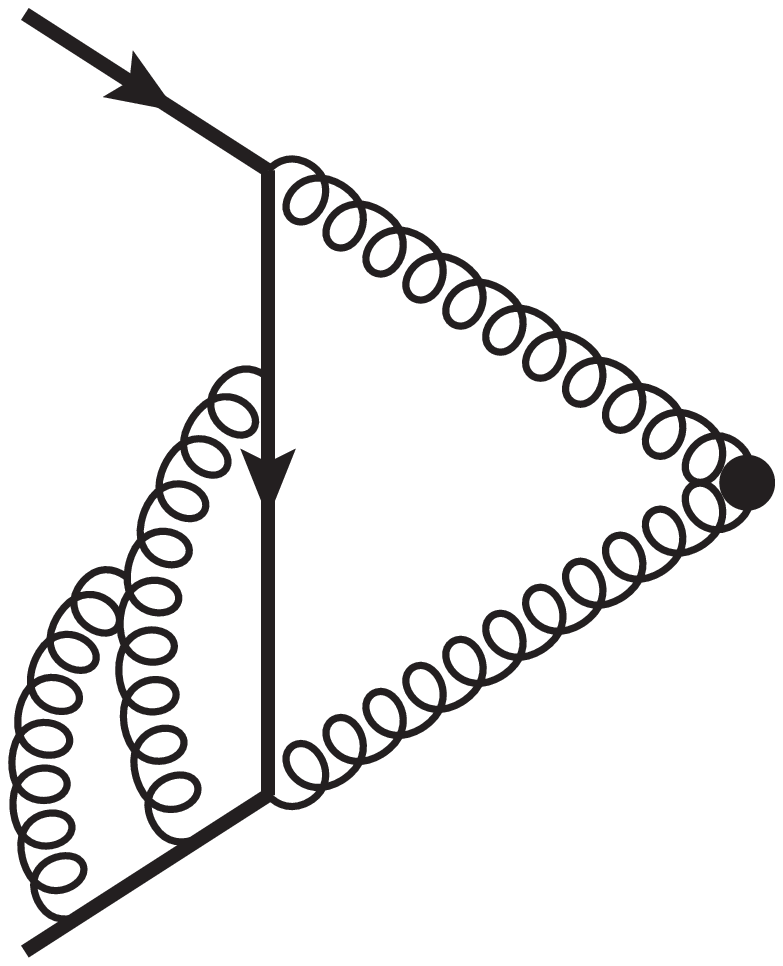}\\
~~~(k) & ~~~(l) & ~~~(m) & ~~~(n) & ~~~(o) \\
~~~ & ~~~ & ~~~ & ~~~ & ~~~ \\
\includegraphics[width=1.8cm]{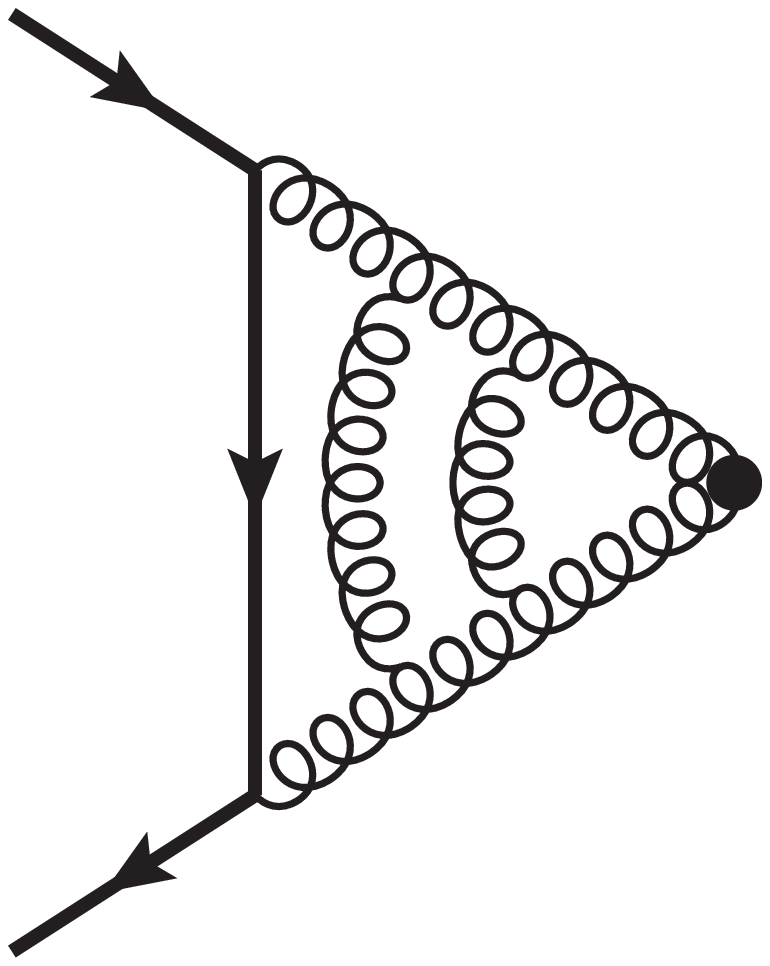} &
\hspace*{03mm}\includegraphics[width=1.8cm]{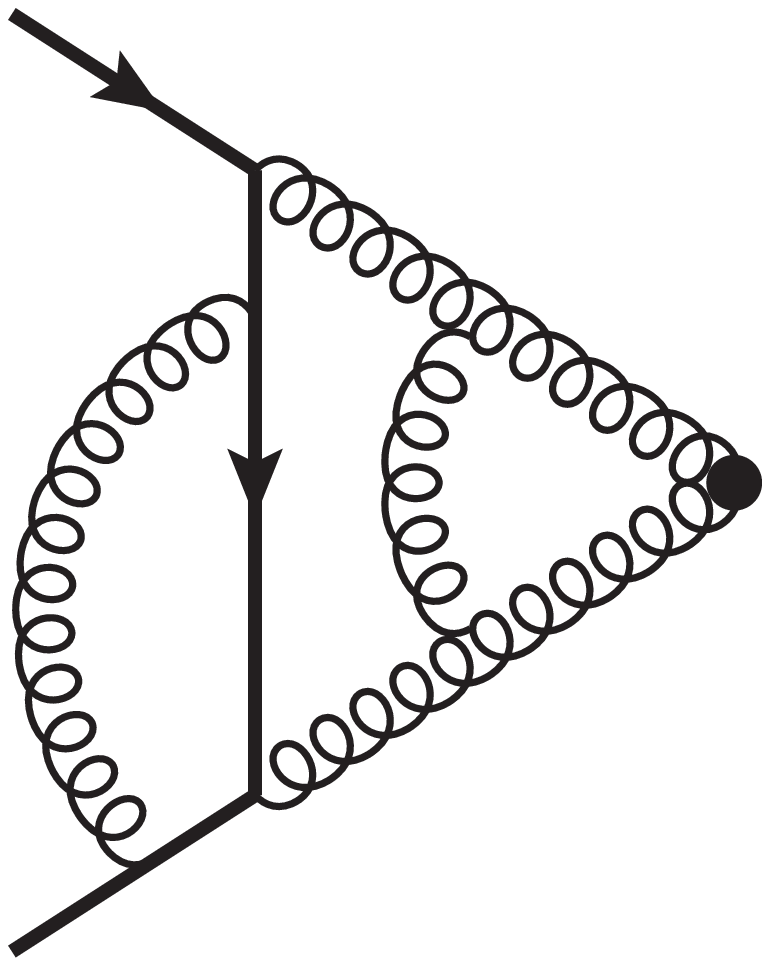} & \hspace*{03mm}\includegraphics[width=1.8cm]{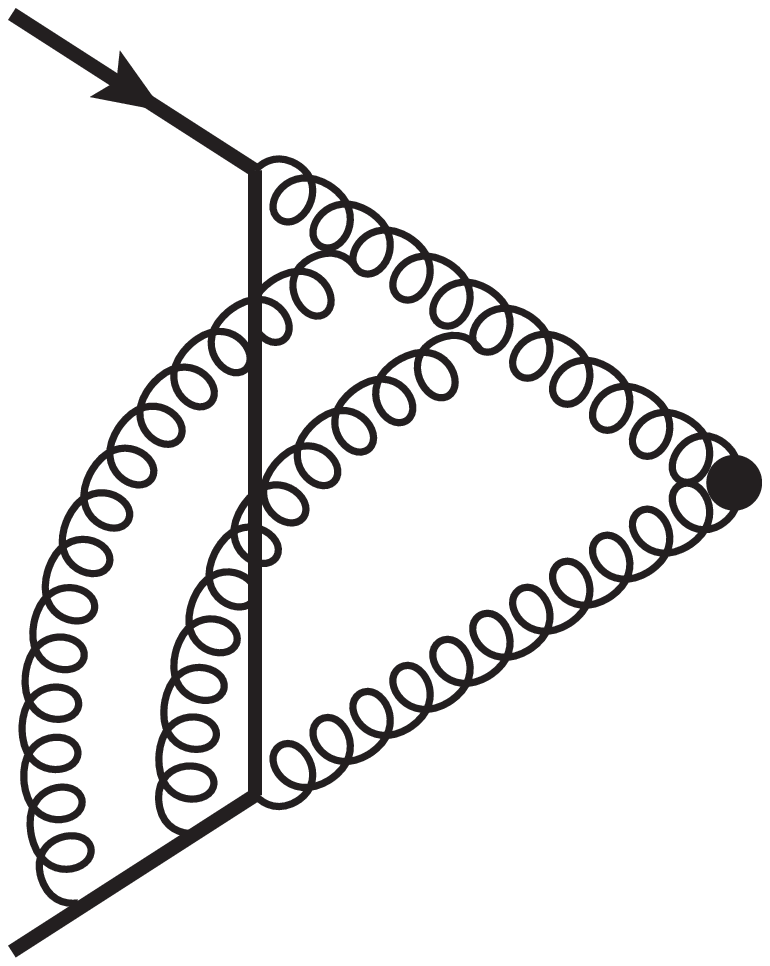} & \hspace*{03mm}\includegraphics[width=1.8cm]{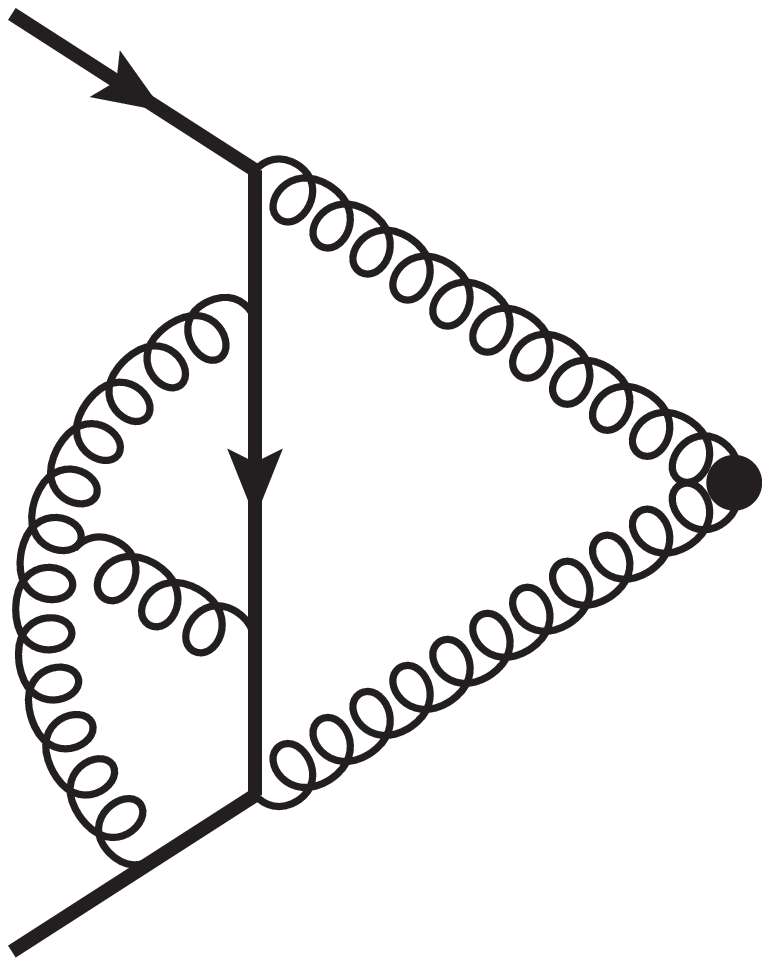} & \hspace*{03mm}\includegraphics[width=1.8cm]{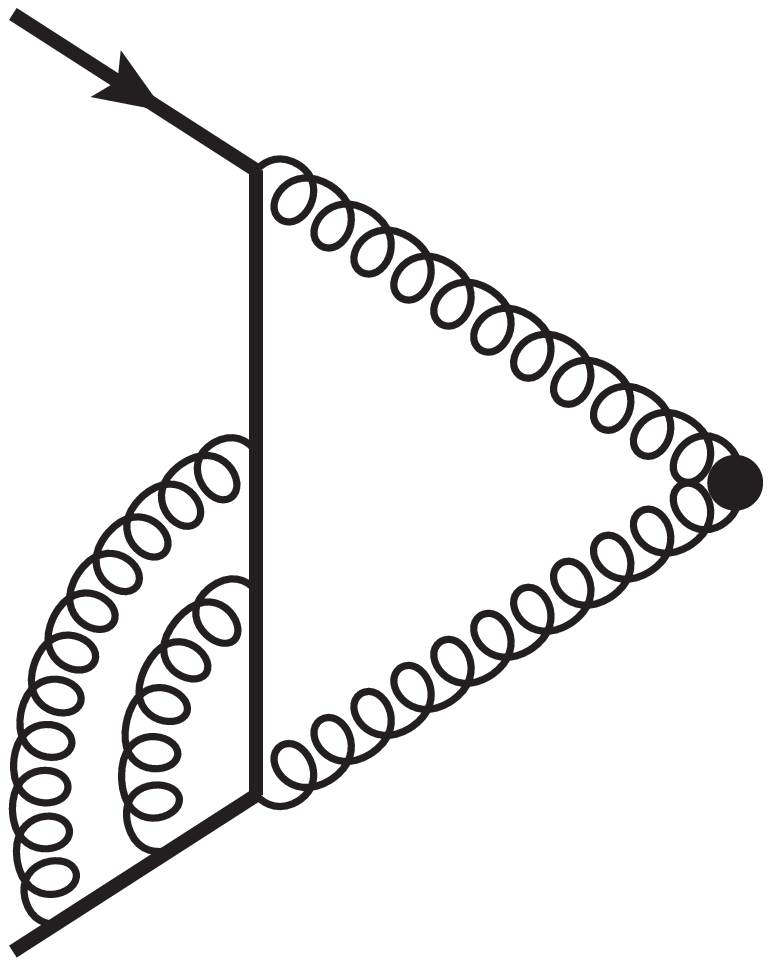}\\
~~~(p) & ~~~(q) & ~~~(r) & ~~~(s) & ~~~(t) \\
~~~ & ~~~ & ~~~ & ~~~ & ~~~ \\
% %
\includegraphics[width=1.8cm]{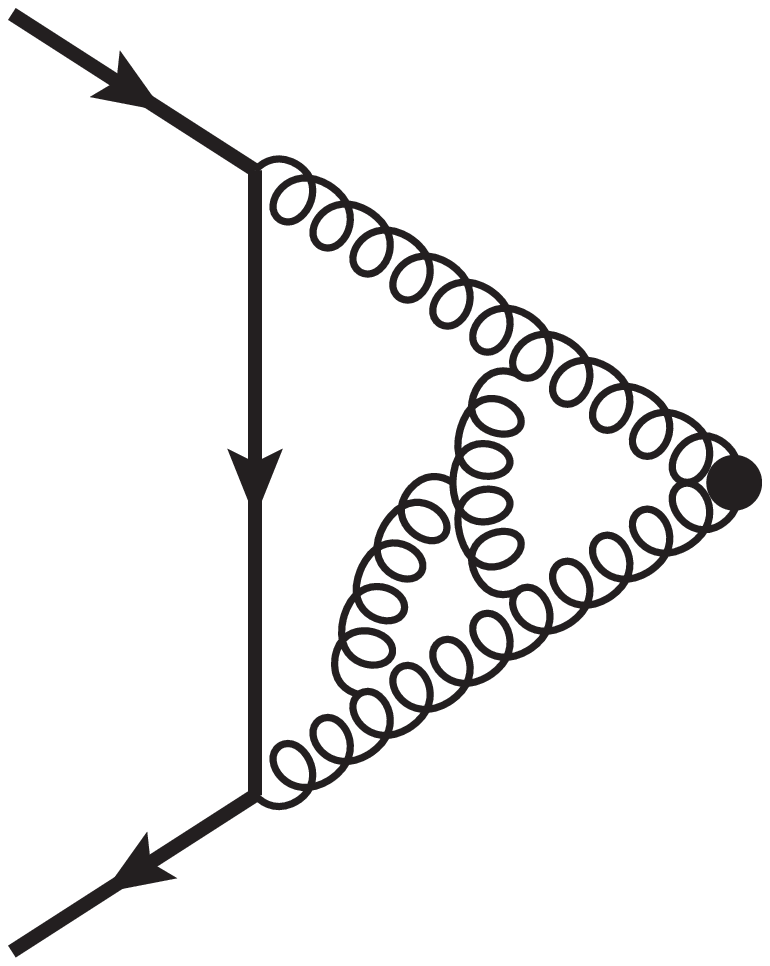} &
\hspace*{03mm}\includegraphics[width=1.8cm]{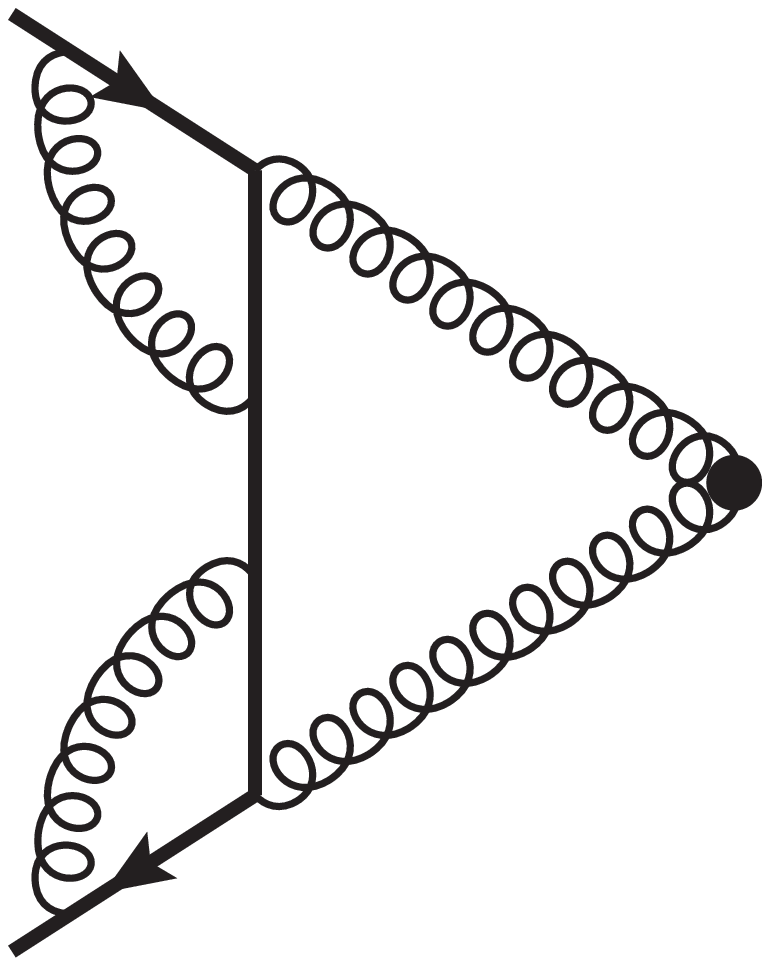} & \hspace*{03mm}\includegraphics[width=1.8cm]{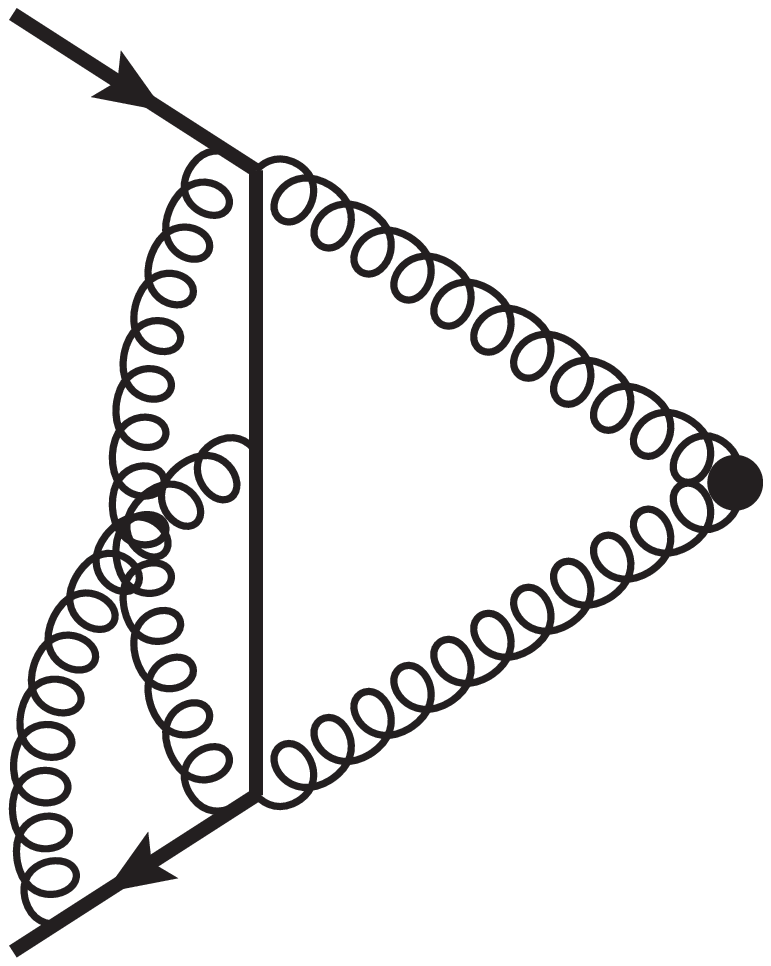} & \hspace*{03mm}\includegraphics[width=1.8cm]{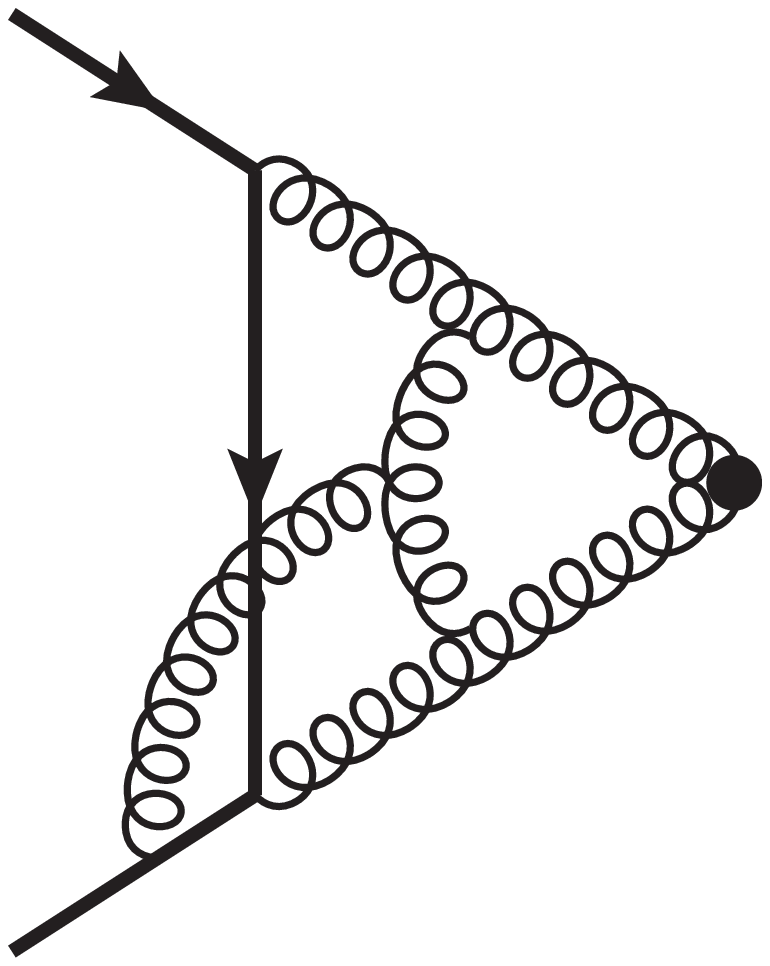} & \hspace*{03mm}\includegraphics[width=1.8cm]{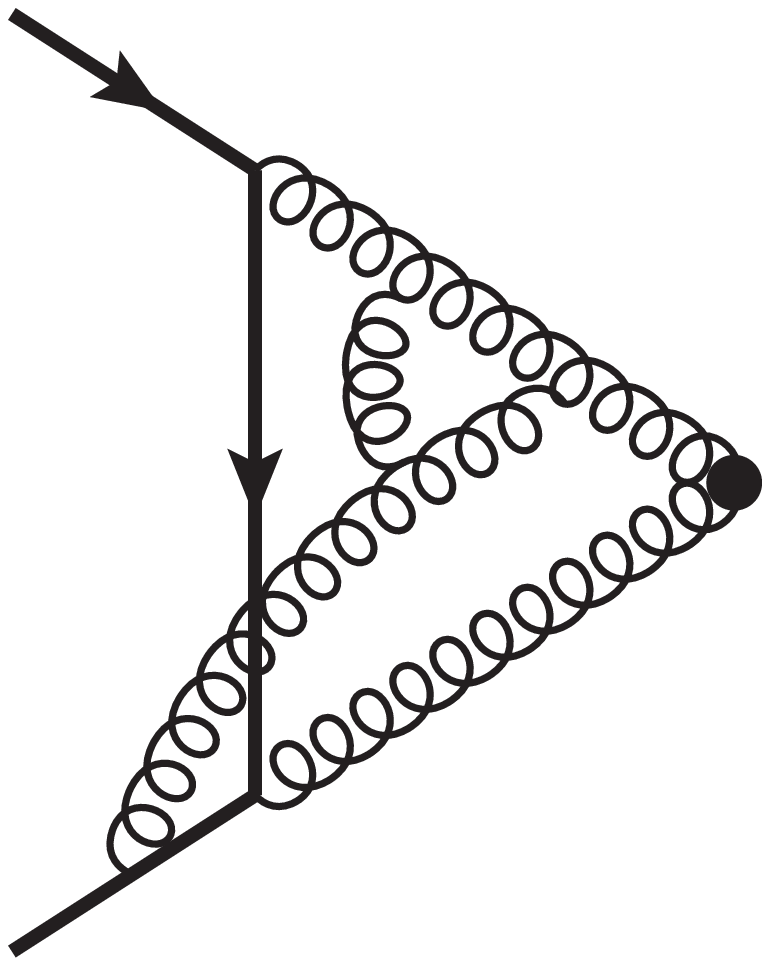}\\
~~~(u) & ~~~(v) & ~~~(w) & ~~~(x) & ~~~(y) \\
~~~ & ~~~ & ~~~ & ~~~ & ~~~ \\
% %
\includegraphics[width=1.8cm]{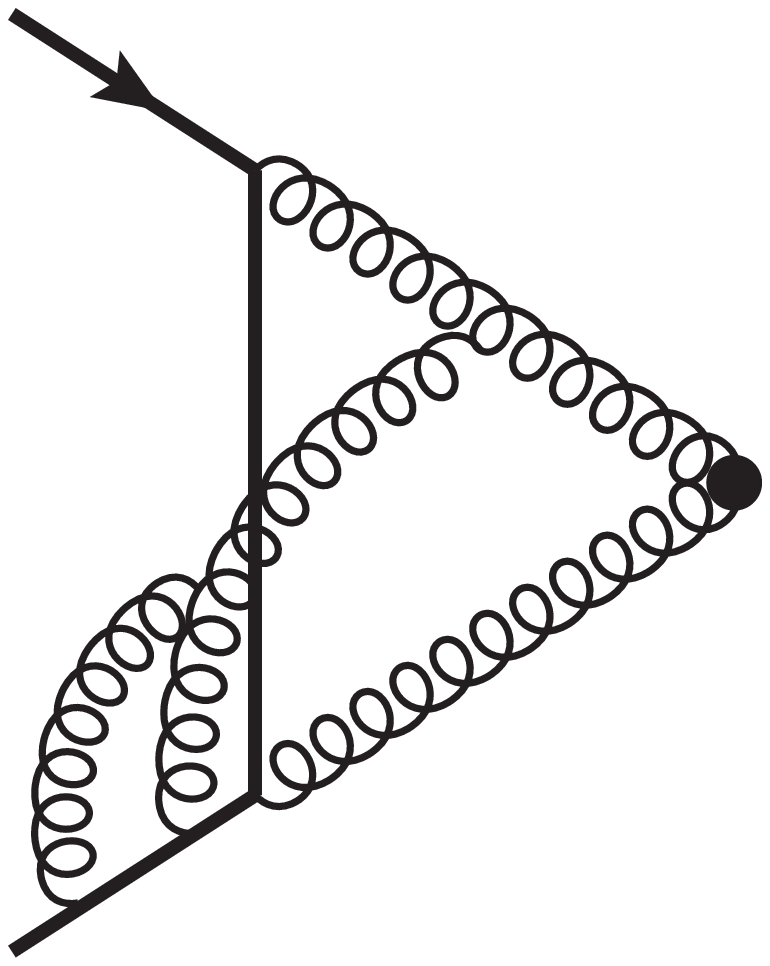} &
\hspace*{03mm}\includegraphics[width=1.8cm]{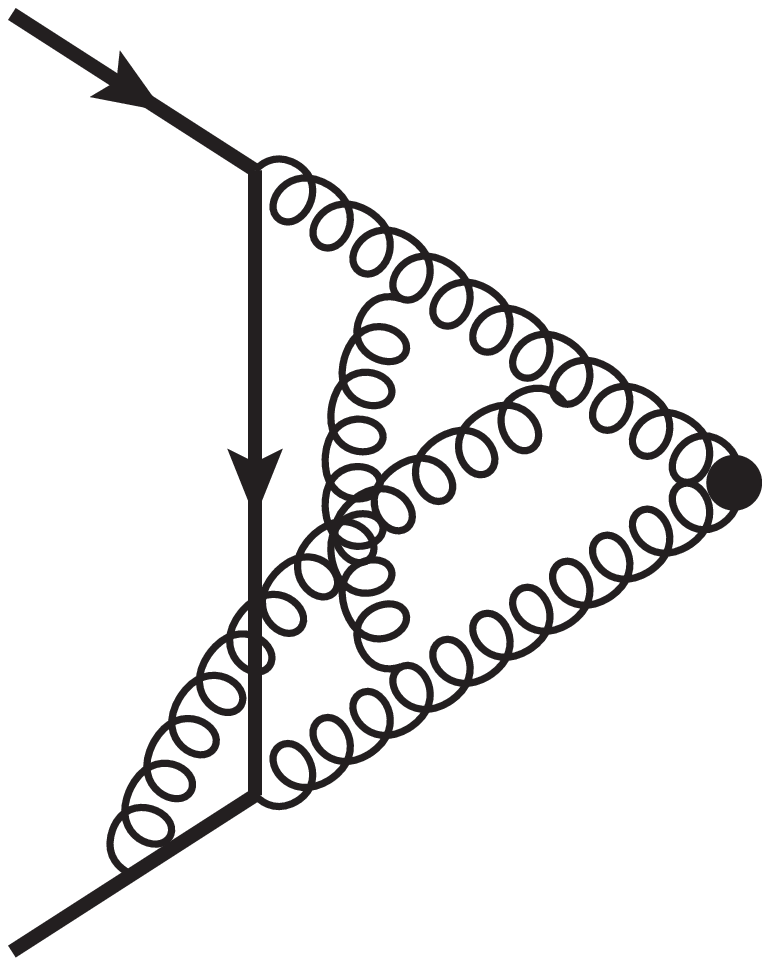} & \hspace*{03mm}\includegraphics[width=1.8cm]{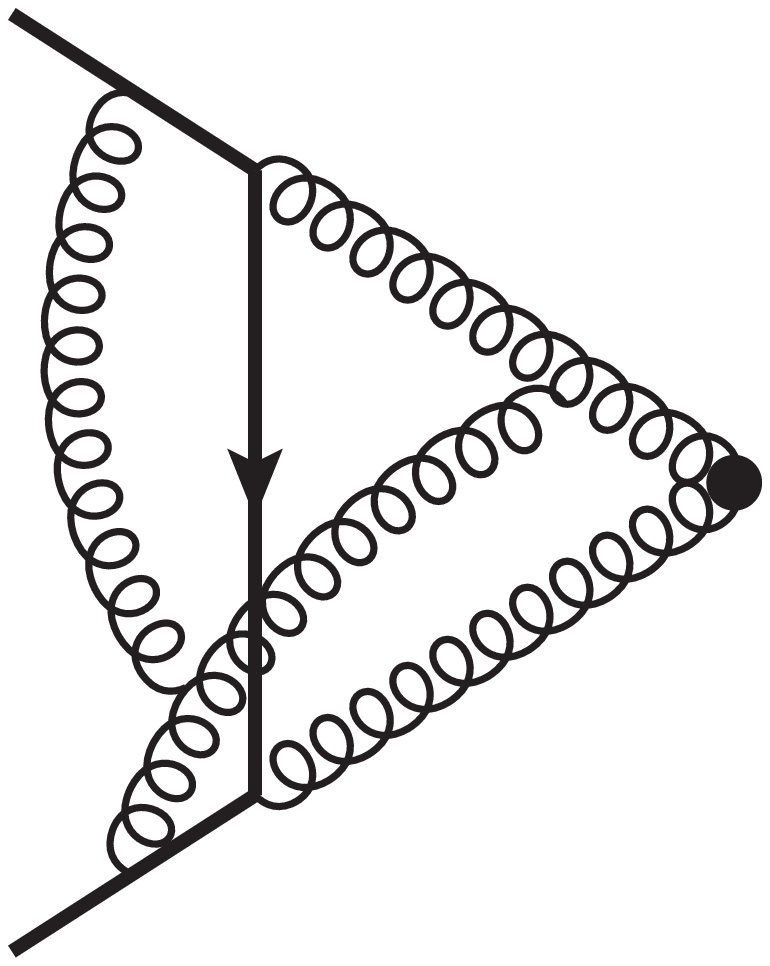} & \hspace*{03mm}\includegraphics[width=1.8cm]{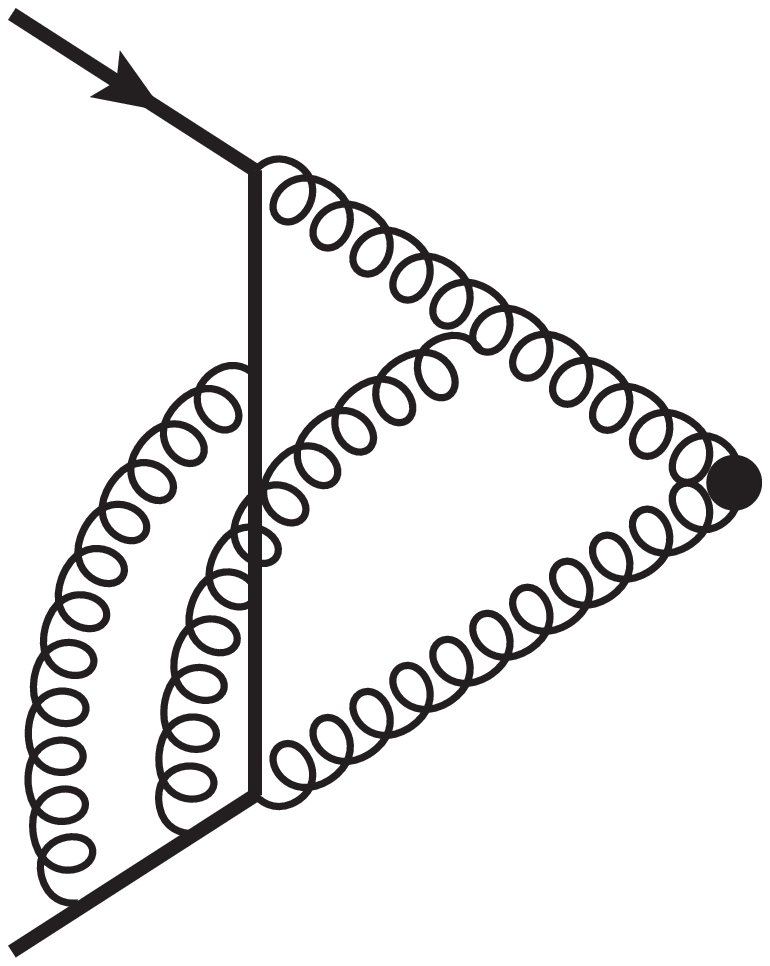} & \\
~~~(z) & ~~~($\rho$) & ~~~($\sigma$) & ~~~($\tau$) &  \\
\end{tabular}
\end{center}
\caption{\label{fig::4}  The three-loop  Feynman diagrams for the quark scattering
by the $(G_{\mu\nu}^a)^2$ vertex  which contribute in the double-logarithmic
approximation. Symmetric diagrams are not shown.}
\end{figure}

\newpage
\begin{table}[h]
%\tiny
\begin{center}
    \begin{tabular}{|c|c|c|}
    \hline
      $\lambda$ &  $w^{(2)}_\lambda$   & $c^{(2)}_\lambda$ \\
      \hline
      a &  $2\xi/3 + \xi^2 + \xi^3/2 + \xi^4/24 + 2\eta/3 - \xi^2 \eta/2 - \xi^3 \eta/6 $     &    $4 C_F^2$ \\
      &  $+ \eta^2 - \xi \eta^2/2 + \xi^2 \eta^2/4 + \eta^3/2 - \xi \eta^3/6 + \eta^4/24  $ &  \\
      b &  $-2\xi/3 - \xi^2/2 + \xi^4/12 - 2\eta/3 + \xi \eta - \xi^3 \eta/3 - \eta^2/2 $ &    $4C_F^2 - 2C_F C_A $ \\
      &$+ \xi^2 \eta^2/2 - \xi \eta^3/3 + \eta^4/12 $    &\\
      c &  $-2\xi/3 - \xi^2/2 + \xi^4/12 - \eta^2/4 + \xi \eta^2/2 - \xi^2 \eta^2/4 - \eta^3/6 $   &    $ C_F C_A $ \\
      &$+ \xi \eta^3/6 $  &\\
      d  &  $-2\eta/3 + \xi^2 \eta/2 + \xi^3 \eta/6 - 5\eta^2/4 + \xi\eta^2 - \xi^2 \eta^2/2 - 5\eta^3/6 $  &    $4C_F^2-2C_F C_A $ \\
      &$+ \xi\eta^3/2 - \eta^4/8 $   &\\
      e &  $-2\xi/3 - 5\xi^2/4 - 5\xi^3/6 - \xi^4/8 + \xi^2\eta + \xi^3 \eta/2 $   &    $2C_F C_A $ \\
      &$+ \xi \eta^2/2 - \xi^2 \eta^2/2 + \xi \eta^3/6 $  &\\
      f  &  $-\xi^2 \eta - \xi^3 \eta/2 - \xi \eta^2 + \xi^2 \eta^2 - \xi \eta^3/2 $     &    $ 4 C_F C_A$ \\
      g  &  $ 2\eta/3 - \xi \eta + \xi^3 \eta/3 + \eta^2/4 + \xi \eta^2/2 - 3\xi^2 \eta^2/4  - \eta^3/6$     &    $ C_F C_A - \frac{1}{2} C_A^2  $ \\
      &  $  + \xi \eta^3/2 - \eta^4/8+2\xi/3 + \xi^2/4 - \xi^3/6 - \xi^4/8 - \xi \eta $           &\\
      &$+ \xi^2 \eta/2 + \xi^3 \eta/2 - 3\xi^2 \eta^2/4 + \xi \eta^3/3 $&\\
      h    &  0     &    $ \frac{1}{4} C_A^2$ \\
      i&  $\eta^2/2 - \xi \eta^2 + \xi^2\eta^2/2 + \eta^3/2 - \xi \eta^3/2 +  \eta^4/8 $     &    $2C_F C_A -  C_A^2 $ \\
      j &  $\xi \eta - \xi^2 \eta/2 - \xi^3 \eta/2 - \xi \eta^2/2 + 5\xi^2 \eta^2/4 - \xi \eta^3/2 $     &    $ 2C_F C_A -  C_A^2 $ \\
      k  &  $\xi \eta^2/2 - \xi^2 \eta^2/2 + \xi \eta^3/3 $     &    $2 C_A^2 $ \\
      l  &  $\eta^2/4 - \xi \eta^2/2 + \xi^2 \eta^2/4 + \eta^3/6 - \xi \eta^3/6 +  \eta^4/24$     &    $4C_F^2-6C_F C_A+2 C_A^2 $ \\
      m &  $ \xi^2 \eta^2/4 $     &    $C_A^2$ \\
      n &  $2\xi/3 + \xi^2/4 - \xi^3/6 - \xi^4/8 - \xi\eta + \xi^2 \eta/2 + \xi^3 \eta/2 $    &    $2C_F C_A- C_A^2$ \\
      &$- 3\xi^2 \eta^2/4 + \xi \eta^3/3 $ &\\
      o &  $\eta^2/4 - \xi \eta^2/2 + \xi^2 \eta^2/4 +  \eta^3/6 - \xi \eta^3/6 $     &    $C_F C_A-\frac{1}{2} C_A^2 $\\
      p  &  $ \xi^2\eta^2/4 $     &    $ 4C_A^2$ \\
      q  &  $ \xi \eta^2 - \xi^2 \eta^2 + \xi \eta^3/2 $     &    $4C_F C_A - 2 C_A^2 $ \\
      r  &  $\eta^2/4 - \xi\eta^2/2 + \xi^2 \eta^2/4 +  \eta^3/3 - \xi \eta^3/3 +  \eta^4/12 $     &    $  C_A^2$ \\
      s &  $ \eta^2/4 - \xi \eta^2/2 + \xi^2 \eta^2/4 + \eta^3/6 - \xi \eta^3/6 + \eta^4/24$     &    $ C_F C_A-\frac{1}{2} C_A^2$ \\
      t &  $\eta^2/4 - \xi \eta^2/2 + \xi^2 \eta^2/4 + \eta^3/3 - \xi \eta^3/3 + \eta^4/12 $     &    $4C_F^2 - 4C_F C_A + C_A^2 $ \\
      u  &  $\xi^2 \eta^2/4 $     &    $ C_A^2 $ \\
      v &  $ \xi \eta - \xi^2 \eta/2 - \xi^3 \eta/2 - \xi \eta^2/2 + 5\xi^2 \eta^2/4 - \xi \eta^3/2 $     &    $4C_F^2 - 4C_F C_A + C_A^2 $ \\
      w &  $2\eta/3 - \xi \eta + \xi^3 \eta/3 + \eta^2/4 + \xi \eta^2/2 - 3\xi^2 \eta^2/4 - \eta^3/6$      &    $4C_F^2 - 4C_F C_A +  C_A^2$ \\
      & $+ \xi \eta^3/2 - \eta^4/8$&\\
      x &  $\xi \eta^2/2 - \xi^2 \eta^2/2 + \xi \eta^3/6	 $     &    $\frac{1}{2} C_A^2 $ \\
      y &  $\eta^2/4 - \xi^2 \eta^2/4 + \eta^3/6 + \eta^4/24 $     &      $\frac{1}{2} C_A^2 $ \\
      z  &  $\eta^2/4 - \xi \eta^2/2 + \xi^2 \eta^2/4 + \eta^3/6 - \xi \eta^3/6 $     &    $\frac{1}{2} C_A^2 $ \\
      $\rho$  &  $\xi^2 \eta/2 + \xi^3 \eta/6 - \xi^2 \eta^2/2 $     &    $ C_A^2 $ \\
      $\sigma$ &  $2\xi/3 + \xi^2/4 - \xi^3/6 - \xi^4/8 - \xi \eta^2/2 + \xi^2 \eta^2/2 - \xi \eta^3/6 $     &    $\frac{1}{2} C_A^2 $ \\
      $\tau$ &  $\eta^2/2 - \xi \eta^2 + \xi^2 \eta^2/2 + \eta^3/2 - \xi \eta^3/2 + \eta^4/8 $     &    $2C_F C_A- C_A^2 $ \\
      \hline
      \end{tabular}
      \end{center}
      \caption{\label{tab::2}
The weights $w^{(2)}_\lambda$ and the color factors $c^{(2)}_\lambda$ for the diagrams in Fig.~\ref{fig::4}.
}
\end{table}

\subsection{Explicit evaluation of the three-loop amplitude}
\label{sec::2.3}
For the calculation of the three-loop logarithmic corrections we use the same
method of Sudakov parameters but now have to integrate over two virtual soft
gluon momenta ${l_g}_1$ and ${l_g}_2$. The relevant Feynman diagrams are
given in Fig.~\ref{fig::4}. The sum of all the diagrams has the infrared
divergent part  described by the Sudakov factor Eq.~(\ref{eq::Zq}). As in the
two-loop case the divergent parts of the individual diagrams  can be
separated  in the Sudakov parameter space by subtracting the factorized
infrared divergent contributions where the upper integration for a given
parameter ${v_{g}}_i$ or ${u_{g}}_j$  is set to $1$ ({\it cf.}
Eq.~(\ref{eq::intlfac})). The  factorized double-pole singular contributions
of the diagrams in Figs.~\ref{fig::4}(a-c) indeed add up to
\begin{equation} {C_F^2\over
2}\left({\alpha_s\over 2\pi}{\ln\rho\over \varepsilon}+x\right)^2{\cal
G}^0\,,
\label{eq::2loopsud2}
\end{equation}
while the factorized single-pole contributions of the diagrams in Figs.~\ref{fig::4}(a-f)
give
\begin{equation}
{C_Fz\over 6}\left({\alpha_s\over 2\pi}{\ln\rho\over \varepsilon}+x\right){\cal G}^0\,,
\label{eq::2loopsud1}
\end{equation}
which agrees with Eq.~(\ref{eq::Zq}-\ref{eq::gseries}). Though
some of the remaining diagrams taken separately are infrared divergent, their
sum is finite and gives the following contribution to the amplitude
\begin{equation}
x^2\left(2\sum_\lambda c^{(2)}_\lambda
\int_0^1 {\rm d}\xi \int_{0}^{1-\xi}{\rm d}\eta\, w^{(2)}_\lambda(\eta,\xi)\right)\,
{\cal G}^0\,,
\label{eq::3loopG}
 \end{equation}
where the  color factors $c^{(2)}_\lambda$ and the weight function $w^{(2)}_\lambda$
resulting from the logarithmic integration over the two soft gluon momenta
are collected in Table~\ref{tab::2}. Note that the  weights  $w^{(2)}_\lambda$
correspond to the infrared subtracted diagrams and the weights for the
symmetric diagrams not shown in Fig.~\ref{fig::4}  are obtained by
interchanging the $\eta$ and $\xi$ variables and should be included into the sum.
It is straightforward to check that the sum in Eq.~(\ref{eq::3loopG}) reduces to
\begin{equation}
z^2
\left(2\int_0^1 {\rm d}\xi \int_{0}^{1-\xi}{\rm d}\eta\,
2\left(\eta\xi\right)^2\right){\cal G}^0
={z^2\over 45}{\cal G}^0\,,
\label{eq::3loopg}
\end{equation}
in full agreement with Eqs.~(\ref{eq::Gfac}-\ref{eq::gseries}).

\section{Higgs boson production  mediated by bottom quark loop}
\label{sec::3}
The analysis and the result of the previous section can be generalized in a
straightforward way to an important case of the bottom quark mediated Higgs
boson production in gluon fusion. We postpone the discussion of the
phenomenological aspects of this process to the last section and focus now on
the structure of the radiative corrections. The leading order contribution is
given by the one-loop diagram in Fig.~\ref{fig::1}(c). Note that the dominant
contribution to the gluon fusion process is given by the same diagram with
the top quark loop and in the formal limit of the large top quark mass
$m_t\gg m_H$ is proportional  to the square of the Higgs boson mass $m_H$. By
contrast for the intermediate bottom quark with $m_b\ll m_H$ the amplitude is
suppressed by the square of the bottom quark mass. Indeed, the  Higgs boson
coupling to the bottom quark is proportional to  $m_b$.  Then the scalar
interaction of the Higgs boson results in a helicity flip at the interaction
vertex and helicity conservation requires the amplitude to vanish in the
limit $m_b\to 0$ even if the Higgs coupling to the bottom quark is kept
fixed. As in the example considered in the previous section the additional
power of $m_b$ originates  from the $t$-channel quark propagator which
effectively becomes scalar and results in double-logarithmic scaling of the
diagram absent for the top quark contribution. By using the explicit one-loop
result the bottom quark mediated amplitude can be written in such a way that
its power suppression and the logarithmic enhancement  is manifest
\begin{equation}
{{\cal M}^b}^{(0)}_{gg\to H}=-{3\over 2}\ln^2\!\rho\,\rho \,
{{\cal M}^{t}}^{(0)}_{gg\to H}\,,
\label{eq::MH0}
\end{equation}
where $\rho=m_b^2/s$  is now a Minkowskian parameter, $s\approx m_H^2$ is the
total energy of colliding gluons, and the result is given in terms of the
heavy top quark  mediated  amplitude ${{\cal M}^{t}}^{(0)}_{gg\to H}$, which
corresponds to a local gluon-gluon-Higgs interaction vertex and has one
independent helicity component.

\begin{figure}
\begin{center}
\begin{tabular}{cccc}
\includegraphics[width=2.2cm]{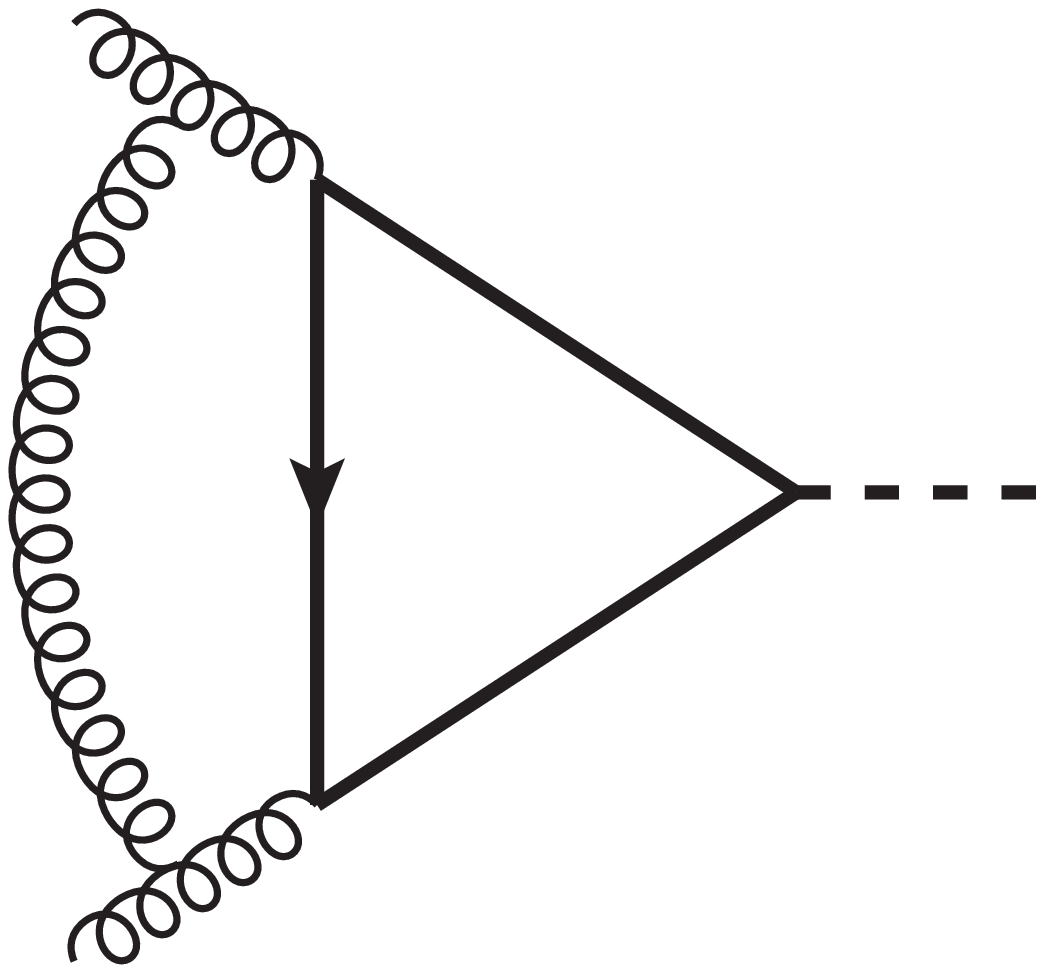} &
\hspace*{03mm}\includegraphics[width=2.2cm]{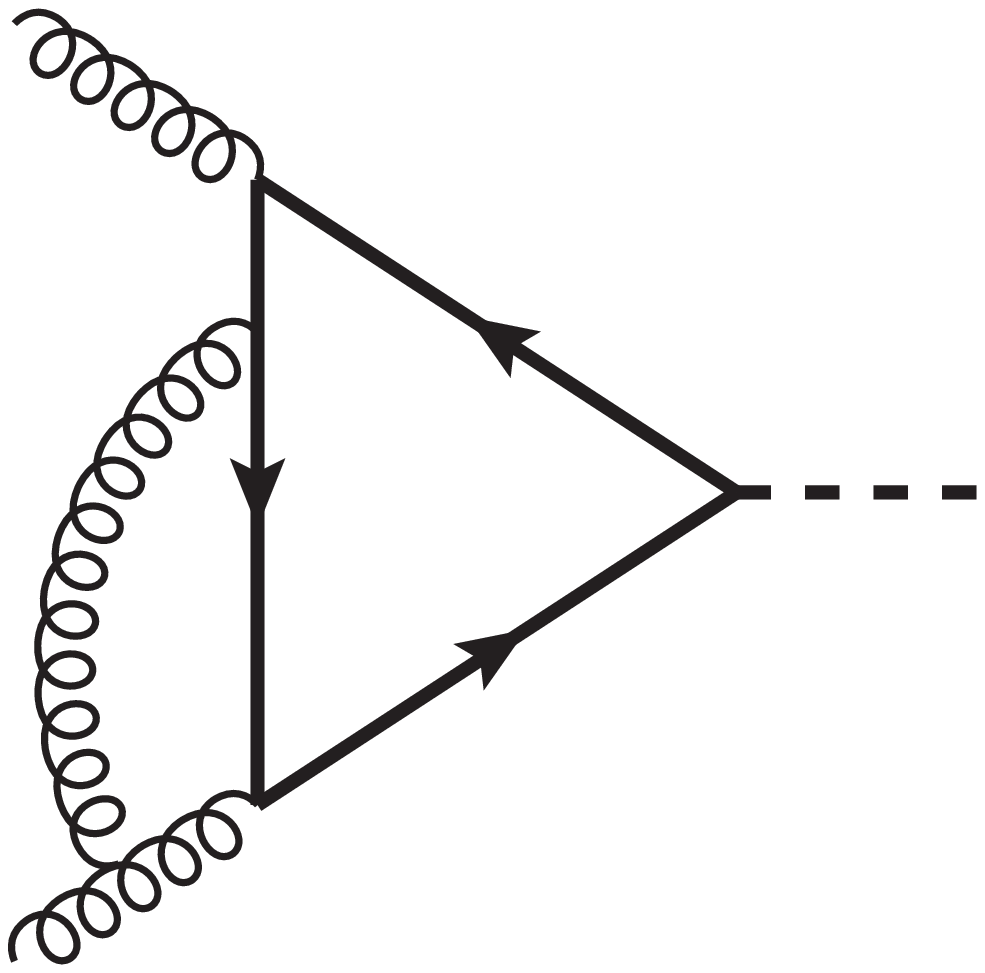} &
\hspace*{03mm} \includegraphics[width=2.2cm]{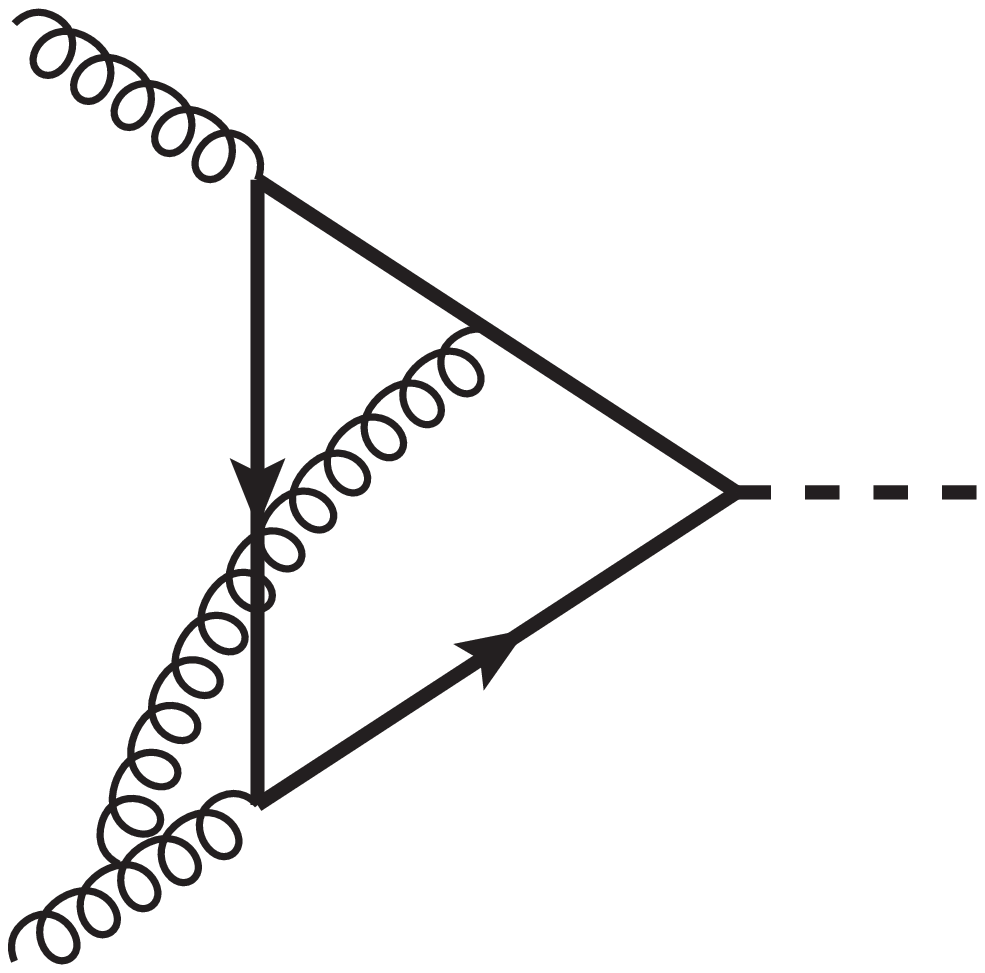} &
\hspace*{03mm}\includegraphics[width=2.2cm]{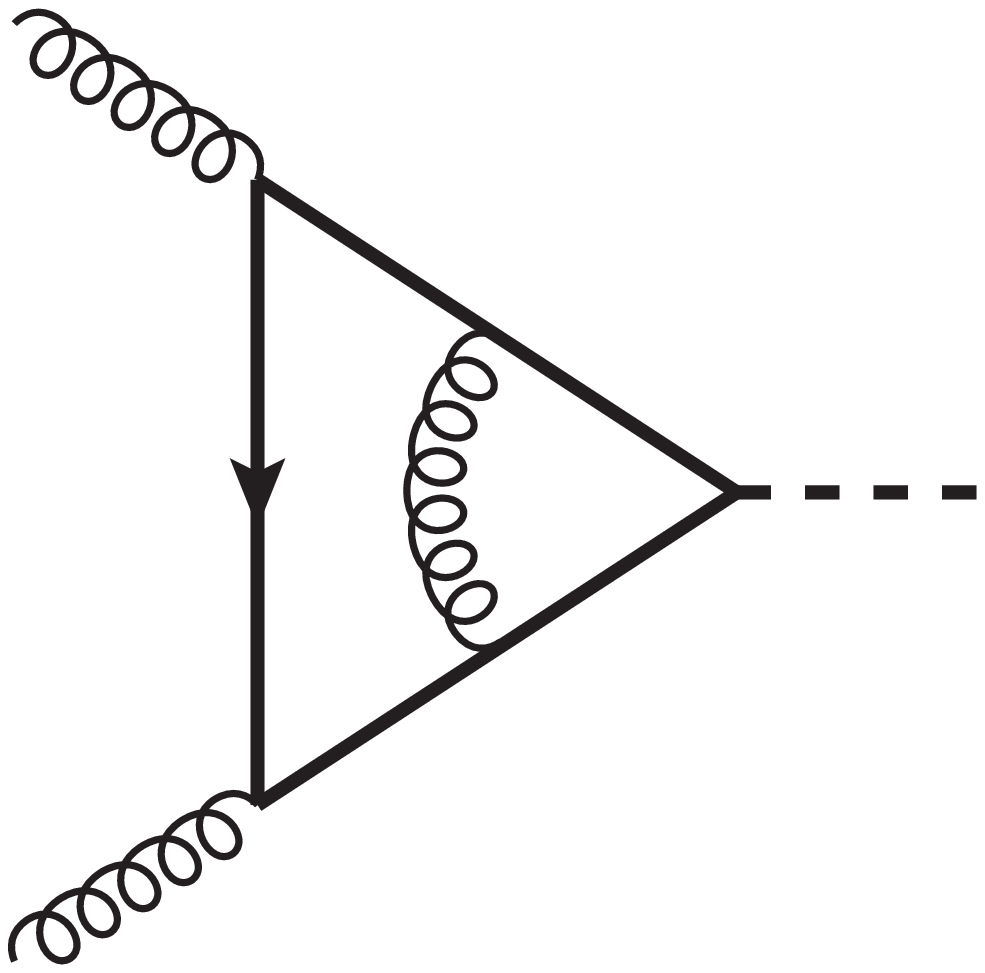}\\
~~~(a) & ~~~(b)  & ~~~(c) & ~~~(d)
\end{tabular}
\end{center}
\caption{\label{fig::5}  The two-loop  Feynman diagrams for the  bottom quark
mediated $gg\to H$ fusion, which contribute in the double-logarithmic
approximation. Symmetric diagrams and the diagrams with the opposite
direction of the closed quark line are not shown.}
\end{figure}

Thus the processes  in Figs.~\ref{fig::1}(a,c) are similar in a few important
aspects: they are mass-suppressed due to  the quark helicity flip, the
double-logarithmic contribution is induced by the soft quark exchange  and
the color charge is not conserved along the eikonal lines.  Moreover, since
the eikonal or Wilson lines are characterized by the momentum and color
charge but not the spin, in the double-logarithmic approximation the
processes are identical up to the color group representation of the external
lines and the direction of the color charge flow to/from an eikonal line  at
the soft quark emission vertex.  Therefore   the factorization structure of
the double-logarithmic corrections found in the Sect.~\ref{sec::2} directly
applies to the case under consideration. In particular the two-loop
non-Sudakov double-logarithmic contribution is given by the  diagram
Fig.~\ref{fig::1}(d) where  the effective gluon exchange has the color weight
$C_F-C_A$ rather than $C_A-C_F$ of the diagram Fig.~\ref{fig::1}(b), and the
higher-order non-Sudakov  double-logarithmic terms are  described by the same
function $g(z)$ with $C_A$ and $C_F$ exchanged, {\it i.e.} with the opposite
sign of the argument. Hence to all orders in $\alpha_s$ we get
\begin{equation}
{\cal M}^b_{gg\to H}= Z^2_{g} g(z)\,{{\cal M}^b}^{(0)}_{gg\to H}\,,
\label{eq::MH}
\end{equation}
where
\begin{equation}
 Z_{g}^2=\exp\left[{-{C_As^{-\varepsilon}\over\varepsilon^2}
 {\alpha_s\over 2\pi}}\right]
\label{eq::Zg}
 \end{equation}
is the Sudakov factor for  a gluon scattering. Let us now demonstrate how the
above factorization is realized for the two-loop corrections. The
relevant diagrams are given in Fig.~\ref{fig::5} and the corresponding
contribution to the amplitude can be written in the form similar to
Eq.~(\ref{eq::G2loop})
\begin{equation}
x\left(2\sum_i c^{(1)}_\lambda \int_0^1 {\rm d}\xi \int_{0}^{1-\xi}
{\rm d}\eta\, \tilde{w}^{(1)}_\lambda(\eta,\xi)\right)\,{{\cal M}^b}^{(0)}_{gg\to H}\,,
\label{eq::MH2loop}
 \end{equation}
with the color factors   $c^{(1)}_\lambda$ and the weights
$\tilde{w}^{(1)}_\lambda$ collected in Table~\ref{tab::3}.  As before the
weights for the symmetric diagrams not shown in Fig.~\ref{fig::5} are
obtained by interchanging the $\eta$ and $\xi$ variables and should be
included into the sum. To make the factorization of the Sudakov logarithms
explicit  we do not subtract the factorized contribution and the functions
$\tilde{w}^{(1)}_i$ correspond to the unsubtracted Feynman integrals over the
soft gluon momentum, which are infrared divergent. They are  regularized  by
introducing a small auxiliary gluon mass $\lambda_g\ll m_b$, which is more
convenient for the calculation in the Sudakov parameter space than
dimensional regularization. In Table~\ref{tab::3} the dependence on the
infrared  regulator is encoded into the parameter
$\tau=\ln^2(\lambda^2_g/s)/\ln^2(m_b^2/s)$. The contributions of the
individual diagrams in  Eq.~(\ref{eq::MH2loop}) combine into the sum of two
terms
\begin{equation}
\left(2\int_0^1 {\rm d}\xi \int_{0}^{1-\xi}{\rm d}\eta\,
\left(2z\eta\xi-C_Ax\tau\right)\right){{\cal M}^b}^{(0)}_{gg\to H}\,.
\label{eq::MH2loopfac}
\end{equation}
The first term in Eq.~(\ref{eq::MH2loop})  coincides with the expression for
the diagram Fig.~\ref{fig::1}(d) and represents the first-order term in the
perturbative expansion of the function $g(z)$. The second term in  the
brackets does not depend on the soft quark momentum variables, {\it i.e.}
here the soft gluon momentum integral factorizes and gives the one-loop
massive gluon Sudakov form factor $-C_Ax\tau=-{C_A\alpha_s\over
4\pi}\ln^2(\lambda^2_g/s)$. After converting to the dimensional
regularization it recovers the first-order term in the perturbative expansion
of $Z_g^2$. Thus the double-logarithmic contributions factorize at the
integrand level as suggested by the Ward identities discussed in the previous
section. Note that the two-loop contribution to Eq.~(\ref{eq::MH}) agrees
with the analytical result  for the amplitude with an arbitrary  value of the
quark mass \cite{Anastasiou:2006hc} expanded in the series in $\rho$.

\begin{table}[t]
\begin{center}
    \begin{tabular}{|c|c|c|c|}
    \hline
       $\lambda$ &  $\tilde w^{(1)}_\lambda$   & $c^{(1)}_\lambda$ \\
      \hline
      a &   $-(\eta +\xi -\tau)^2$              &    $\frac{1}{2}C_A$\\
      b & $\eta^2 + 2\eta \xi- 2\eta \tau $     &     $\frac{1}{4}C_A$ \\
      c &    $\eta^2 + 2\eta \xi- 2\eta \tau $  &    $\frac{1}{4}C_A$ \\
      d &   $ 2\eta\xi $                        &    -$\frac{1}{2}C_F$ \\
      \hline
      \end{tabular}
     \caption{\label{tab::3} The weights $\tilde w^{(1)}_\lambda$
     and the color factors $c^{(1)}_\lambda$  for the diagrams in Fig.~\ref{fig::5}.}
\end{center}
\end{table}

\section{Quark form factors beyond the leading-power approximation}
\label{sec::4}

In this section we consider  the asymptotic behavior of the leading
mass-suppressed  contribution to the amplitude of quark scattering in an
external field. The problem is more complex since in contrast to the
amplitudes considered in the previous sections the quark form factors do get
the leading-power contribution which does not vanish in $m_q\to 0$ limit. In
Refs.~\cite{Penin:2014msa,Penin:2016wiw}  it was shown  within   the
expansion by regions framework  that  a soft gauge boson exchange responsible
for the standard  Sudakov logarithms does not generate the leading
mass-suppressed double-logarithmic  contribution. Such a contribution results
from the soft fermion pair exchange between the eikonal lines.  Therefore the
approach elaborated in the previous section can be naturally extended to the
quark form factors.   We start  with the analysis of the  external vector
field and consider the  scalar  field case next.

\subsection{Vector form factor}
\label{sec::4.1}

\begin{figure}[t]
\begin{center}
\begin{tabular}{ccccc}
\includegraphics[width=1.8cm]{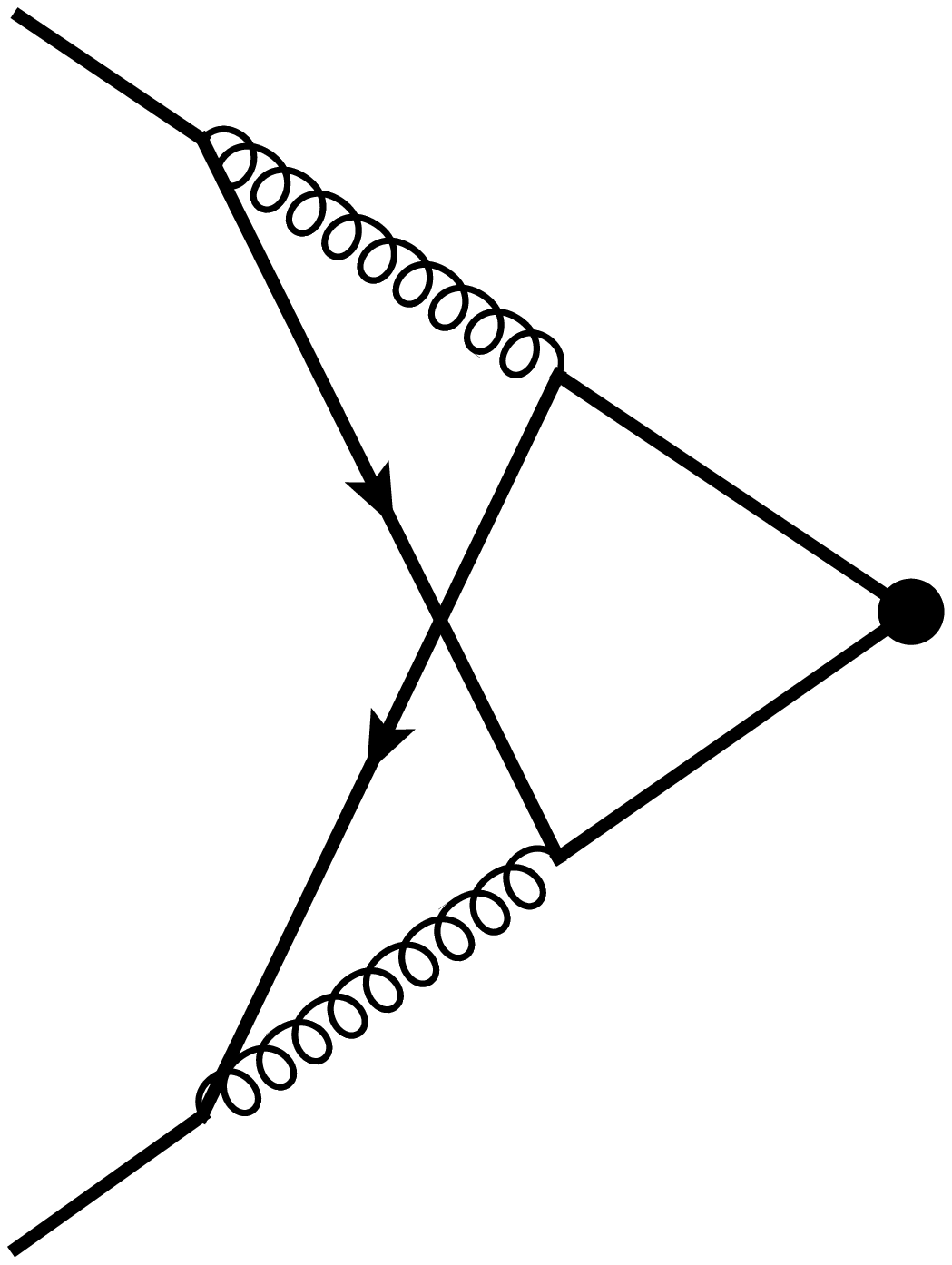}&
\hspace*{5mm}\includegraphics[width=1.8cm]{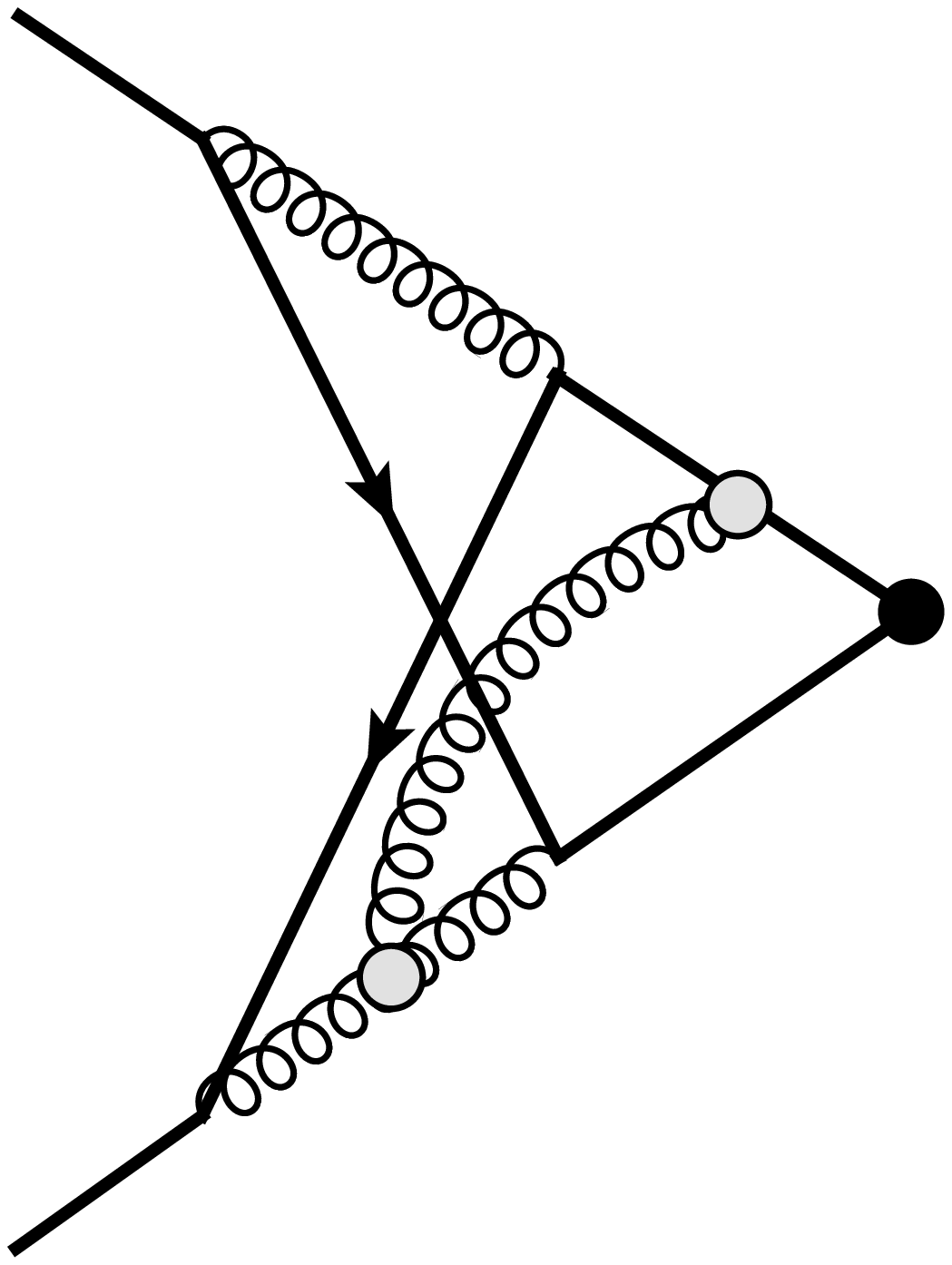}&
\hspace*{5mm}\includegraphics[width=1.8cm]{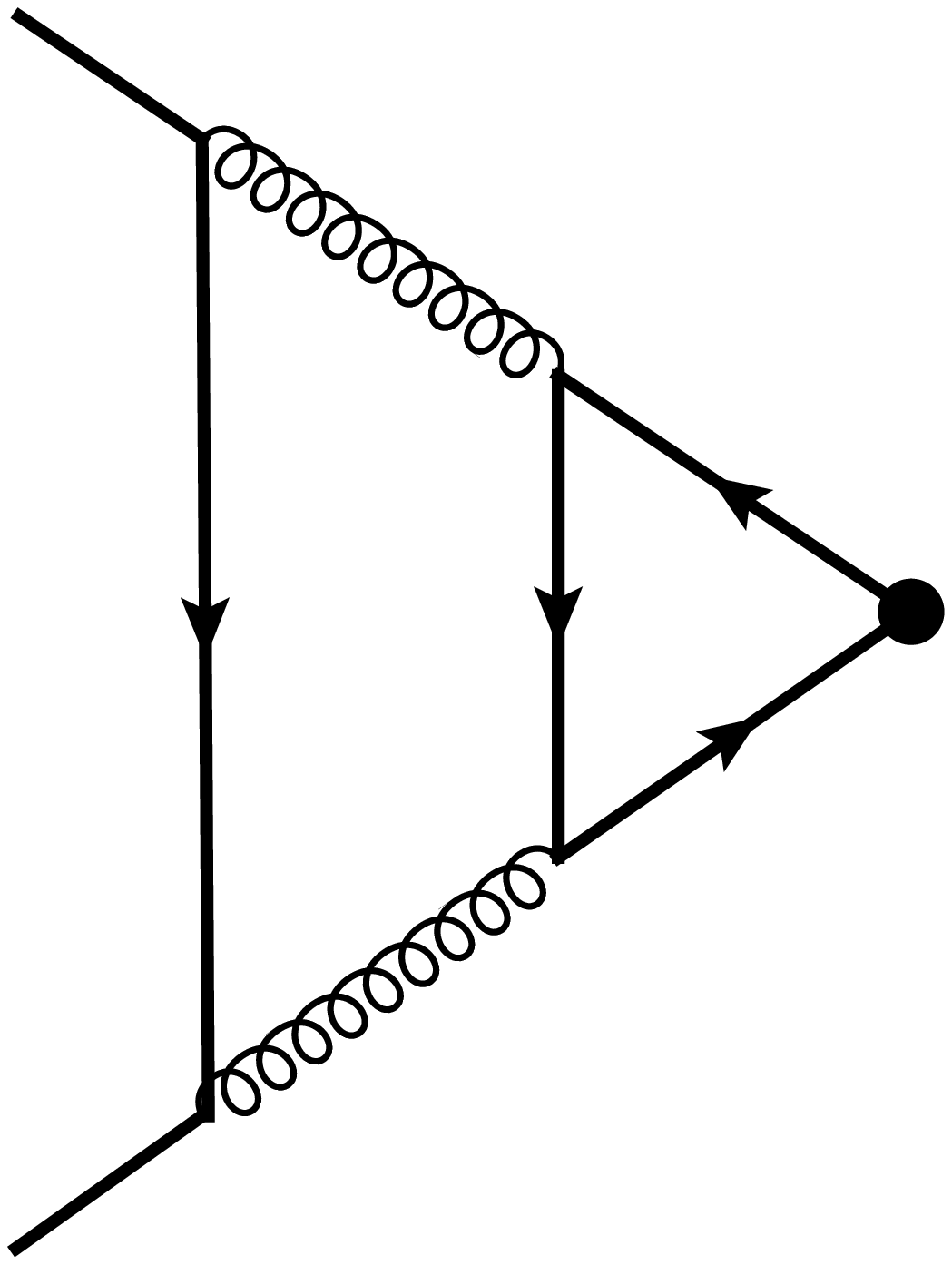}&
\hspace*{5mm}\includegraphics[width=1.8cm]{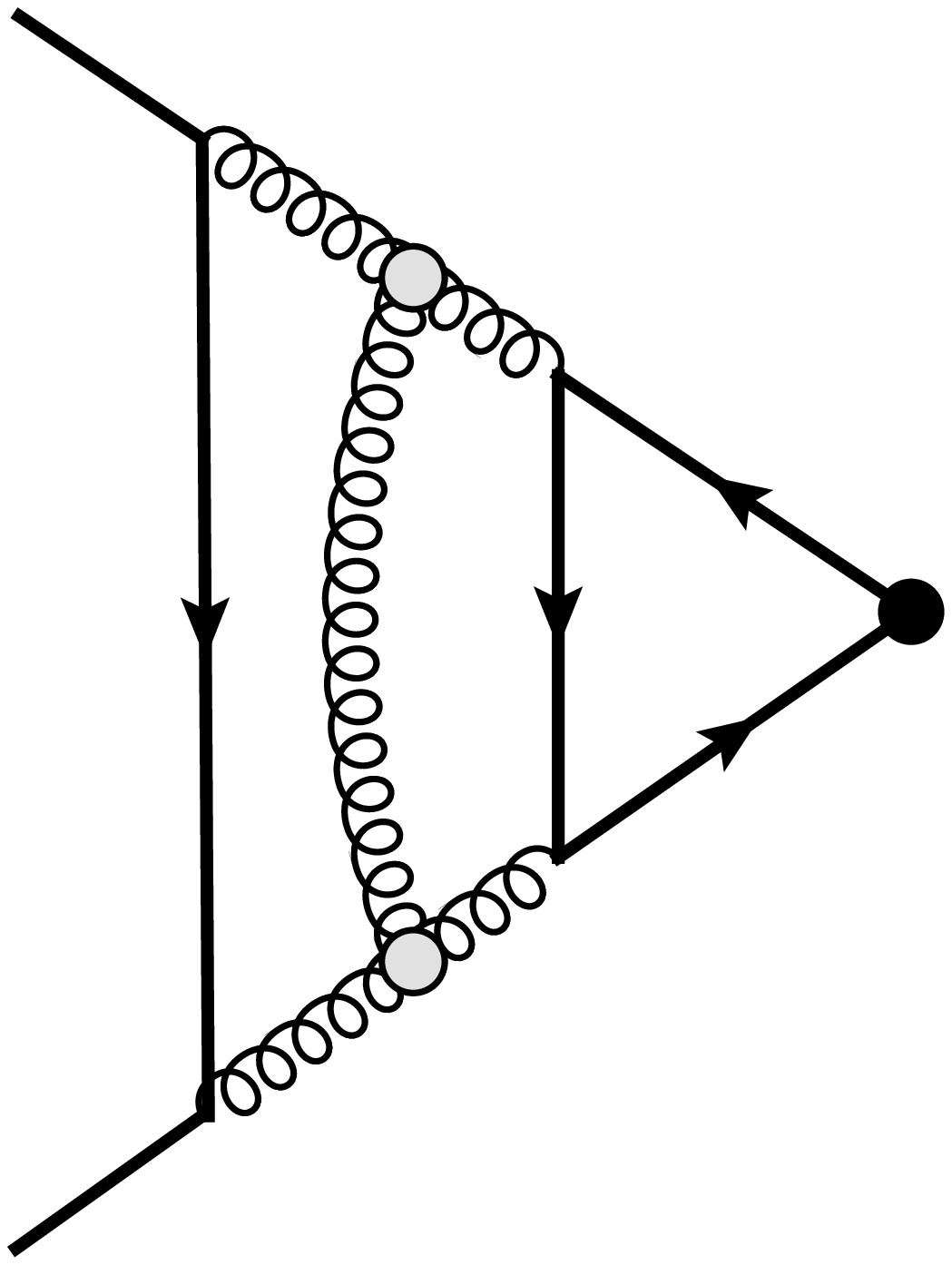}&
\hspace*{5mm}\includegraphics[width=1.8cm]{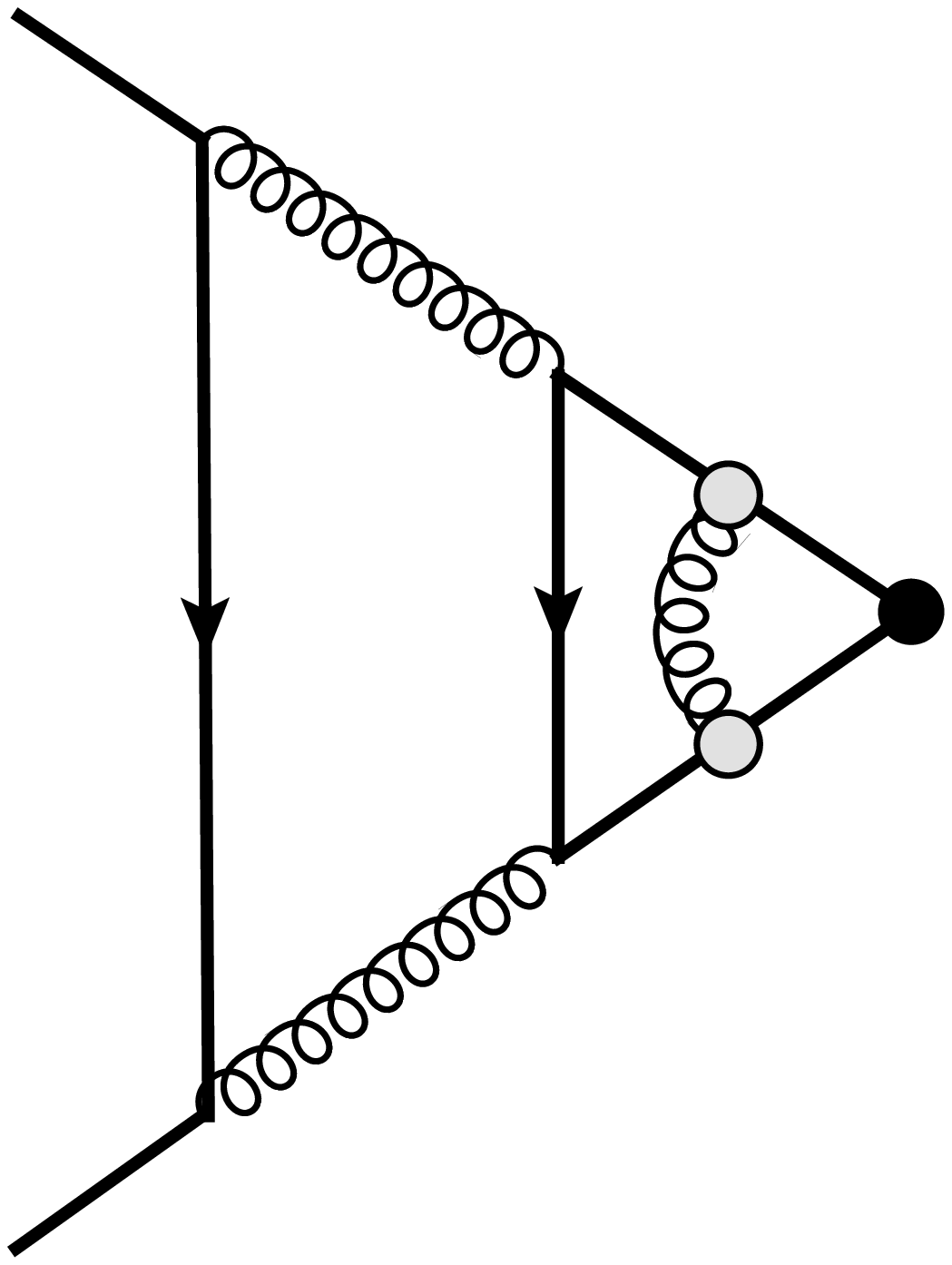}\\
(a)&\hspace*{5mm}(b)&\hspace*{5mm}(c)&\hspace*{5mm}(d)&\hspace*{5mm}(e)\\
\end{tabular}
\end{center}
\caption{\label{fig::6} The leading order two-loop Feynman diagrams for  (a)
vector form factor $F_1^{(1)}$  and (c) scalar form factor $F_S^{(1)}$ in the
double-logarithmic approximation. The diagrams with an effective soft gluon
exchange which incorporate the non-Sudakov double-logarithmic corrections to
(b) vector and (d,e) scalar form factor. Symmetric diagrams and the diagrams
with the opposite  direction of the closed quark line are not shown.}
\end{figure}

\begin{figure}
\begin{center}
\begin{tabular}{ccc}
\includegraphics[width=1.8cm]{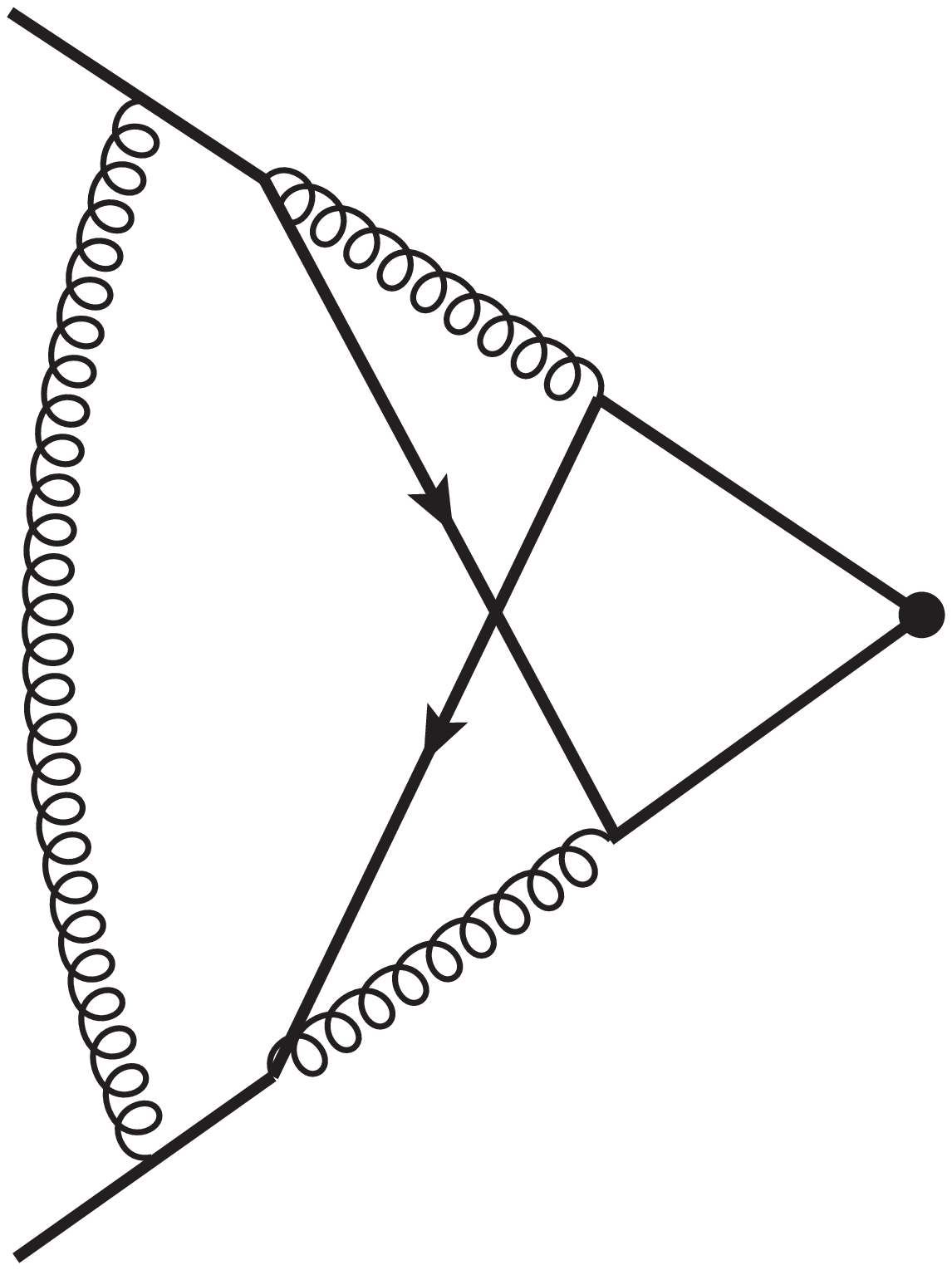}&
\hspace*{10mm}\includegraphics[width=1.8cm]{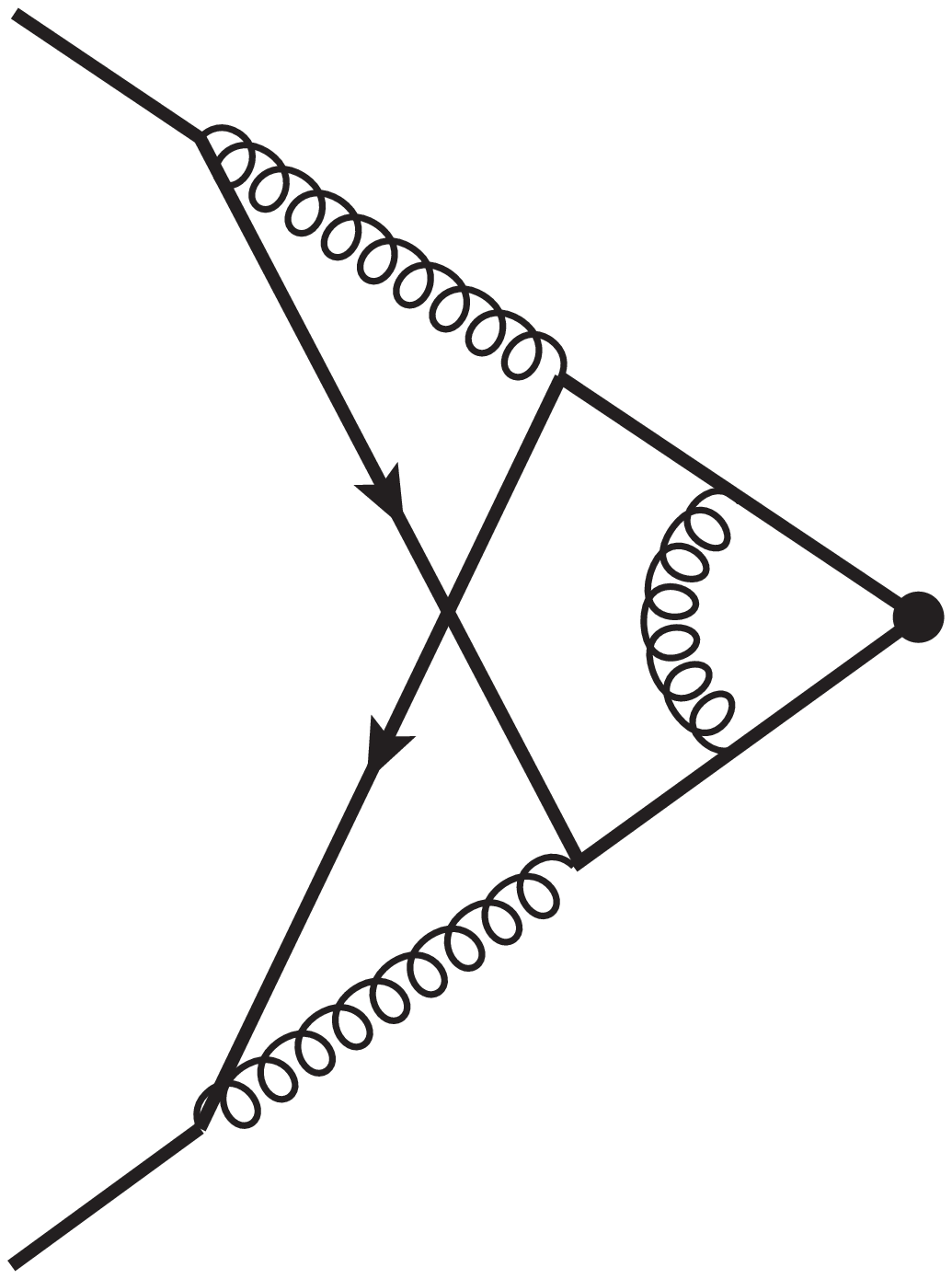}&
\hspace*{10mm}\includegraphics[width=1.8cm]{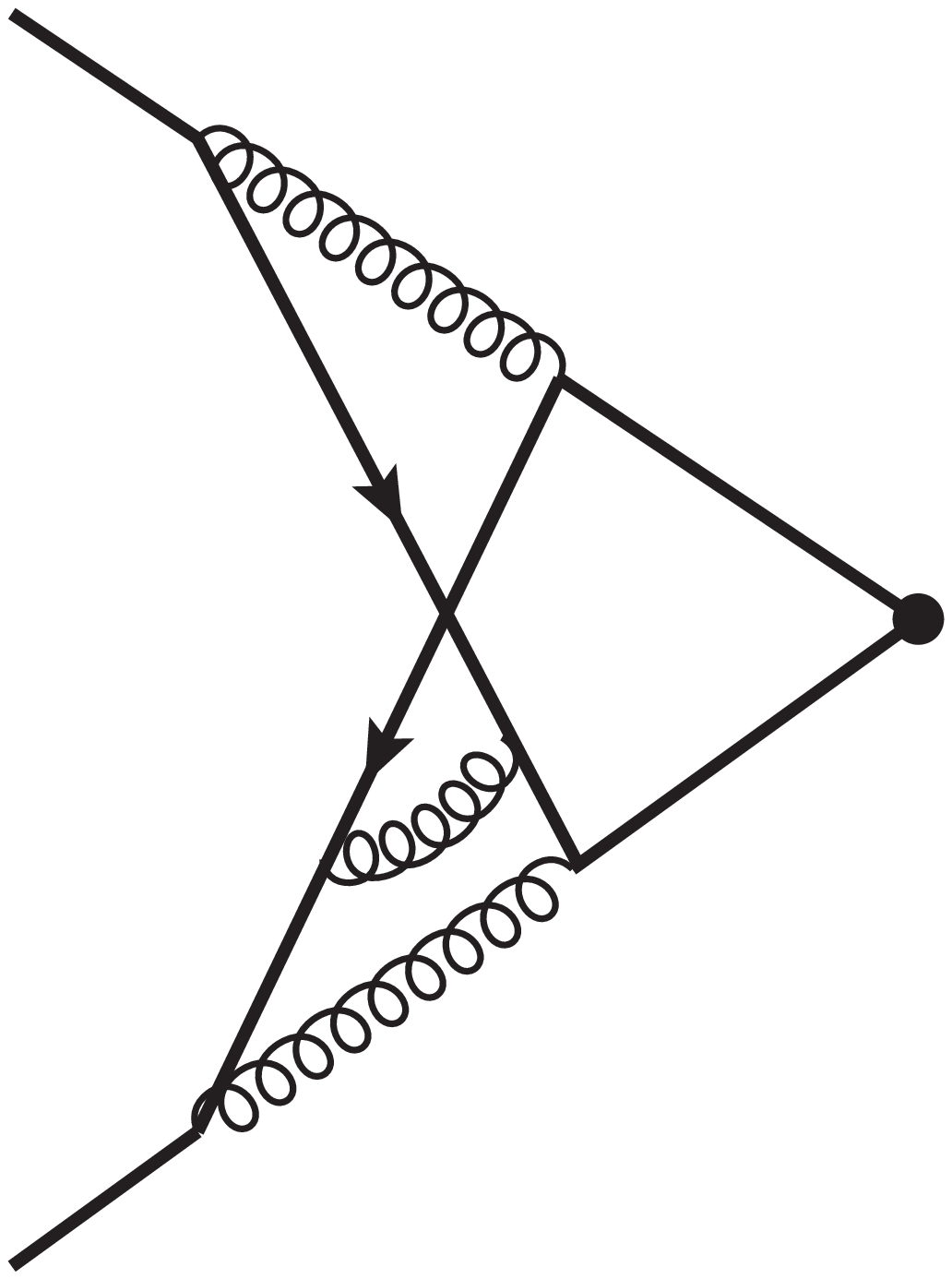}\\
(a)&\hspace*{10mm}(b)&\hspace*{10mm}(c)\\
\includegraphics[width=1.8cm]{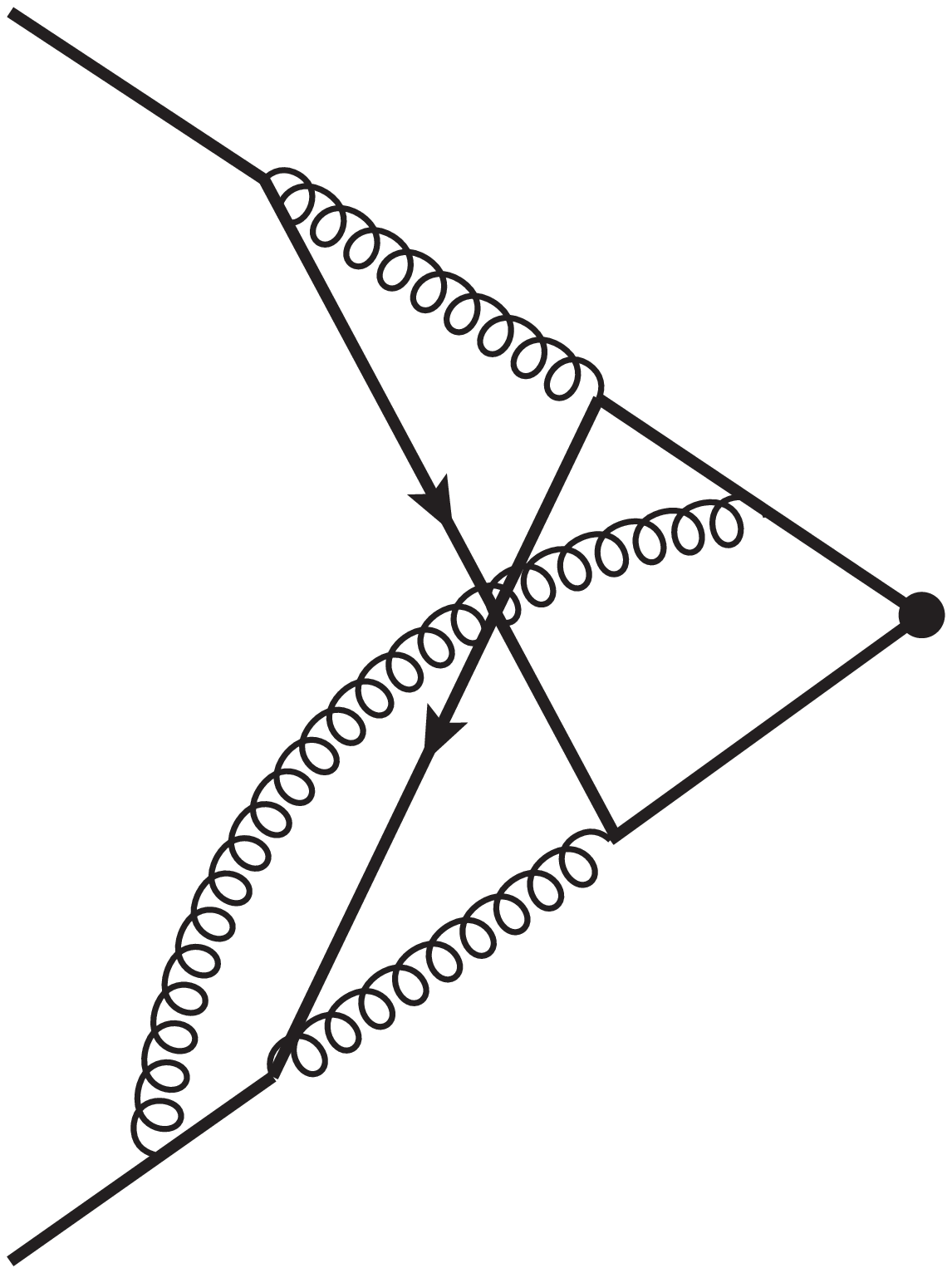}&
\hspace*{10mm}\includegraphics[width=1.8cm]{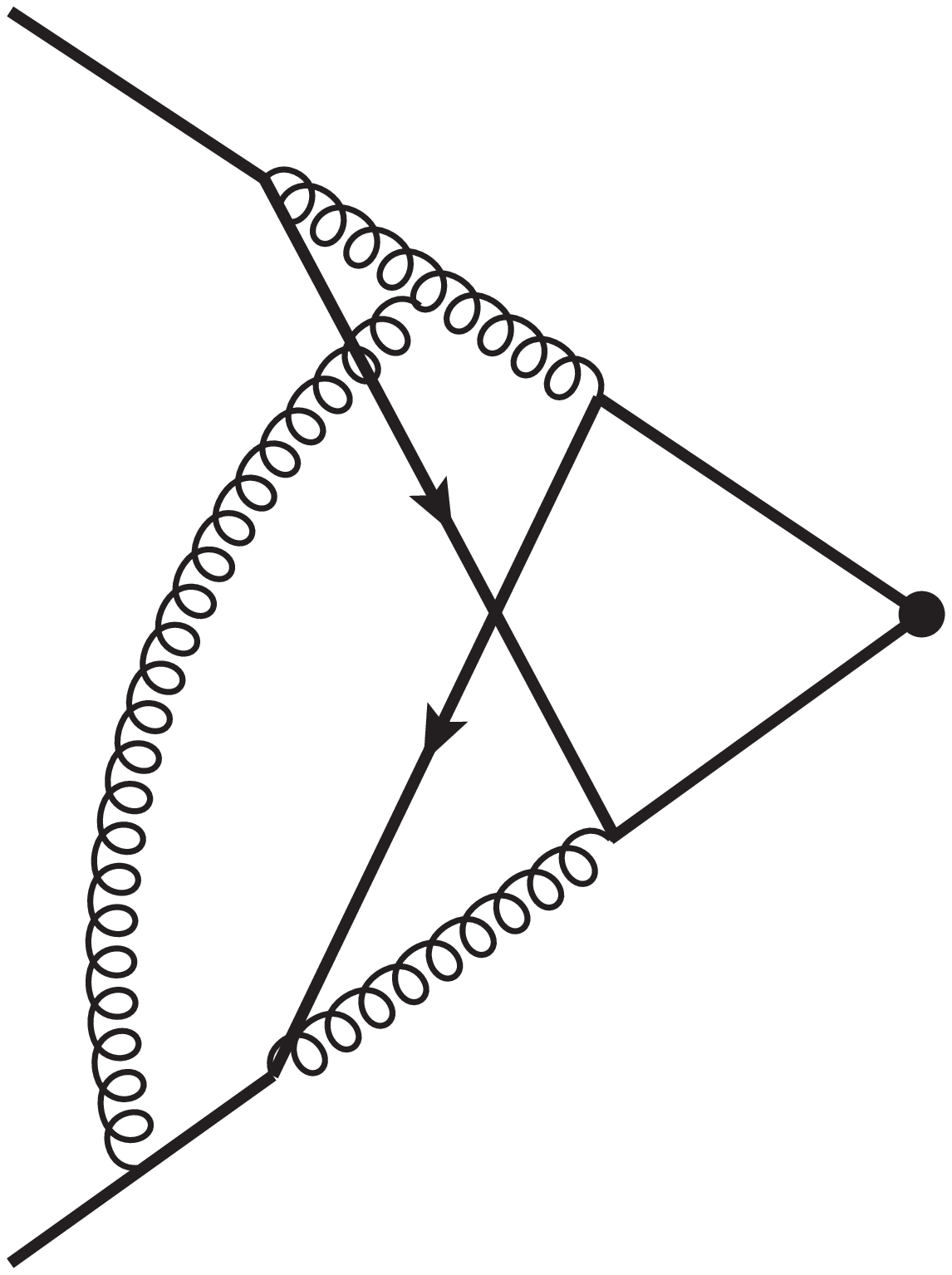}&
\hspace*{10mm}\includegraphics[width=1.8cm]{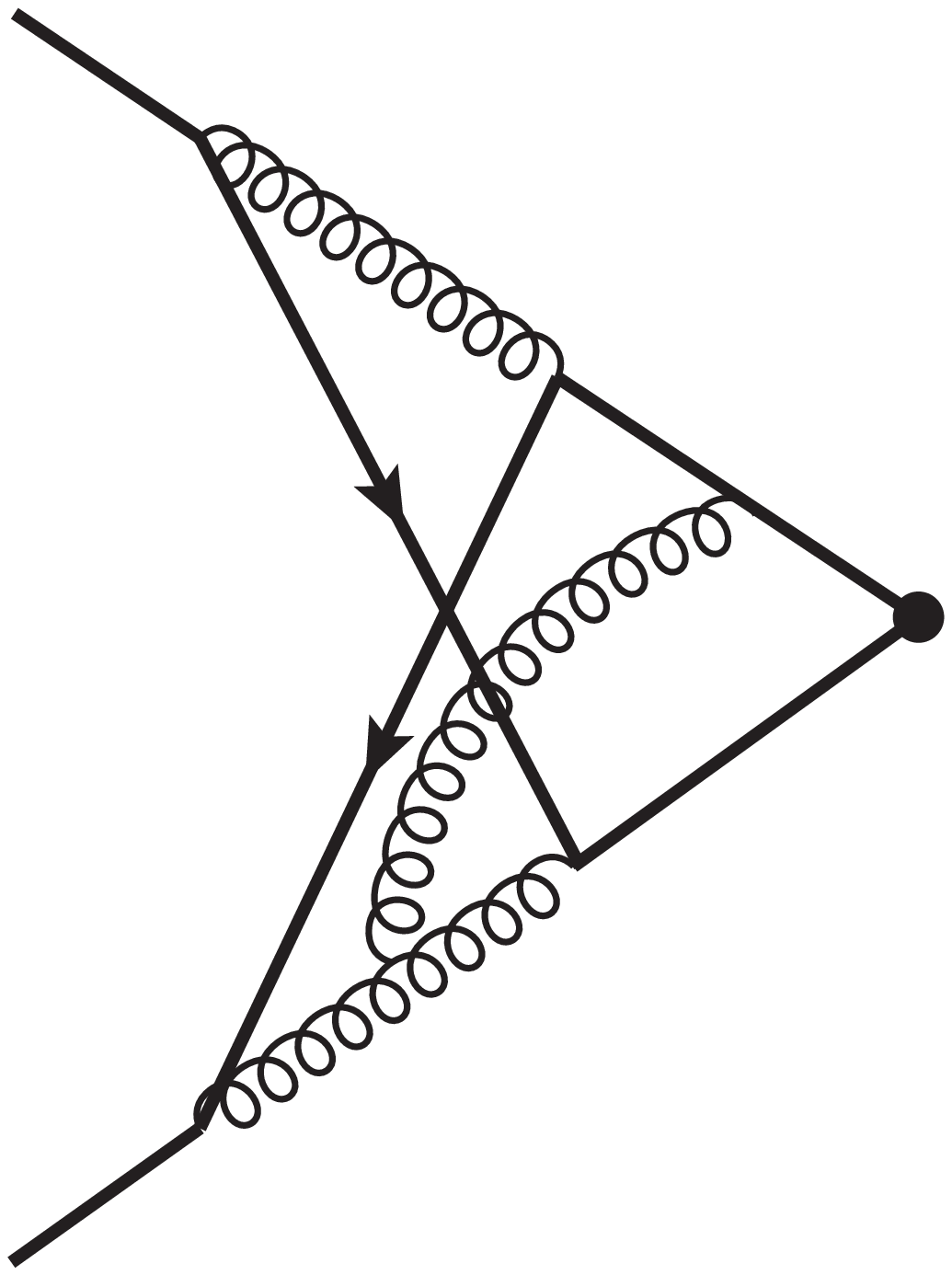}\\
(d)&\hspace*{10mm}(e)&\hspace*{10mm}(f)\\
\includegraphics[width=2.0cm]{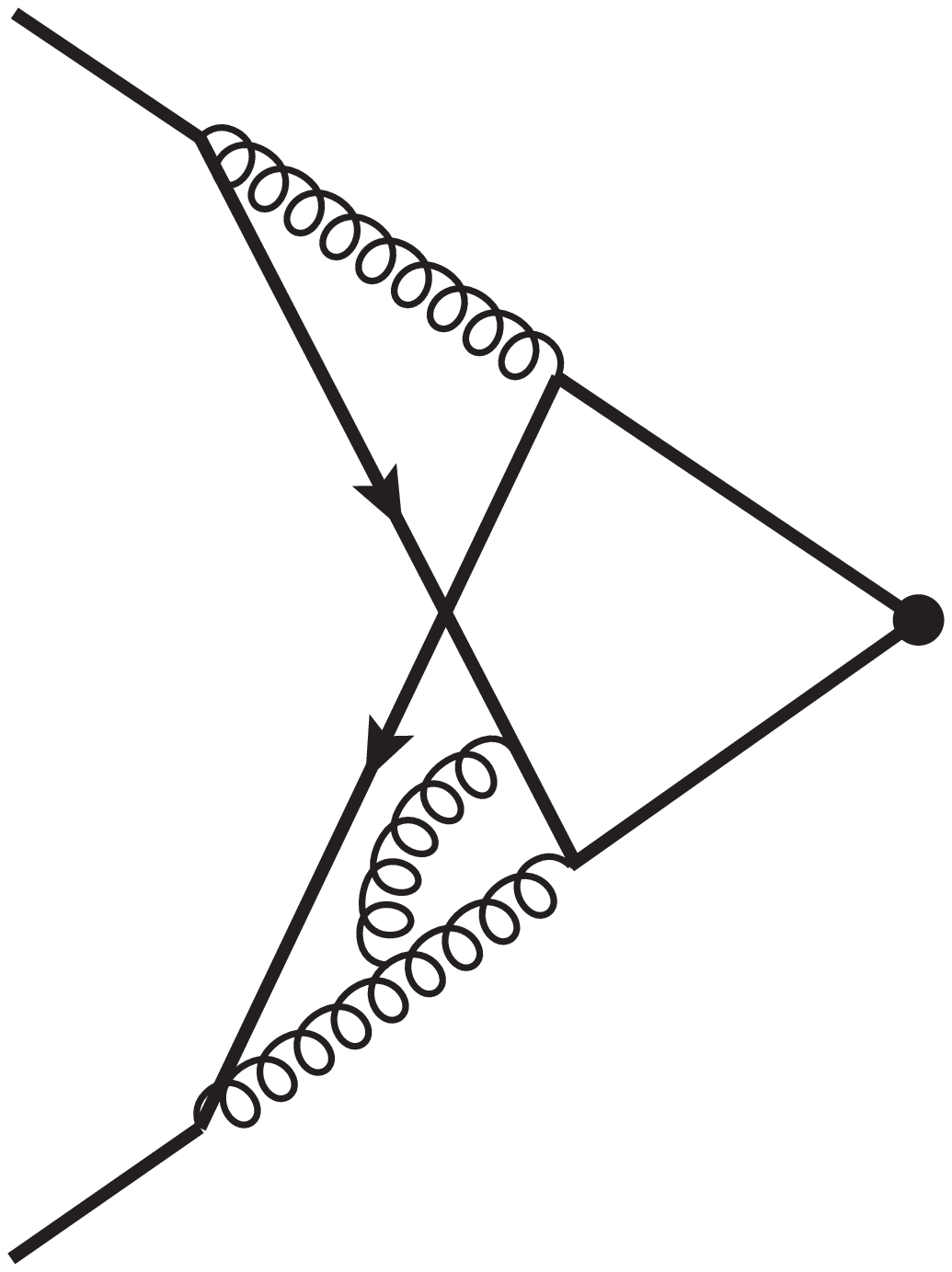}&
\hspace*{10mm}\includegraphics[width=1.8cm]{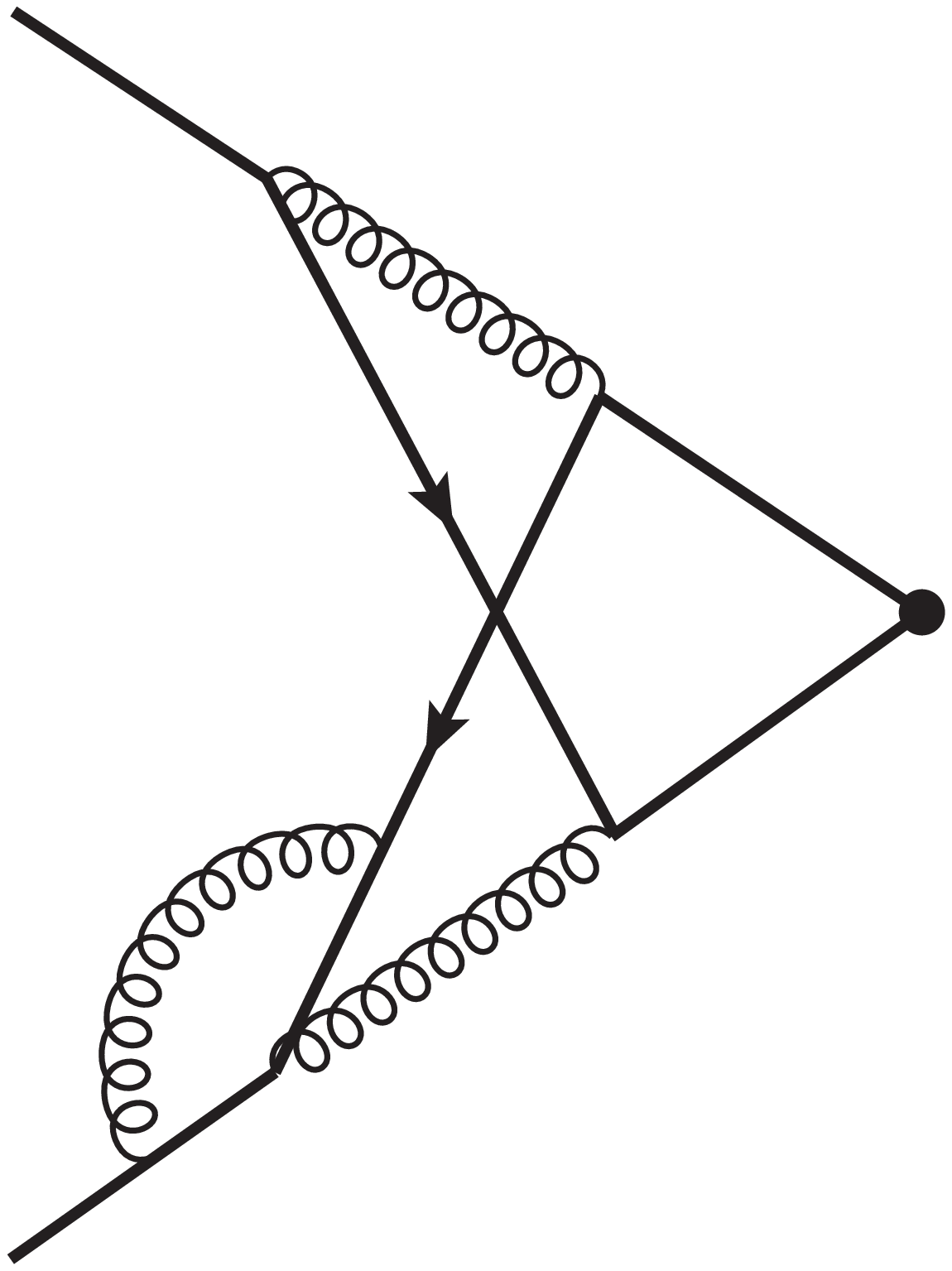}&
\hspace*{10mm}\includegraphics[width=1.8cm]{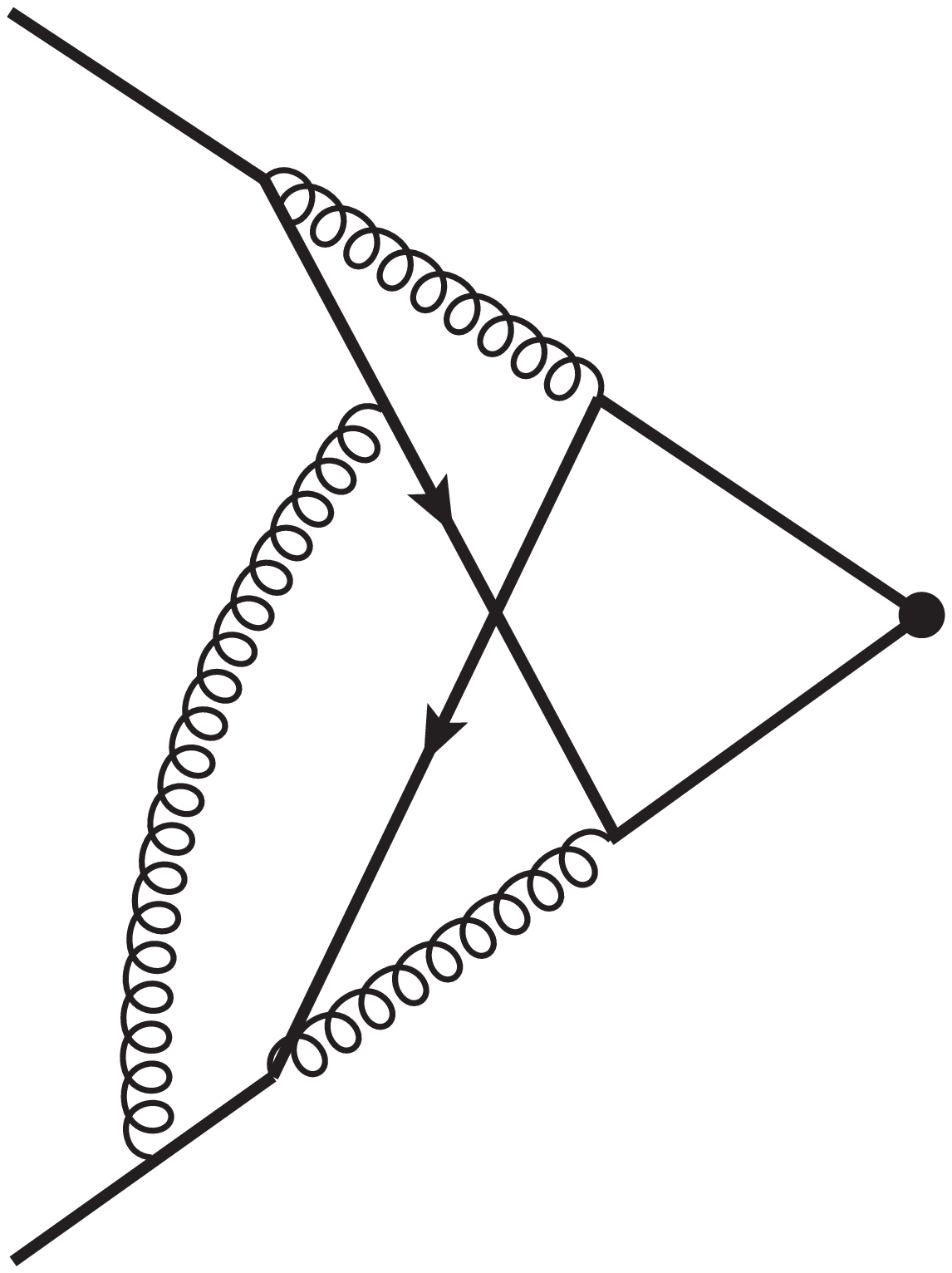}\\
(g)&\hspace*{10mm}(h)&\hspace*{10mm}(i)\\
\end{tabular}
\end{center}
\caption{\label{fig::7}  The three-loop diagrams contributing to the vector
form factor $F^{(1)}_1$ in the double-logarithmic approximation. Symmetric
diagrams are not shown.  The remaining diagrams either do not have the
double-logarithmic integration region or have vanishing color factor.}
\end{figure}

The  amplitude of a quark scattering in an external color-singlet vector
field can be parameterized in the standard way  by the  Dirac and Pauli form
factors. The Pauli form factor contribution to the amplitude at high energy
is suppressed by the first power of $\rho$ but does not acquire the
double-logarithmic corrections in the approximation discussed in this paper.
Indeed, the leading-order one-loop Pauli form factor $F_2$ is finite so the
higher-order Sudakov double logarithms will give a subleading contribution to
the scattering amplitude suppressed by an additional power of the coupling
constant. Thus we focus on the  high-energy behavior of the Dirac form-factor
$F_1$ described by an  asymptotic series in $\rho$
\begin{equation}
F_1=Z_{q}^2\sum_{n=0}^\infty \rho^n F^{(n)}_1\,,
\label{eq::F1series}
\end{equation}
where $F^{(n)}_1$ are given by the power series in $\alpha_s$ with the
coefficients depending on $\rho$ only logarithmically, and we use the same
notations and kinematics as in Sect.~\ref{sec::2}.  Since the Sudakov
corrections in Eq.~(\ref{eq::F1series}) are factored out, in the
double-logarithmic approximation the leading term of  the expansion   is just
the Born value $F_1^{(0)}=1$, and the double-logarithmic corrections  to the
leading power-suppressed term $F_1^{(1)}$ are purely non-Sudakov. According
to the results of  Refs.~\cite{Penin:2014msa,Liu:2017axv} such corrections
are induced by the nonplanar soft quark pair exchange, Fig.~\ref{fig::6}(a),
and start with the two-loop contribution. In contrast  to the previously
considered cases the vector interaction conserves helicity and require  a
helicity flip on each of the soft quark lines which become sufficiently
singular to develop the double-logarithmic scaling.  The corresponding
Feynman integral reads
\begin{eqnarray}
&&\left({2iQ^2\over \pi^2 }\right)^2
\int\left({{d^4l_1}\over (l^2_1-m_q^2) (p_2+l_1)^2
((p_1 +l_1+l_2)^2-m_q^2)}\right.
\nonumber\\
&&\left.\times{{d^4l_2}\over (l_2^2-m_q^2) (p_1+l_2)^2
((p_2 +l_1+l_2)^2-m_q^2)}\right)\,.
\label{eq::intf}
\end{eqnarray}

\begin{figure}
\begin{center}
\begin{tabular}{ccccc}
\includegraphics[width=1.8cm]{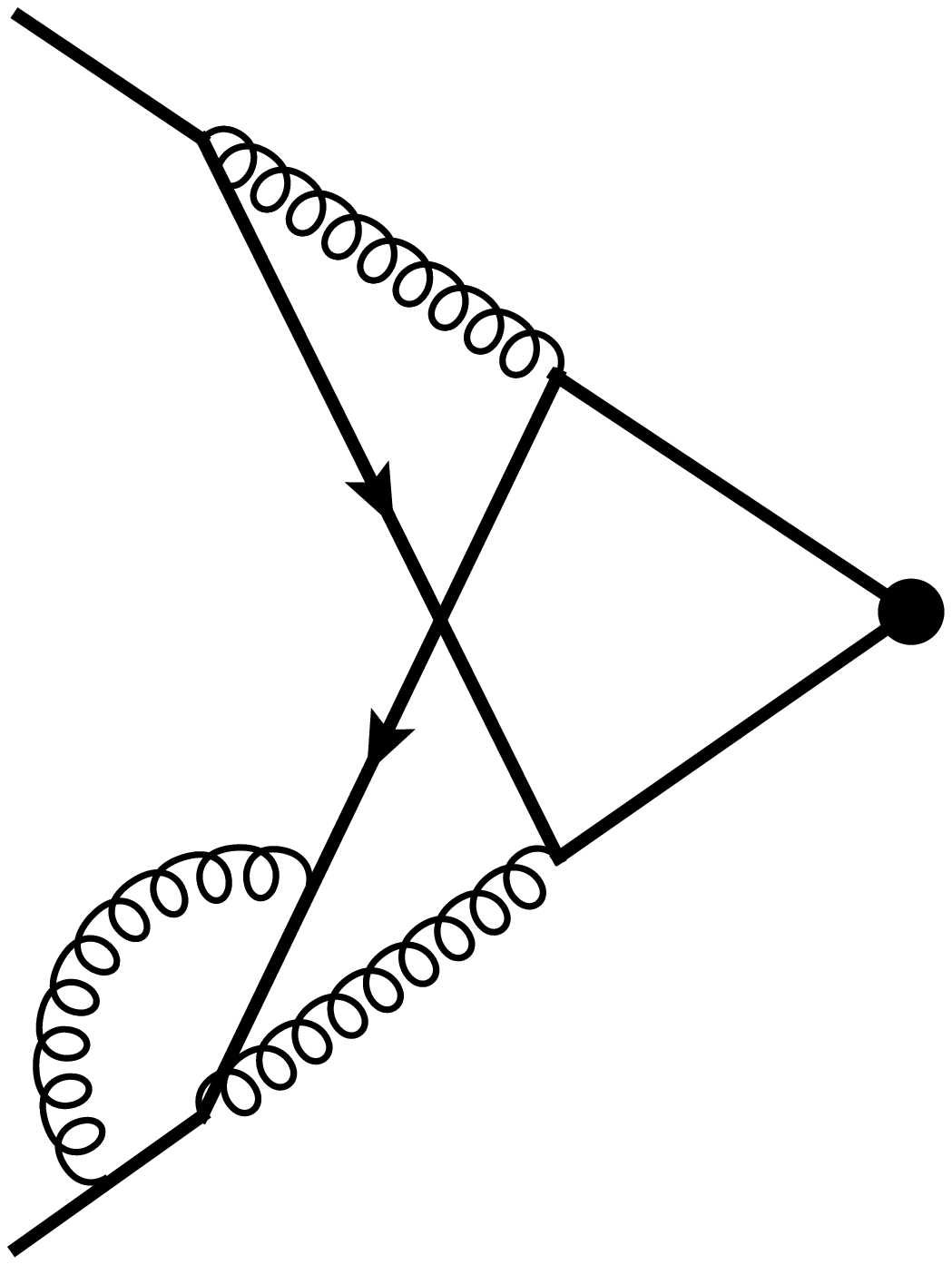}
\hspace*{5mm}\raisebox{11.5mm}{$\bfm{\to}$}&
\hspace*{-5mm}\includegraphics[width=1.8cm]{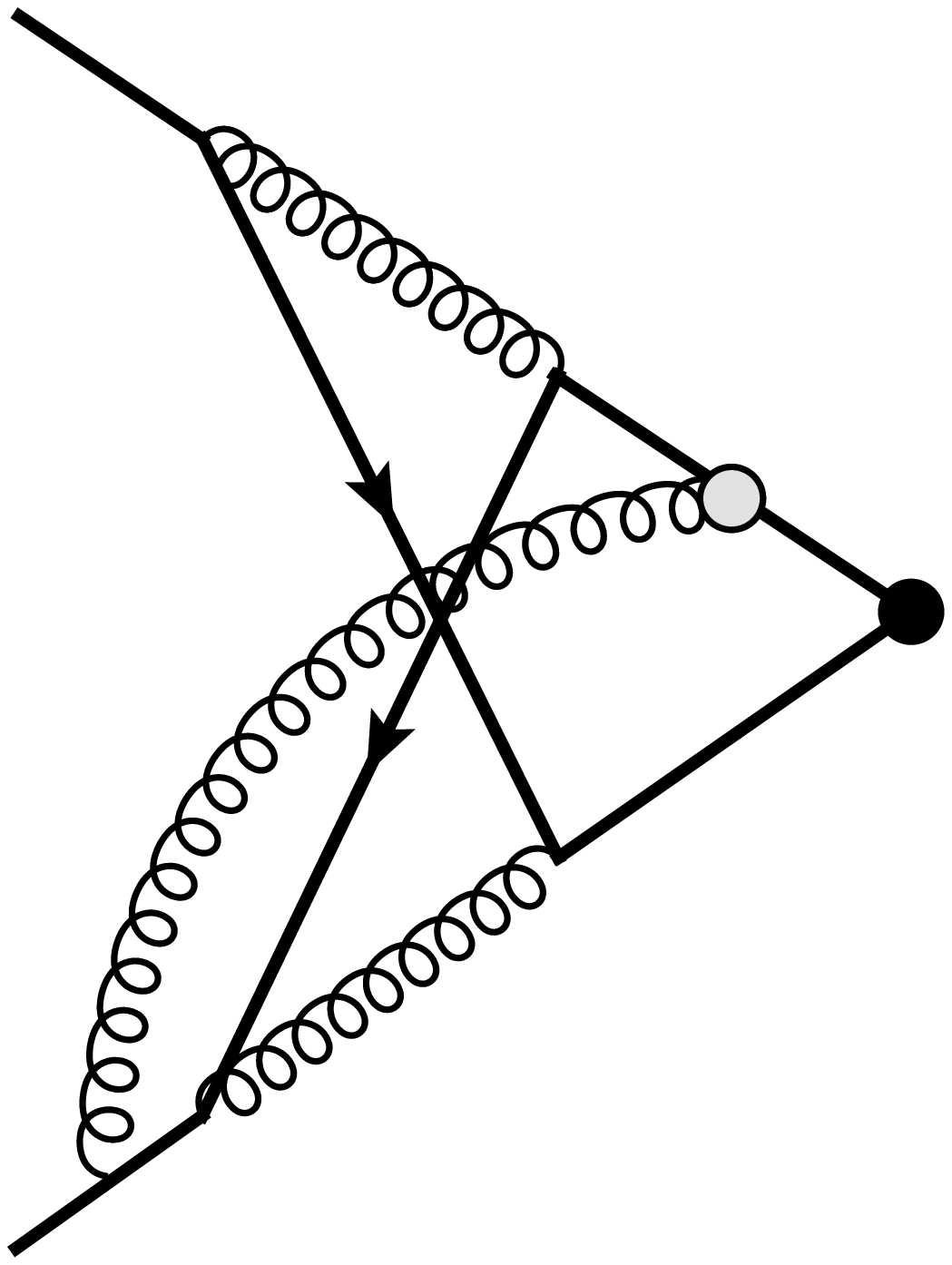}& &
\hspace*{00mm}\includegraphics[width=1.8cm]{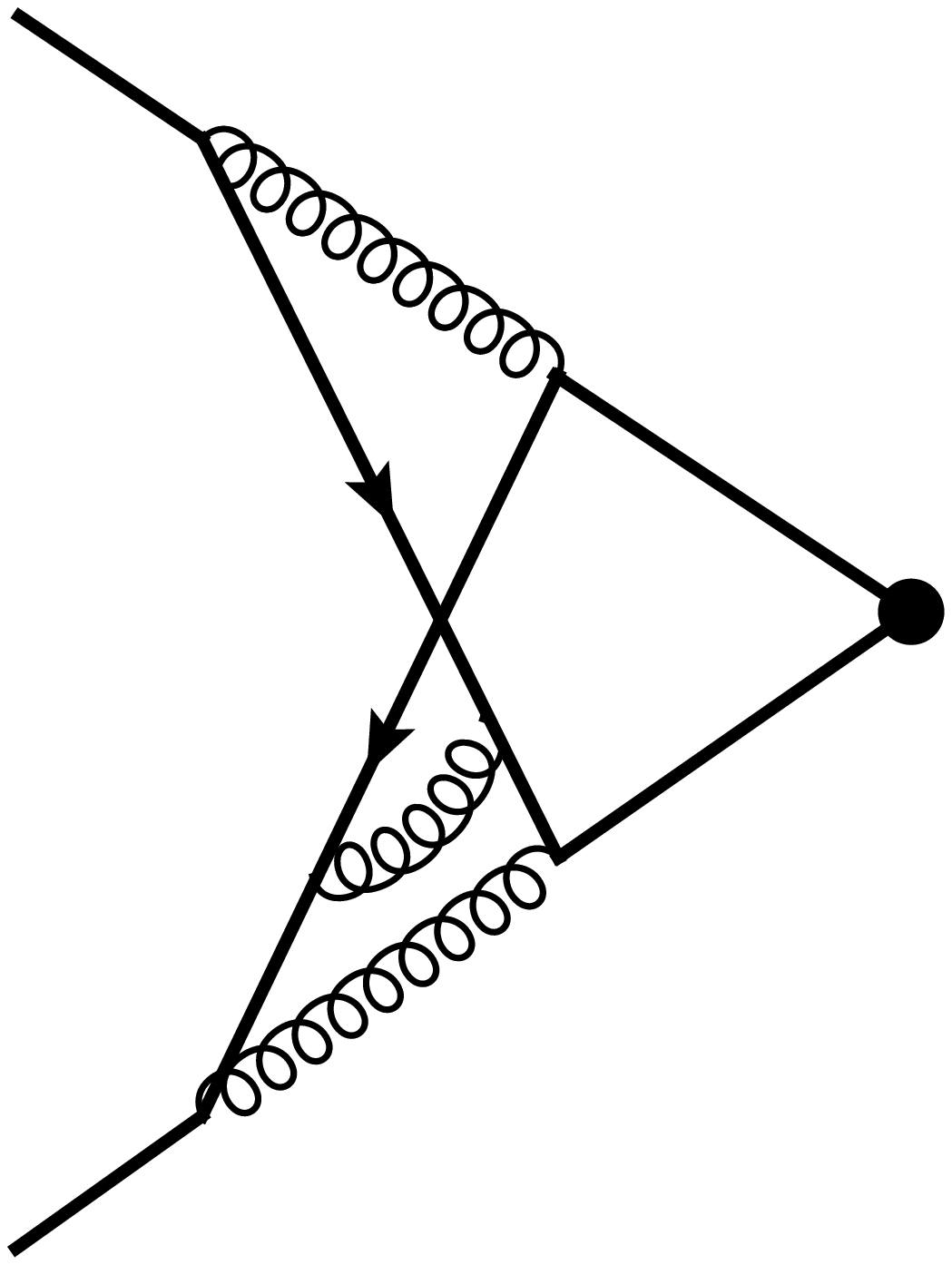}
\hspace*{5mm}\raisebox{11.5mm}{$\bfm{\to}$}&
\hspace*{-2mm}\includegraphics[width=1.8cm]{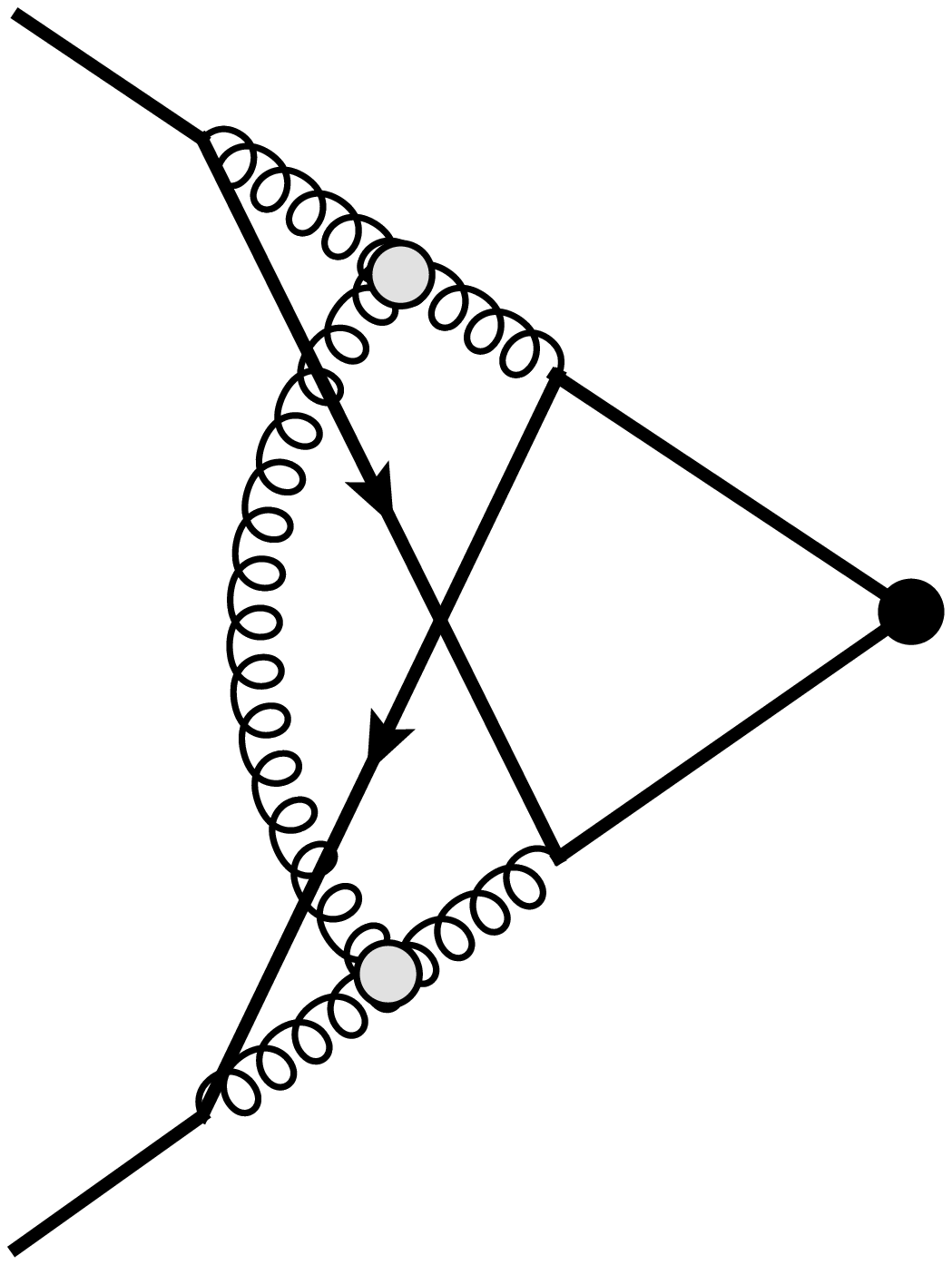}\\
\hspace*{-7mm}(a)&\hspace*{-3mm}(b)&&\hspace*{-5mm}(c)&
\hspace*{00mm} (d)\\[5mm]&
\includegraphics[width=1.8cm]{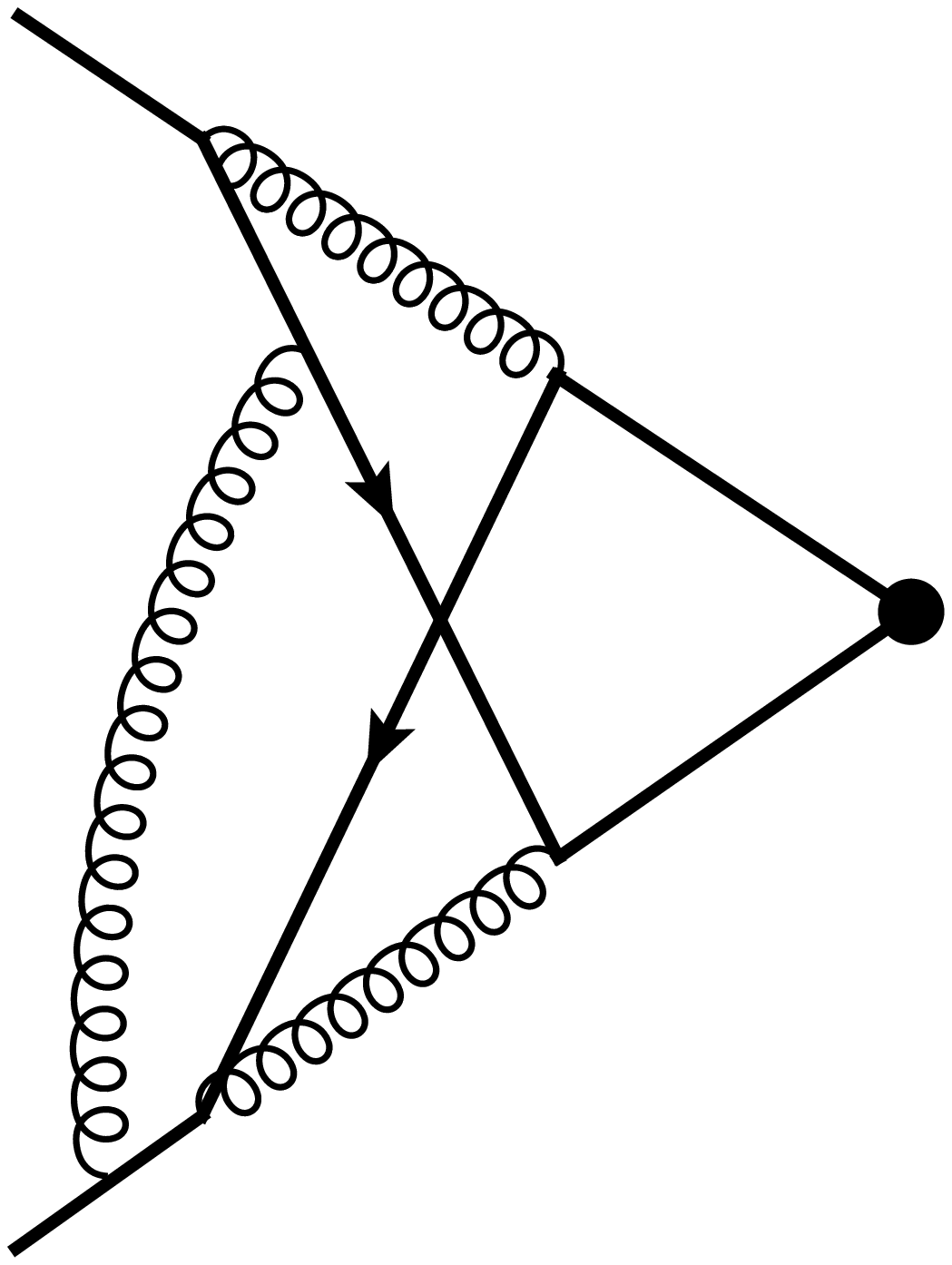}
\hspace*{5mm}\raisebox{11.5mm}{$\bfm{+}$}&
\hspace*{-2mm}\includegraphics[width=1.8cm]{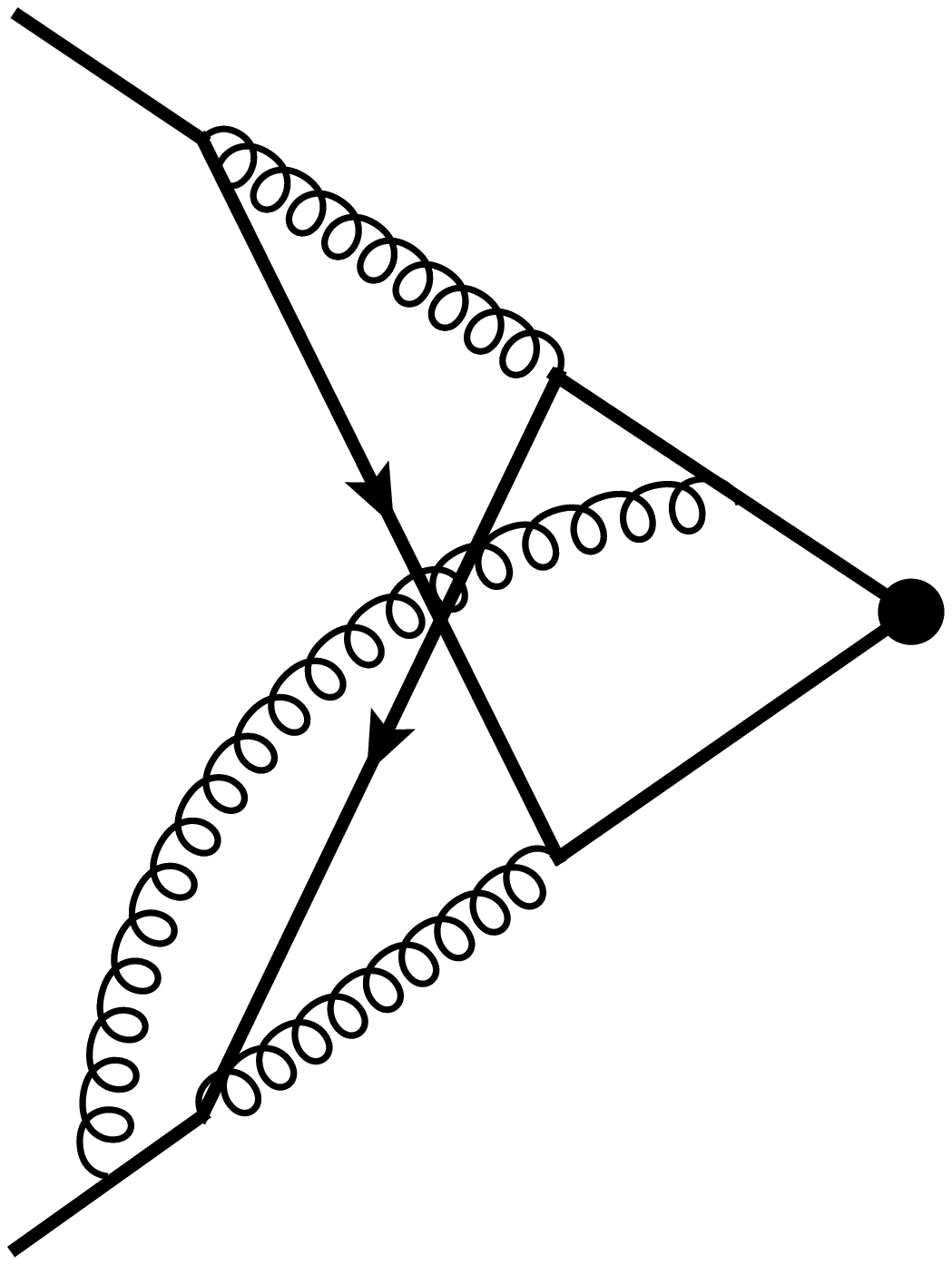}
\hspace*{5mm}\raisebox{11.5mm}{$\bfm{\to}$}&
\hspace*{-5mm}\includegraphics[width=1.8cm]{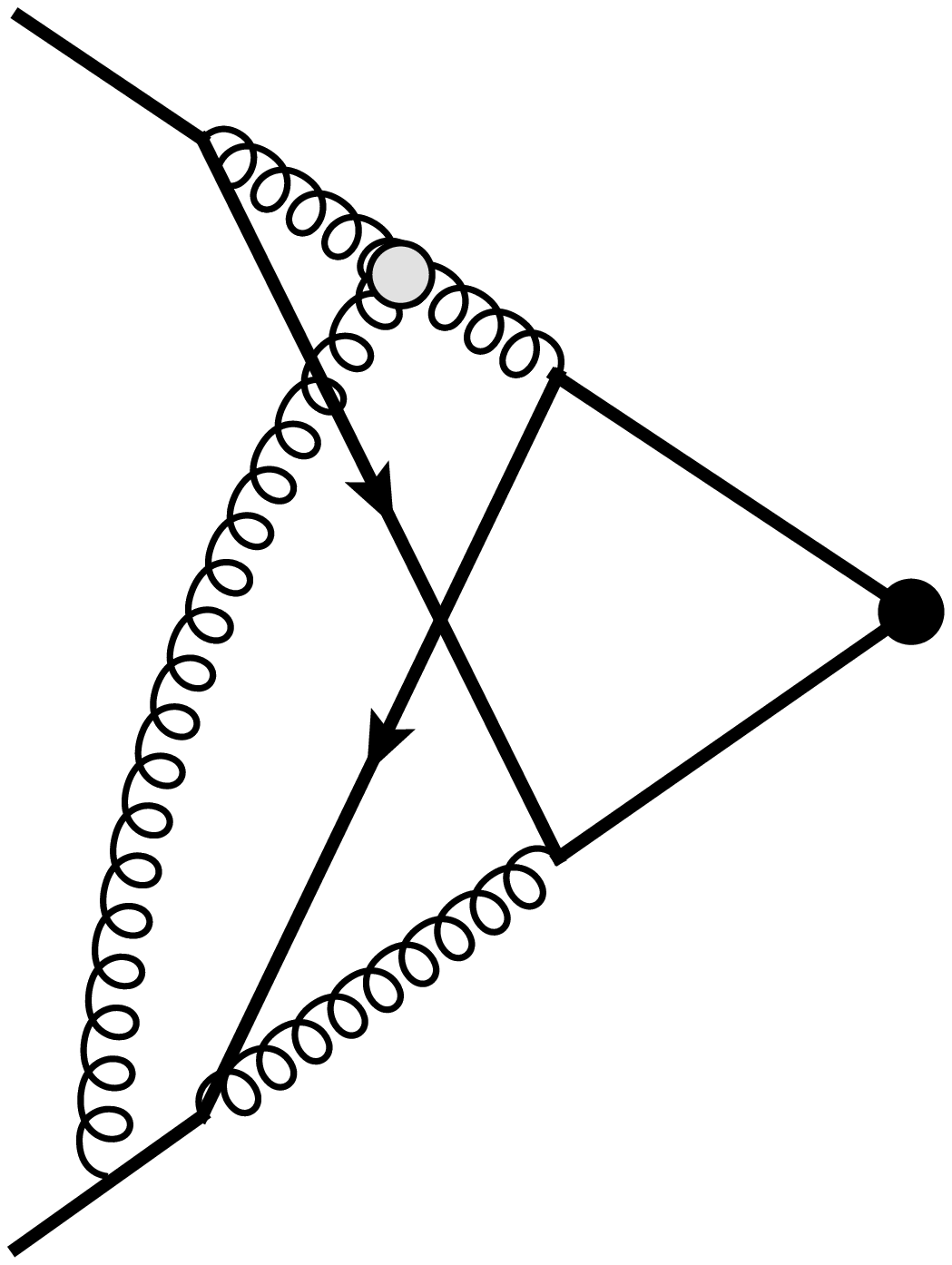}&\\
&\hspace*{-6mm}(e)&\hspace*{-8mm}(f)&\hspace*{-3mm}(g)& \\
\end{tabular}
\end{center}
\caption{\label{fig::8}  Diagramatic representation of the transformation
which moves the soft gauge boson vertex from a soft quark to an eikonal
line.}
\end{figure}

\noindent
The integration over the soft quark momenta $l_i$ in this case is
double-logarithmic when  $l_2p_1<  l_1p_1$,  $l_1p_2< l_2p_2$ and the
corresponding Sudakov parameters are ordered along the eikonal lines
$v_2<v_1 $,  $u_1< u_2$. With the additional kinematical constraints
$u_iv_i>\rho$ the integral over the Sudakov parameters reduces to
\begin{equation}
4\ln^4\!\rho\int_0^1{\rm d}\eta_1\int_{\eta_1}^{1}{\rm d}\eta_2
\int_0^{1-\eta_2}{\rm d}\xi_2\int_{\xi_2}^{1-\eta_1}
 {\rm d}\xi_1= {\ln^4\!\rho\over 3}\,,
\label{eq::intfres}
\end{equation}
which corresponds to the leading-order result
\begin{equation}
\left[F_1^{(1)}\right]_{2-loop}={C_F(C_A-2C_F)\over 6}x^2\,,
\label{eq::F12loop}
\end{equation}
in agreement with \cite{Bernreuther:2004ih}. To derive the factorization
formula for the higher-order double-logarithmic terms let us consider the
three-loop corrections, Fig.~\ref{fig::7}, and start with the abelian QED
case.  Following the algorithm described in Sect.~\ref{sec::2.2} we use the
sequence of Ward identities and the soft quark momentum shifts to move the
soft photon vertex from a soft quark to an eikonal line. The result of the
transformation is shown in Fig.~\ref{fig::8}, where the effective vertex is
proportional to the charge of the corresponding external line.  The first two
relations are rather straightforward to derive. In the third relation the
presence of  the additional diagram Fig.~\ref{fig::8}(f) needs to be
clarified. First we note that this diagram has an opposite sign with respect
to Fig.~\ref{fig::8}(b) since the photon couples to the  antiquark rather
than the incoming quark line. Thus the  diagram Fig.~\ref{fig::8}(f) is not
relevant for the eikonal factorization of the soft photon ladder. On the
contrary, its role is similar to   the  diagram Fig.~\ref{fig::2}(c) which
cancels the color space commutator when the soft gluon vertex is moved in
Fig.~\ref{fig::2}(b). Though now  we consider the abelian case,  such
commutator appears in  the transformation of Fig.~\ref{fig::8}(e) since the
electric  charge is not conserved  along the  eikonal line. A set of the
resulting ladder diagrams Figs.~\ref{fig::8}(b,d,g) together with
Figs.~\ref{fig::7}(a,b) needs the last missing permutation of the soft photon
vertex, Fig.~\ref{fig::6}(b), to complete the eikonal factorization. After
adding this diagram the soft photon exchange factorizes into the one-loop
Sudakov  factor in Eq.~(\ref{eq::F1series}). Thus the  remaining soft photon
contribution to $F_1^{(1)}$ is given by the negative of
Fig.~\ref{fig::6}(b), and the symmetric one. A characteristic feature of the
diagram Fig.~\ref{fig::6}(b) is that the soft gluon connects two eikonal
lines determined by the same soft quark momentum. Such a diagram cannot  be
obtained by the transformation of another diagram described  in
Sect.~\ref{sec::2.2} and has to be added by hand to complete the Sudakov
logarithms factorization. Hence the above property can be a guiding principle
for selecting the diagrams which define the  non-Sudakov double-logarithmic
contribution.  As in the example discussed in  Sect.~\ref{sec::2.2} the QCD
result is obtained from the QED one by substituting   $-e_q^2/(4\pi)$ with
$(C_A-C_F)\alpha_s$ in the above diagram. In analogy with
Eq.~(\ref{eq::G2loop}) we can write the three-loop contribution to the form
factor as follows
\begin{equation}
x\left(12\sum_i c^{(1)}_\lambda \int_0^1{\rm d}\eta_1\int_{\eta_1}^{1}{\rm d}\eta_2
\int_0^{1-\eta_2}{\rm d}\xi_2\int_{\xi_2}^{1-\eta_1}{\rm d}\xi_1
w^{(1)}_\lambda(\eta,\xi)\right)\, \left[F_1^{(1)}\right]_{2-loop}\,,
\label{eq::F13loop}
\end{equation}
where $ c^{(1)}_\lambda$ and  $w^{(1)}_\lambda$ are listed in
Table~\ref{tab::4}. The weights for the symmetric diagrams not shown in
Fig.~\ref{fig::7}  are obtained in this case by the replacement
$\eta_1\leftrightarrow \xi_2$ and  $\eta_2\leftrightarrow \xi_1$.
Eq.~(\ref{eq::F13loop}) sums up to
\begin{equation}
-z\left(12\int_0^1{\rm d}\eta_1\int_{\eta_1}^{1}{\rm d}\eta_2
\int_0^{1-\eta_2}{\rm d}\xi_2\int_{\xi_2}^{1-\eta_1}{\rm d}\xi_1
\, \left(2\eta_1(\xi_1-\xi_2)+2\xi_2(\eta_2-\eta_1)\right)\right)
\left[F_1^{(1)}\right]_{2-loop}\,,
\label{eq::2loopfna}
\end{equation}
which can be recognized as the contribution of the effective soft gluon
exchange in  Fig.~\ref{fig::6}(b)  and in the symmetric diagram.

\begin{table}[t]
\begin{center}
    \begin{tabular}{|c|c|c|}
     \hline
       $\lambda$ &  $w^{(1)}_\lambda$  & $c^{(1)}_\lambda$ \\
    \hline
      a  & $-\eta_2(\eta_2 + 2)-\xi_1(\xi_1-2\eta_2+2) $ &   $-C_F$ \\
      b & $2\xi_2\eta_1$                                 &   $-C_F$ \\
      c & $2(\xi_1-\xi_2)(\eta_2-\eta_1)$                &   $C_A-C_F$ \\
      d & $-\eta_1(\eta_1-2\xi_1+2) $                    &  $C_A-C_F$ \\
      e & $(\eta_2-\eta_1)( \eta_1+\eta_2- 2\xi_1 +2)$   &   $-\frac{C_A}{2}$ \\
      f & $2\eta_1(\xi_1 - \xi_2)$                       &   $-\frac{C_A}{2}$ \\
      g & $2\eta_2(\xi_1-\xi_2)$                         &   $-\frac{C_A}{2}$ \\
      h & $\eta_1(\eta_1-2\xi_1+2) $                     &   $\frac{C_A}{2}-C_F$  \\
      i & $\eta_2(\eta_2-2\xi_1+2)$                      &   $\frac{C_A}{2}-C_F$  \\
    \hline
    \end{tabular}
\end{center}
\caption{\label{tab::4}  The weights $w^{(1)}_\lambda$  and the color
         factors $c^{(1)}_\lambda$ for the diagrams in Fig.~\ref{fig::7}.}
\end{table}

Now the factorization formula  for the leading power-suppressed
contribution to the vector form factor can be written as follows
\begin{equation}
F_1^{(1)}={C_F(C_A-2C_F)\over 6}x^2f(-z)\,,
\label{eq::F1result}
\end{equation}
where the function $f(-z)$ incorporates the non-Sudakov contribution of
Fig.~\ref{fig::6}(b) with an arbitrary number of the effective soft gluon
exchanges and is normalized to the two-loop result $f(0)=1$.  This function
is  obtained by exponentiating the single effective soft gluon exchange in
Eq.~(\ref{eq::2loopfna}) and therefore has the following  integral
representation
\begin{equation}
f(z)=12\int_0^1{\rm d}\eta_1\int_{\eta_1}^{1}{\rm d}\eta_2
\int_0^{1-\eta_2}{\rm d}\xi_2\int_{\xi_2}^{1-\eta_1}
 {\rm d}\xi_1\,e^{2z\eta_1(\xi_1-\xi_2)} e^{2z\xi_2(\eta_2-\eta_1)}\,.
\label{eq::f}
\end{equation}
It is difficult  to solve the  four-fold integral Eq.~(\ref{eq::f}) in a closed
analytic form. However,  the coefficients of the series
$f(z)=1+\sum_{n=1}^\infty c_n z^n $ can be computed for any given $n$
corresponding to the $(n+2)$-loop double-logarithmic contribution and have the
following large-$n$ behavior  $c_n\sim {\ln n\over n! 2^{n} n^{5/2}}$. The first
ten coefficients of the series are listed in Table~\ref{tab::5}.  The asymptotic
behavior of the function at $z\to \infty$ reads
\begin{equation}
f(-z)\sim   C_-\left({\ln{z} \over z}\right)^2\!,
\quad f(z)\sim C_+\ln z\left({e^{z}\over z^5}\right)^{1/2}\!\!,
\label{eq::fasym}
\end{equation}
where the constant $C_-=3.6\ldots$, $C_+=14.8\ldots$ are found numerically.
The result Eq.~(\ref{eq::F1result}) vanishes for $N_c\to\infty$,
which is consistent with the explicit evaluation of the three-loop
massive form factor in this limit \cite{Henn:2016tyf}.

\begin{table}[t]
  \begin{center}
    \begin{tabular}{|c|c|c|c|c|c|c|c|c|c|c|}
     \hline
      $n$ & $1$ & $2$ & $3$ & $4$ & $5$ & $6$ & $7$  &$8$ &$9$ &$10$  \\
      \hline
      $2^nn^2n!c_n$ &  ${2\over 5}$ & ${88\over  105}$ & ${8\over 7}$&
      ${70144\over 51975}$ & ${640\over 429}$ &
      ${25344\over 15925}$ & ${2727424\over 1640925}$&
      ${1868824576\over 1091215125}$&
      ${8994816\over 5143775}$ &
      ${27430420480\over 15460335891}$\\
      \hline
    \end{tabular}
   \end{center}
    \caption{\label{tab::5}
      The   normalized coefficients of the Taylor series for the function
      $f(z)$, Eq.~(\ref{eq::f}), up to  $n=10$.}
\end{table}

\subsection{Scalar form factor}
\label{sec::4.2}
The  quark scattering  in the external scalar field is
parametrized by a single form factor $F_S$ which has
 the high-energy asymptotic  expansion similar to Eq.~(\ref{eq::F1series})
\begin{equation}
F_S=Z_{q}^2\sum_{n=0}^\infty \rho^n F^{(n)}_S\,,
\label{eq::FSseries}
\end{equation}
with the Born result normalized to  $F^{(0)}_S=1$. The general arguments of
the previous section on the origin of  the leading mass-suppressed
double-logarithmic corrections are equally applicable to the scalar form
factor. However, the contribution of the nonplanar diagram
Fig.~\ref{fig::6}(a) in this case vanishes. Indeed, the scalar vertex induces
an additional helicity flip along the quark line and requires an odd number
of soft quark exchanges. At the same time the planar soft quark pair
exchange Fig.~\ref{fig::6}(c) with a closed quark line, which vanishes for
the external vector field by Furry theorem, does contribute in the scalar
case.  The relevant Feynman integral reads
\begin{eqnarray}
&&\left({2iQ^2\over \pi^2 }\right)^2
\int\left({{d^4l_1}\over (l^2_1-m_q^2) (p_2+l_1)^2
(p_1 +l_1)^2}\right.
\nonumber\\
&&\left.\times{{d^4l_2}\over (l_2^2-m_q^2) ((p_1+l_1+l_2)^2-m_q^2)
((p_2 +l_1+l_2)^2-m_q^2)}\right)\,,
\label{eq::intfs}
\end{eqnarray}
with the double-logarithmic integration region  $l_1p_1<  l_2p_1$,  $l_1p_2<
l_2p_2$, or $v_1<v_2 $,  $u_1< u_2$. It reduces to
\begin{equation}
4\ln^4\!\rho\int_0^1{\rm d}\eta_1\int_{0}^{1-\eta_1}{\rm d}\xi_1
\int_{\eta_1}^{1-\xi_1}{\rm d}\eta_2\int_{\xi_1}^{1-\eta_2}
 {\rm d}\xi_2={\ln^4\rho\over 6}\,
\label{eq::intfsres}
\end{equation}
corresponding to the leading-order form factor
\begin{equation}
\left[F_S^{(1)}\right]_{2-loop}=-{C_FT_F\over 3}x^2\,,
\label{eq::FS2loop}
\end{equation}
where $T_F=1/2$, in agreement  with \cite{Bernreuther:2005gw}. Let us now
discuss the factorization of the double-logarithmic corrections. By using the
same procedure as for the vector form factor one can reduce the non-Sudakov
part of the corrections to the contribution of the diagrams in
Figs.~\ref{fig::6}(d,e) with the  effective soft gluon exchange between the
eikonal lines determined by the same soft quark momentum. The diagram
Fig.~\ref{fig::6}(d) has  the color factor $C_A-C_F$ while the diagram
Fig.~\ref{fig::6}(e) is proportional to $C_F-C_A$ as dictated by the
variation of the color charge along the eikonal lines in each case.  The only
subtlety is related to the fact that in the diagram Fig.~\ref{fig::6}(d) the
soft gluon momentum integral factors out from the inner quark loop in the
same way as the Sudakov corrections factor out in the $gg\to H$ amplitude
discussed in Sect.~\ref{sec::3}. The corresponding correction  to the form
factor reads
\begin{equation}
-z\left(24\int_0^1{\rm d}
\eta_1\int_{0}^{1-\eta_1}{\rm d}\xi_1
\int_{\eta_1}^{1-\xi_1}{\rm d}\eta_2\int_{\xi_1}^{1-\eta_2}
\, \left(2\eta_2\xi_2-2\eta_1\xi_1\right)\right)
\left[F_S^{(1)}\right]_{2-loop}\,,
\label{eq::2loopfsna}
\end{equation}
where the first and the second terms in the brackets represent the
contributions of the diagrams in Fig.~\ref{fig::6}(d) and Fig.~\ref{fig::6}(e),
respectively.

\begin{figure}
\begin{center}
\begin{tabular}{cccc}
\includegraphics[width=1.8cm]{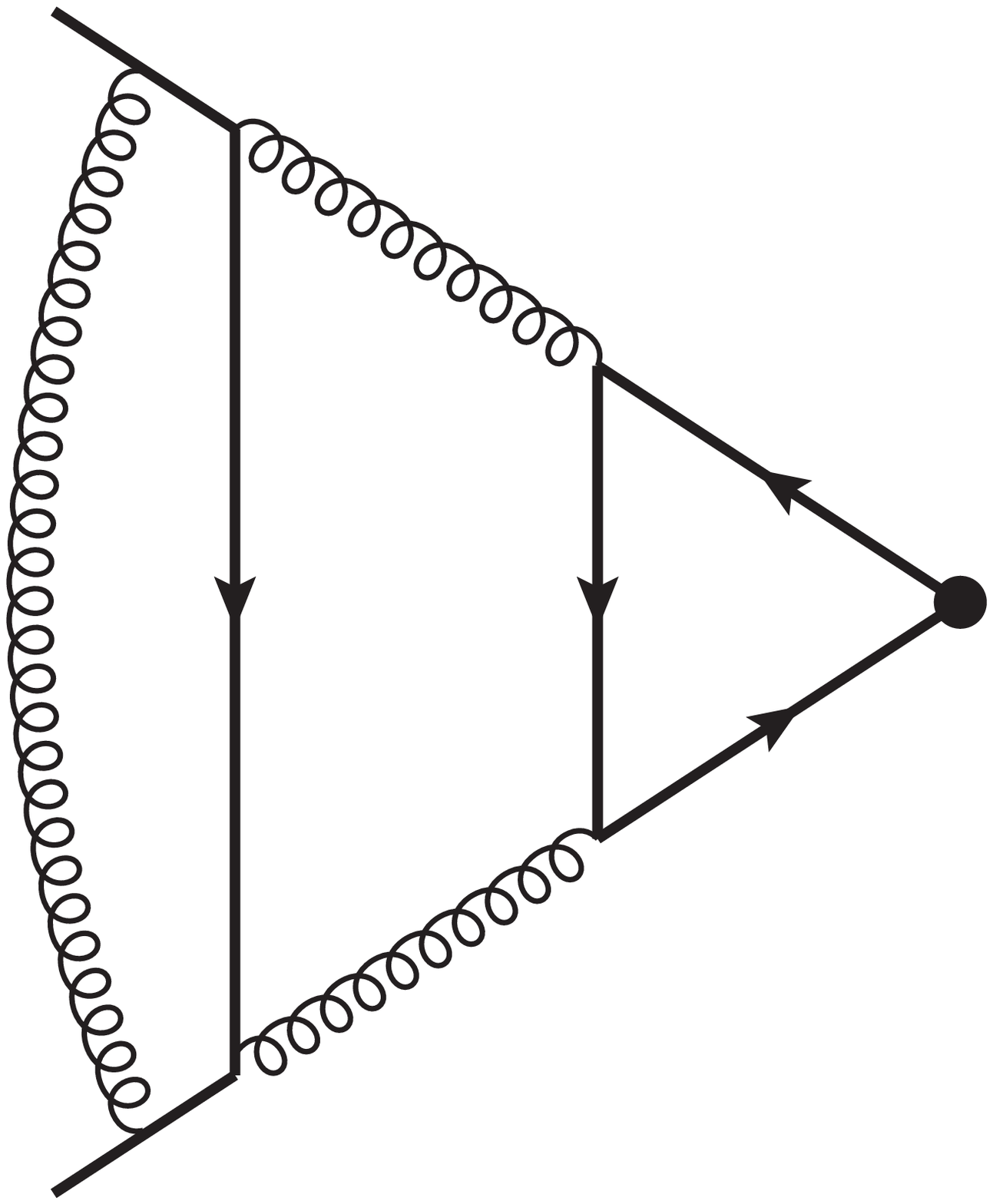} &\hspace*{10mm}
\includegraphics[width=1.8cm]{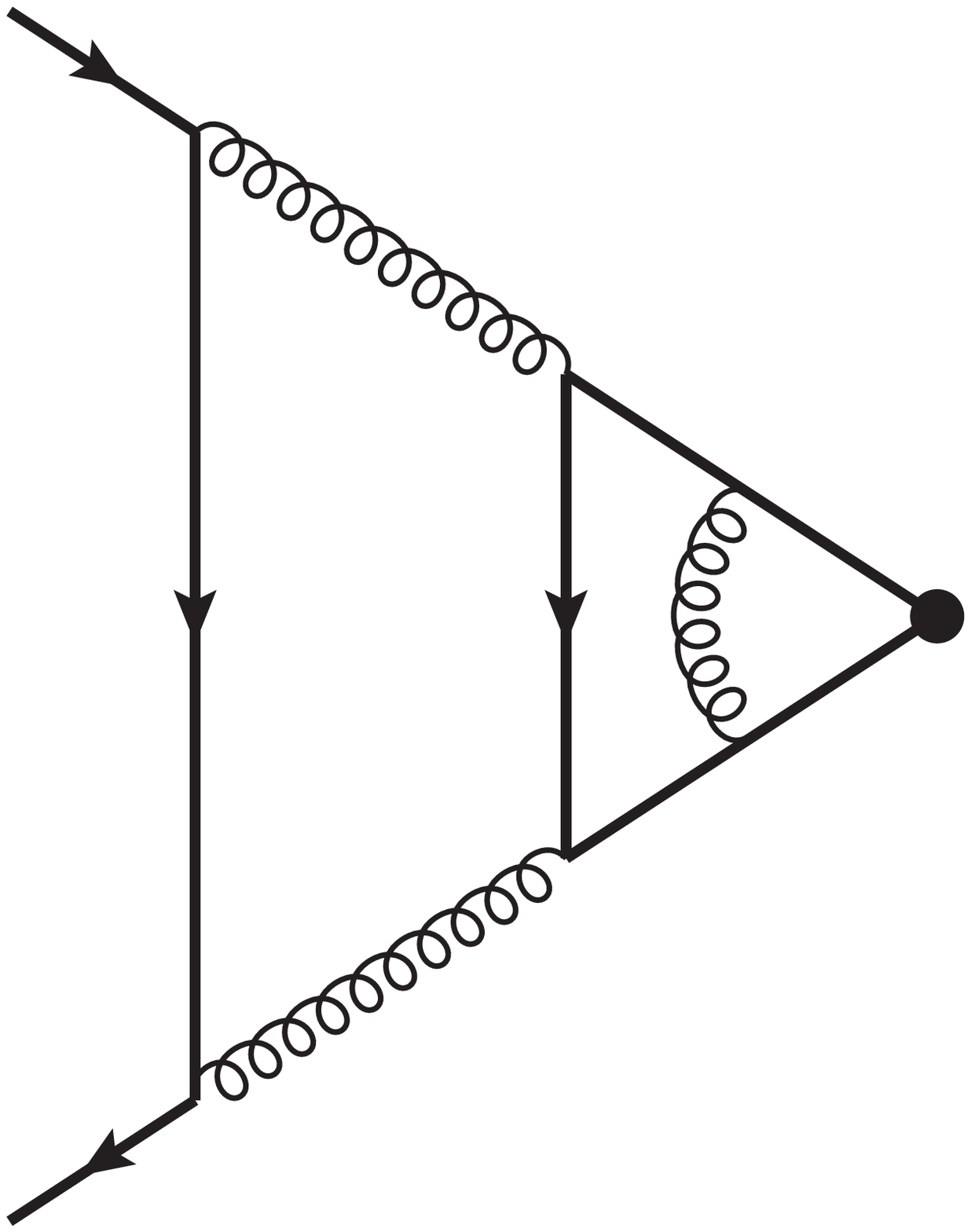} & \hspace*{10mm}
\includegraphics[width=1.8cm]{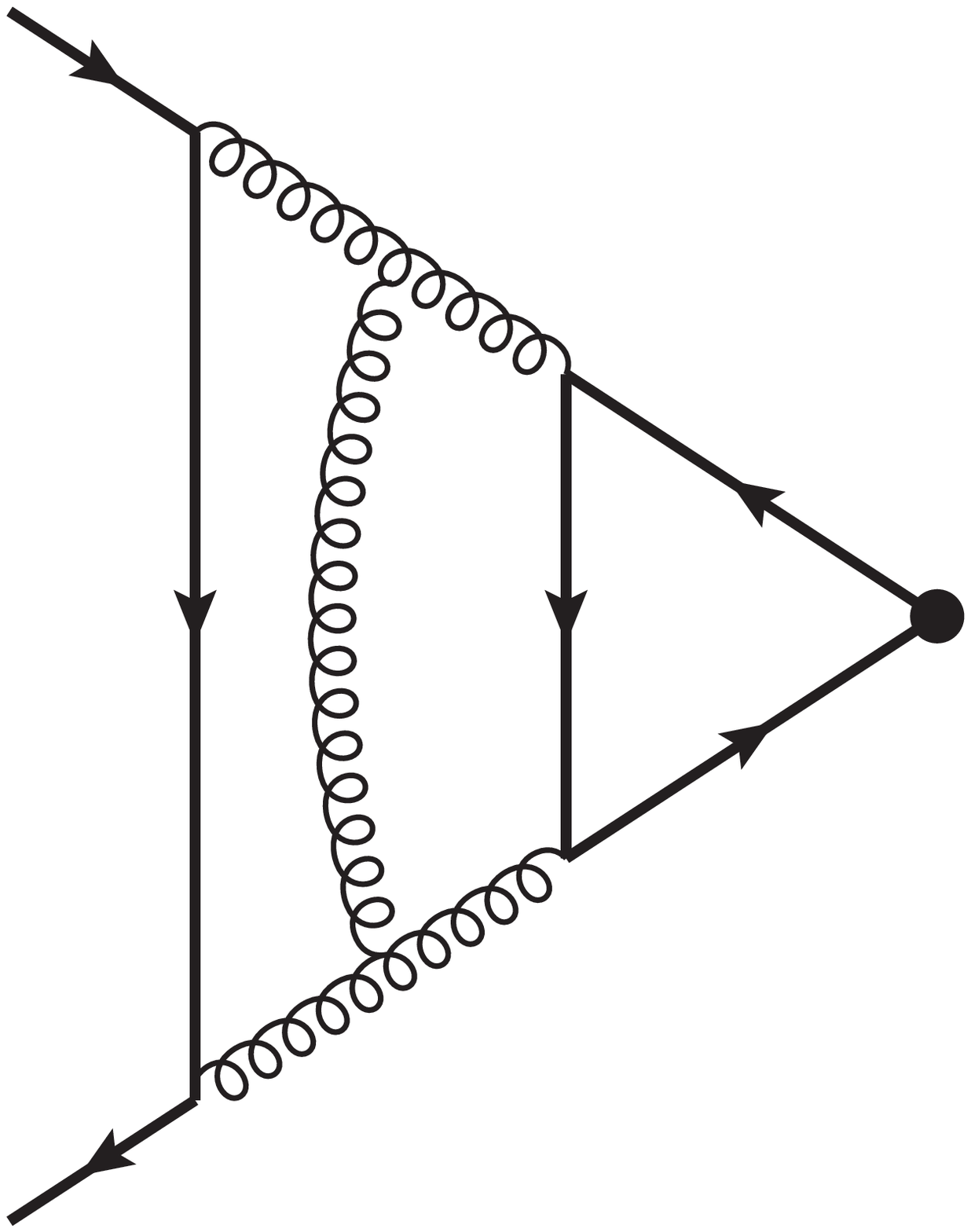} & \hspace*{10mm}
\includegraphics[width=1.8cm]{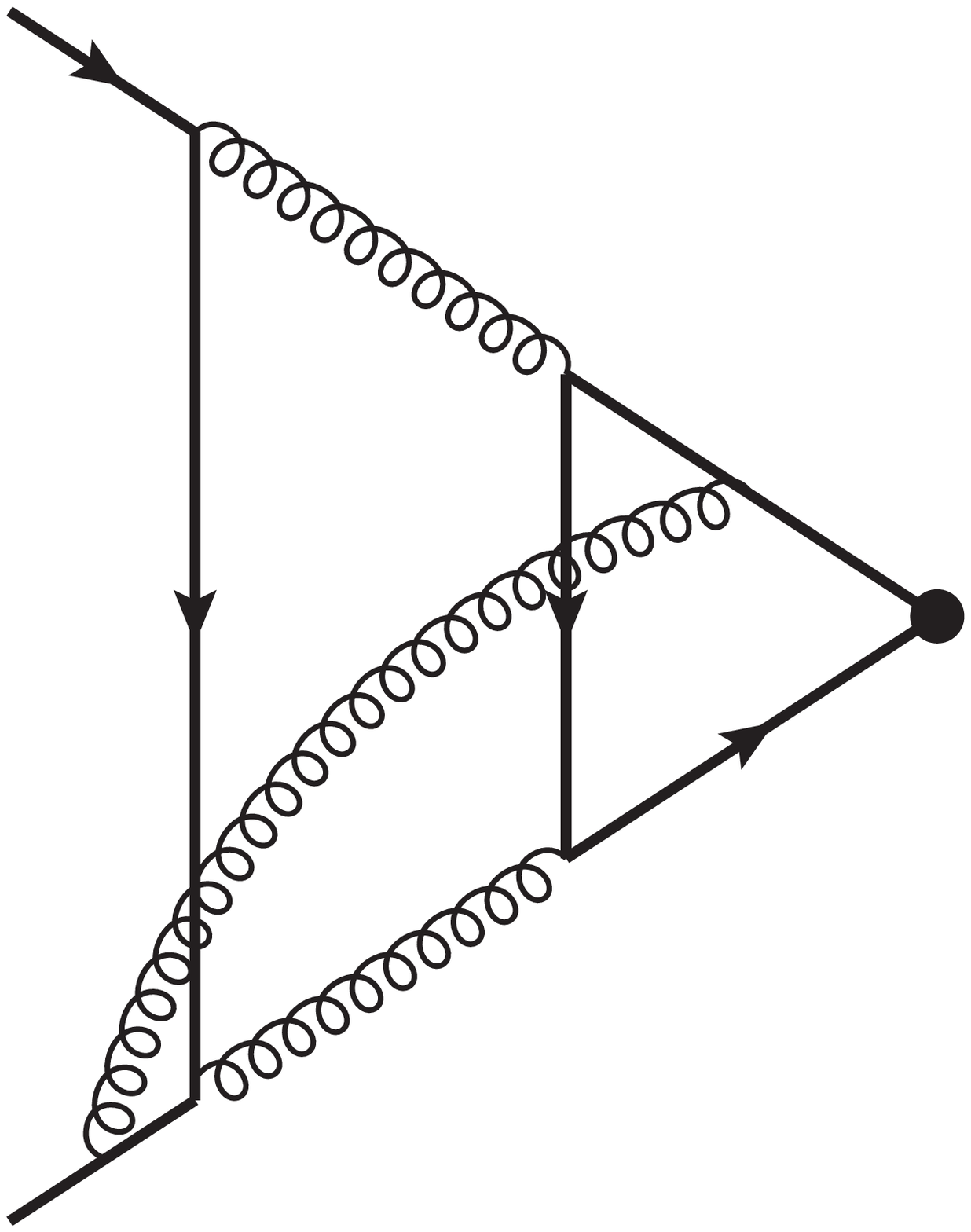} \\
(a) & \hspace*{10mm}(b) & \hspace*{10mm}(c) & \hspace*{10mm}(d)  \\
\includegraphics[width=1.8cm]{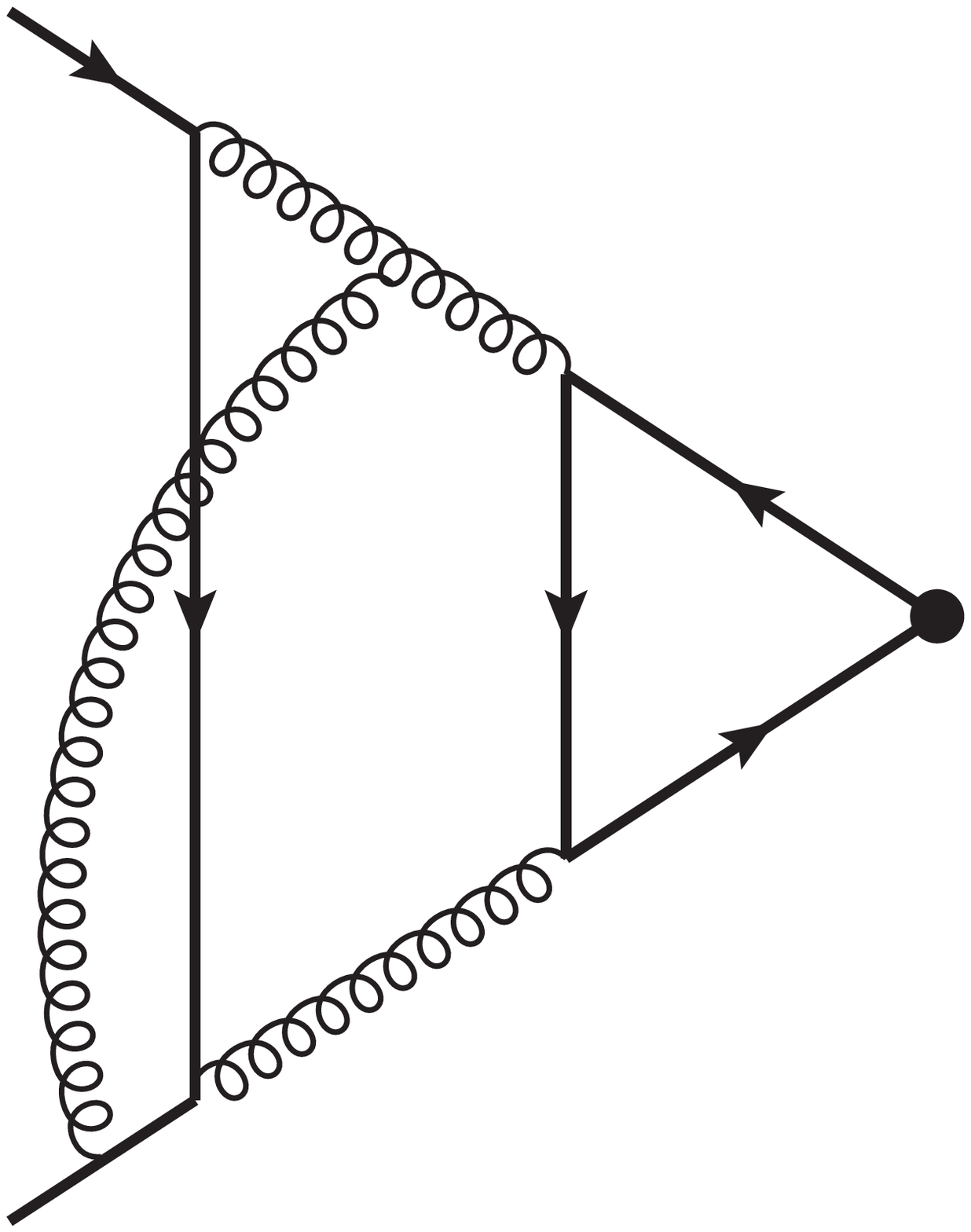} & \hspace*{10mm}
\includegraphics[width=1.8cm]{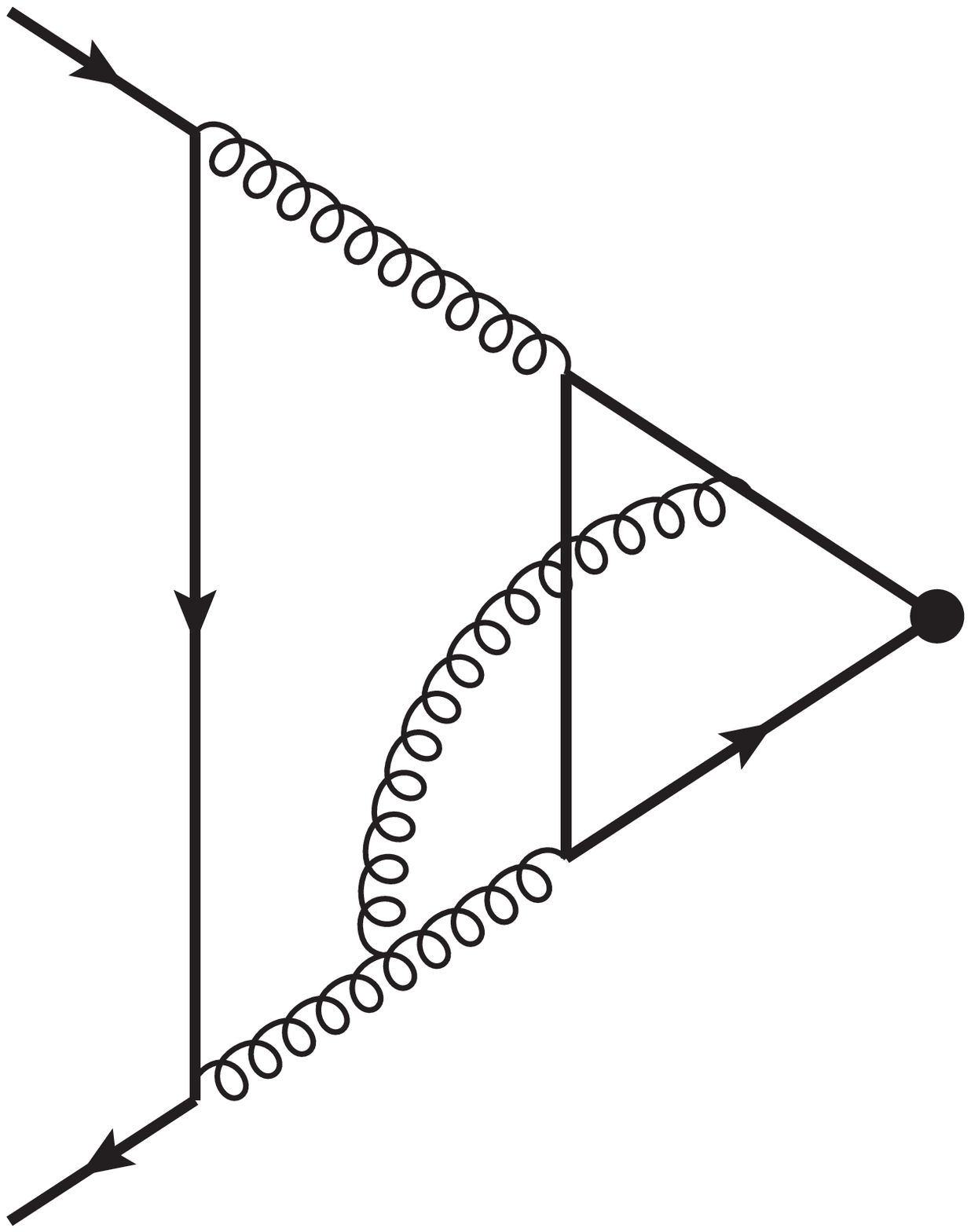} & \hspace*{10mm}
\includegraphics[width=1.8cm]{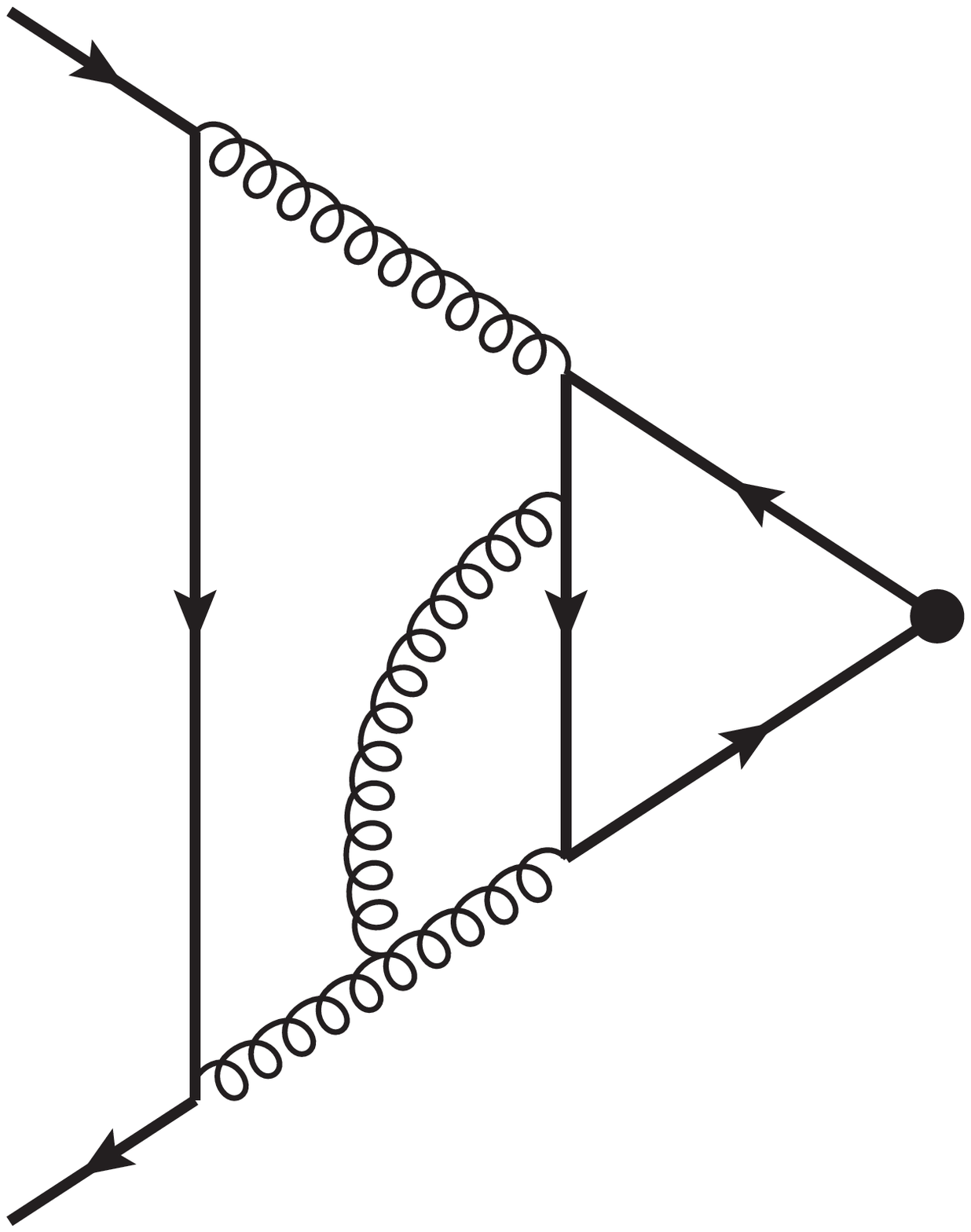} & \hspace*{10mm}
\includegraphics[width=1.8cm]{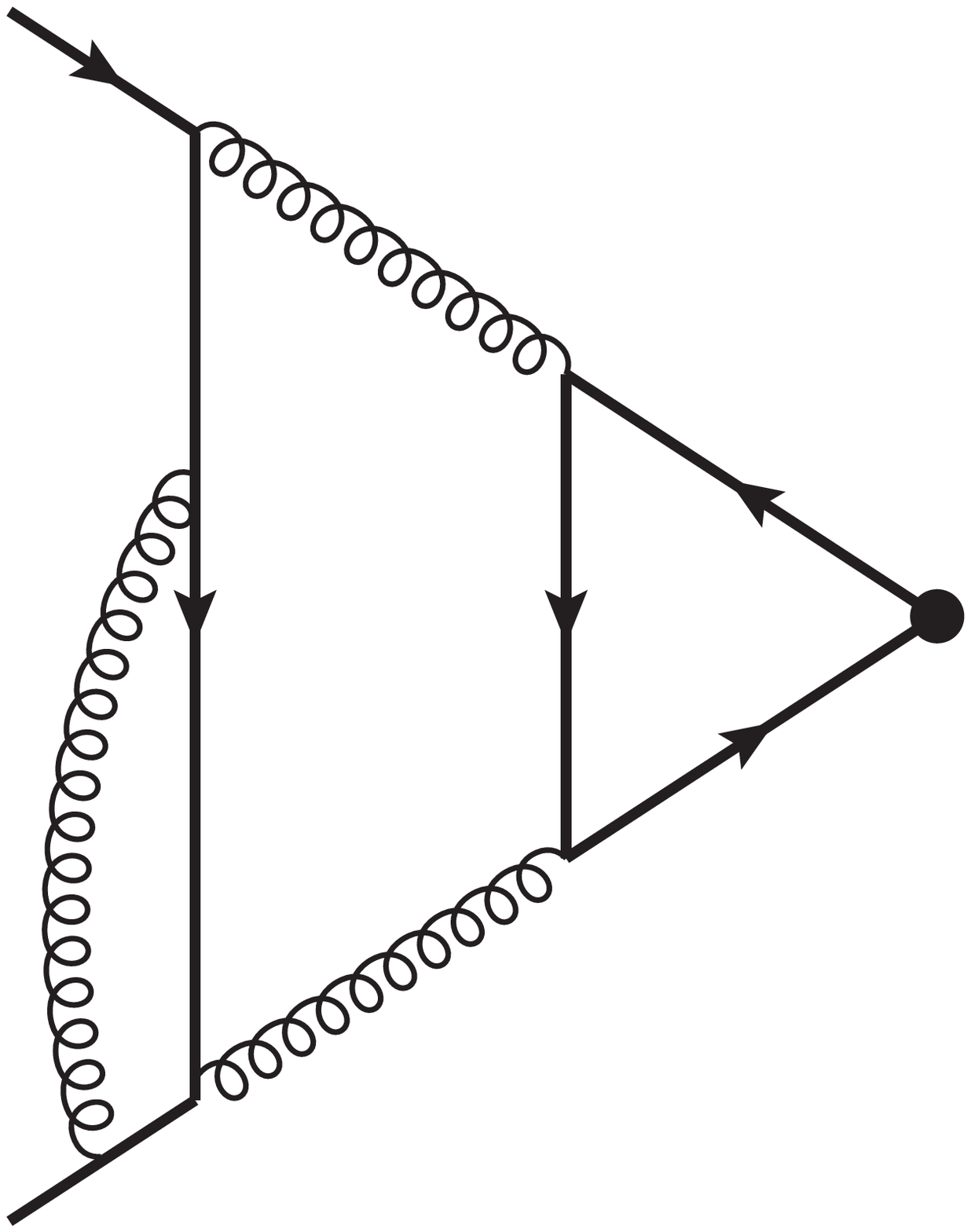} \\
(e) & \hspace*{10mm}(f) & \hspace*{10mm}(g) & \hspace*{10mm}(h)\\
\end{tabular}
\end{center}
\caption{\label{fig::9}  The three-loop diagrams contributing to the scalar
form factor $F^{(1)}_S$. Symmetric diagrams and the diagrams with the
opposite  direction of the closed quark line are not shown. The remaining
diagrams either do not have the double-logarithmic integration region or have
vanishing color factor.}
\end{figure}

The three-loop non-Sudakov double-logarithmic corrections can of course be
evaluated explicitly. The relevant three-loop diagrams are given in
Fig.~\ref{fig::9} and the corresponding contribution to the form factor
written in the standard form reads
\begin{equation}
x\left(24\sum_i c^{(1)}_\lambda \int_0^1{\rm d}
\eta_1\int_{0}^{1-\eta_1}{\rm d}\xi_1
\int_{\eta_1}^{1-\xi_1}{\rm d}\eta_2\int_{\xi_1}^{1-\eta_2}
w^{(1)}_\lambda(\eta,\xi)\right)\, \left[F_S^{(1)}\right]_{2-loop}\,,
\label{eq::FS3loop}
\end{equation}
where $ c^{(1)}_\lambda$ and  $w^{(1)}_\lambda$ are listed in
Table~\ref{tab::5}. The weights for the symmetric diagrams not shown in
Fig.~\ref{fig::9}  are obtained in this case by the replacement
$\eta_1\leftrightarrow \xi_1$ and  $\eta_2\leftrightarrow \xi_2$. As we
expect the sum in Eq.~(\ref{eq::FS3loop}) reduces to Eq.~(\ref{eq::2loopfsna})
confirming the above factorization. Thus we get the following expression
describing the   asymptotic behavior of the leading mass-suppressed
contribution to the  scalar form factor
\begin{equation}
F_S^{(1)}=-{C_FT_F\over 3}x^2f_S(-z)\,,
\label{eq::FSresult}
\end{equation}
where the function
\begin{equation}
f_S(z)=24\int_0^1{\rm d}\eta_1\int_{0}^{1-\eta_1}{\rm d}\xi_1
\int_{\eta_1}^{1-\xi_1}{\rm d}\eta_2\int_{\xi_1}^{1-\eta_2}
 {\rm d}\xi_2
\,e^{2z\eta_2\xi_2}e^{-2z\eta_1\xi_1}
\label{eq::fS}
\end{equation}
is determined by the diagrams in Figs.~\ref{fig::6}(d,e) with the
corresponding exponential factors given separately. Eq.~(\ref{eq::FSresult})
is consistent with the expansion of the exact result for  the three-loop
massive form factor in the large-$N_c$ limit \cite{Lee:2018rgs}.

Amazingly, though the topology of the diagrams in Fig.~\ref{fig::6}(b) and
Figs.~\ref{fig::6}(d,e) is completely different, Eqs.~(\ref{eq::f}) and
(\ref{eq::fS}) describe  the {\it same function}
\begin{equation}
f_S(z)\equiv f(z)\,,
\label{eq::magic}
\end{equation}
as it can be easily verified. It is straightforward to extend the analysis to
the axial $F_A$ and the pseudoscalar $F_P$ form factors, for which we obtain
the result in the form of Eq.~(\ref{eq::F1result}) and
Eq.~(\ref{eq::FSresult}) with $f_A(z)=-f(z)$ and $f_P(z)=f(z)$, respectively.

\begin{table}[t]
\begin{center}
    \begin{tabular}{|c|c|c|c}
      \hline
       $\lambda$ &  $w^{(1)}_\lambda$   & $c^{(1)}_\lambda$ \\
      \hline
      a &  $ \eta_2 ( \eta_2 + 2) + \xi_2 ( \xi_2 -2 \eta_2+2) $  & $C_F $ \\
      b &  $-2\xi_1 \eta_1 $     &    $C_F $ \\
      c &  $2(\xi_1- \xi_2) (\eta_1- \eta_2)$     &    $ -C_A$ \\
      d &  $\eta_1(\eta_1  -2 \xi_2 +2)$     &    $-\frac{1}{2}C_A $ \\
      e &  $(\eta_2- \eta_1)(\eta_2+\eta_1 -2\xi_2 +2) $     &    $-\frac{1}{2}C_A $ \\
      f  &  2$\eta_1 (\xi_1- \xi_2) $     &    $\frac{1}{2}C_A $  \\
      g &  $ 2\eta_1 (\xi_1 - \xi_2) $     &    $ \frac{1}{2}C_A$ \\
      h &  $ \eta_2(\eta_2  -2\eta_2 \xi_2 +2)$     &    $ \frac{1}{2}C_A-C_F $ \\
      \hline
    \end{tabular}
\end{center}
 \caption{\label{tab::6} The weights $w^{(1)}_\lambda$ and  the color factors
 $c^{(1)}_\lambda$ for the diagrams in Fig.~\ref{fig::8}. The contributions of
 the diagrams with the opposite  direction of the closed quark line are
 included.}
\end{table}

\section{Summary and discussion}
\label{sec::5}
We have presented  the details of the first systematic analysis of the
high-energy asymptotic behaviour of the  QCD amplitudes beyond the
leading-power approximation and  derived  all-order double-logarithmic result
for the leading mass-suppressed terms in typical two-scale problems. In
contrast to the  Sudakov logarithms, the mass-suppressed double-logarithmic
corrections are induced by a soft quark exchange. The structure of the
corrections and the asymptotic behavior of the amplitudes in this case
crucially depend on the color flow in a given process and are determined by
the eikonal color charge nonconservation. After separating the standard
Sudakov factors the remaining non-Sudakov double-logarithmic corrections are
described by two universal functions $g(\pm z)$ and $f(\pm z)$,
Eqs.~(\ref{eq::gseries}) and (\ref{eq::f}), of the variable $z={\alpha_s\over
4\pi}(C_A-C_F)\ln^2(m_q^2/Q^2)$ for the processes with single and double soft
quark exchange, respectively.  These functions play the role of ``Sudakov
exponent'' for the non-Sudakov double-logarithmic corrections. They grow as
$e^{z/2}$  {\it i.e.} are exponentially {\it  enhanced} for large positive
values of the argument and are power suppressed for the large negative values.
Our result reveals  highly nontrivial relations  between the asymptotic
behavior of different amplitudes and the amplitudes in different gauge
theories.  In particular, if  a QCD amplitude gets the exponential
enhancement at high energy, an amplitude with the inverted  color charge flow
from the eikonal line defined by a scattering particle, or the same amplitude
in QED are   suppressed by a power of the large logarithm,\footnote{In some
cases such as a QED quark form factor this suppression can be  cancelled
by the energy dependence of the leading order result} and {\it vice versa}.

In general the amplitudes with larger number of scattering  particles, such
as Bhabha scattering in QED \cite{Penin:2016wiw}, get contributions from both
single and double soft fermion exchange.  The factorization structure in this
case can be more complex  and the corresponding asymptotic expressions may
involve new functions besides  $g(z)$ and $f(z)$.

One of the most interesting phenomenological applications of our analysis is
an estimate of the high-order corrections to the bottom quark mediated Higgs
boson production in gluon fusion, which is one of the  main sources of
uncertainty in the theoretical predictions for the Higgs cross section at the
Large Hadron Collider.  The effective expansion parameter in this case is
$\ln^2(m_b^2/m_H^2)\alpha_s \approx 40 \alpha_s$ rather than $\alpha_s$, and
the resummation of the double-logarithmic corrections is  mandatory for a
reliable theoretical estimate. From the result of Sect.~\ref{sec::3} we can
immediately get such an estimate for the exclusive Higgs boson production
cross section with a veto on the jet transverse momentum of the order of the
bottom quark mass. In this case the bottom quark loop induced interaction is
local with respect to the soft emission and therefore results in an overall
correction factor to the leading order cross section. The dominant
contribution is due to its interference with the top-loop mediated amplitudes
which reads
\begin{equation}
\delta\sigma_{gg\to H}=-3\rho
\ln^2\!\rho
\left(1+{z\over 6}+{z^2\over 45}+{z^3\over 420}+\ldots\right)\sigma_{gg\to H}\,,
\label{eq::Hsigma}
\end{equation}
where the series in $z$ is the Taylor expansion of the function $g(z)$. Up to
the  next-to-leading order the exact  dependence of  the cross section on the
bottom quark mass without the expansion in $\rho$  has been known  for a
while \cite{Spira:1995rr}. The next-to-next-to-leading ${\cal O}(z^2)$ term
in Eq.~(\ref{eq::Hsigma}) is new. Numerically for $m_H=125$~GeV, $m_b=5$~GeV,
and $\alpha_s(m_b)=0.21$  we get $z\approx 1.2$ which in general is not a
good expansion parameter.  However for this value of $z$ the above series
becomes $1+0.19+0.030+0.0037+\ldots$ and converges sufficiently fast. The
${\cal O}(z^2)$ term results in  about $0.6$\% decrease of the cross section,
which can be considered as an estimate of the bottom quark loop effect in the
next-to-next-to-leading order. An interesting and important problem is the
generalization of our result for more inclusive observables such as Higgs
plus jet production cross section and the Higgs boson transverse  momentum
distribution,   which can be significantly affected by the bottom quark
contribution. Only the abelian part of the corresponding double-logarithmic
corrections is known so far \cite{Melnikov:2016emg}.

\acknowledgments
The work of A.P. is supported in part by NSERC, Institute for Theoretical
Physics, ETH Z\"urich, and Perimeter Institute for Theoretical Physics. The
work of T.L. is supported by NSERC.

\end{document}